\documentclass[ba,preprint]{imsart}  

\usepackage[english]{babel}
\usepackage[T1]{fontenc}
\usepackage[latin9]{inputenc}
\usepackage{amsthm}
\usepackage{amsmath}
\usepackage{amssymb}
\usepackage{natbib}
\usepackage[colorlinks,citecolor=blue,urlcolor=blue,filecolor=blue,backref=page]{hyperref}
\usepackage{graphicx}
\usepackage{color}
\usepackage{array}
\usepackage{verbatim}
\usepackage{float}
\usepackage{units}
\usepackage{url}
\usepackage{multirow}
\usepackage{microtype}

\pubyear{2023}
\volume{TBA}
\issue{TBA}
\arxiv{2301.08276}
\firstpage{1}
\lastpage{1}

\startlocaldefs
\numberwithin{equation}{section}
\theoremstyle{plain}
\providecommand{\tabularnewline}{\\}
\theoremstyle{plain}
\newtheorem{thm}{Theorem}
\theoremstyle{plain}
\newtheorem{prop}[thm]{Proposition}
\theoremstyle{plain}
\newtheorem{lem}[thm]{Lemma}
\theoremstyle{plain}
\newtheorem{cor}[thm]{Corollary}
\theoremstyle{definition}
\newtheorem{defn}[thm]{Definition}
\DeclareMathOperator{\ARX}{\mathsf{ARX}\!}
\DeclareMathOperator{\eljpd}{eljpd}
\DeclareMathOperator{\elppd}{elppd}
\DeclareMathOperator{\elpd}{elpd}
\endlocaldefs

\begin{document}

\begin{frontmatter}
\title{Cross-validatory model selection for Bayesian autoregressions with exogenous regressors}
\runtitle{Cross-validation for Bayesian autoregressions}
\begin{aug}
    \author{\fnms{Alex} \snm{Cooper}\thanksref{mon_addr,t1}\ead[label=e1]{alexander.cooper@monash.edu}},
    \author{\fnms{Dan} \snm{Simpson}\thanksref{dan_addr}\ead[label=e2]{dp.simpson@gmail.com}}
    \author{\fnms{Lauren} \snm{Kennedy}\thanksref{mon_addr,adl_addr}\ead[label=e3]{lauren.kennedy@adelaide.edu.au}}
    \author{\fnms{Catherine} \snm{Forbes}\thanksref{mon_addr,t2}\ead[label=e4]{catherine.forbes@monash.edu}}
    \author{\fnms{Aki} \snm{Vehtari}\thanksref{aki_addr,t3}\ead[label=e5]{aki.vehtari@aalto.fi}}
\address[mon_addr]{Department of Econometrics and Business Statistics, Monash University, Clayton, Australia
\printead{e1}
\printead*{e4}}
\address[adl_addr]{School of Computer and Mathematical Sciences, University of Adelaide, Adelaide, Australia
\printead{e3}}
\address[aki_addr]{Department of Computer Science, Aalto University, Helsinki, Finland
\printead{e5}}
\address[dan_addr]{\printead{e2}}
\runauthor{A. Cooper et al.}
\end{aug}

\begin{abstract}
Bayesian cross-validation (CV) is a popular method for predictive model assessment
that is simple to implement and broadly applicable. A wide range of
CV schemes is available for time series applications, including generic
leave-one-out (LOO) and K-fold methods, as well as specialized approaches
intended to deal with serial dependence such as leave-future-out (LFO),
$h$-block, and $hv$-block.

Existing large-sample results show that both specialized and generic methods
are applicable to models of serially-dependent data. However, large sample consistency
results overlook the impact of sampling variability on accuracy in finite samples. 
Moreover, the accuracy of a CV scheme depends on many aspects of the procedure. 
We show that poor design choices can lead to elevated
rates of adverse selection.

In this paper, we consider the problem of identifying the regression
component of an important class of models of data with serial dependence, autoregressions of order $p$
with $q$ exogenous regressors ($\ARX(p,q)$), under the logarithmic scoring
rule. We show that when serial dependence is present, scores computed
using the joint (multivariate) density have lower variance and better
model selection accuracy than the popular pointwise estimator.
In addition, we present a detailed case study of the special case of $\ARX$
models with fixed autoregressive structure and variance. For this class, we
derive the finite-sample distribution of the CV estimators and the model
selection statistic. We conclude with recommendations for practitioners.
\end{abstract}



\end{frontmatter}

\section{Overview}

Many workflows for constructing predictive Bayesian models require
the practitioner to choose the best model among a number of candidates
according to their predictive power for the task at hand. Although many
predictive model selection methods are available \citep{Vehtari2012a},
among the most popular is cross-validation \citep[CV;][]{Geisser1975}. 
CV is flexible and applicable to a wide variety of statistical applications.

In finite samples, CV-based selection objectives are biased
and subject to sampling variation, which leads to uncertainty about
model predictive ability and the possibility of adverse model selection
\citep{Arlot2010,Sivula2020a}. In the Bayesian context, there is
a large literature on the frequency properties (i.e. variability
across multiple realizations of a dataset) of model selection
rules using information criteria such as the widely-available information
criterion (WAIC) and the Bayes factor \citep[e.g.,][]{Ward2008a,Schad2022}.
Despite its popularity in Bayesian
applications, however, less is known about the frequency
properties of dependent CV procedures for Bayesian model selection
under log-predictive loss.

Recent work by \citet{Sivula2020a} analyzed the frequency properties
of leave-one-out CV (LOO-CV) for Bayesian regression models of
exchangeable data. The authors identify
at least three scenarios that lead to elevated uncertainty in CV model
selection, and therefore to an increased probability of adverse model
choice. These pathological cases include comparisons between candidate models that
produce similar predictions, where models are badly misspecified,
and where training data sizes are small.

In this study, we extend the analysis of \citet{Sivula2020a} to Bayesian models of 
serially dependent data. We aim to characterize CV model selection uncertainty
for a simple but important class of models: autoregressions of order $p$ with $q$ exogenous
regressors, $\ARX(p,q)$. Our goal is to identify the regression
component of the model under the logarithmic scoring rule, leaving to one
side the related task of identifying the autoregressive component. In this context,
a scoring rule is a loss function for assessing the quality of probabilistic
predictions \citep{Gneiting2007}. While many scoring rules are available,
we focus on the logarithmic scoring rule, for which a measure of predictive
performance is the expected log predictive density (elpd) described in
Section~\ref{subsec:Model-selection}.

\begin{figure}
\begin{centering}
    \includegraphics[width=1.0\textwidth]{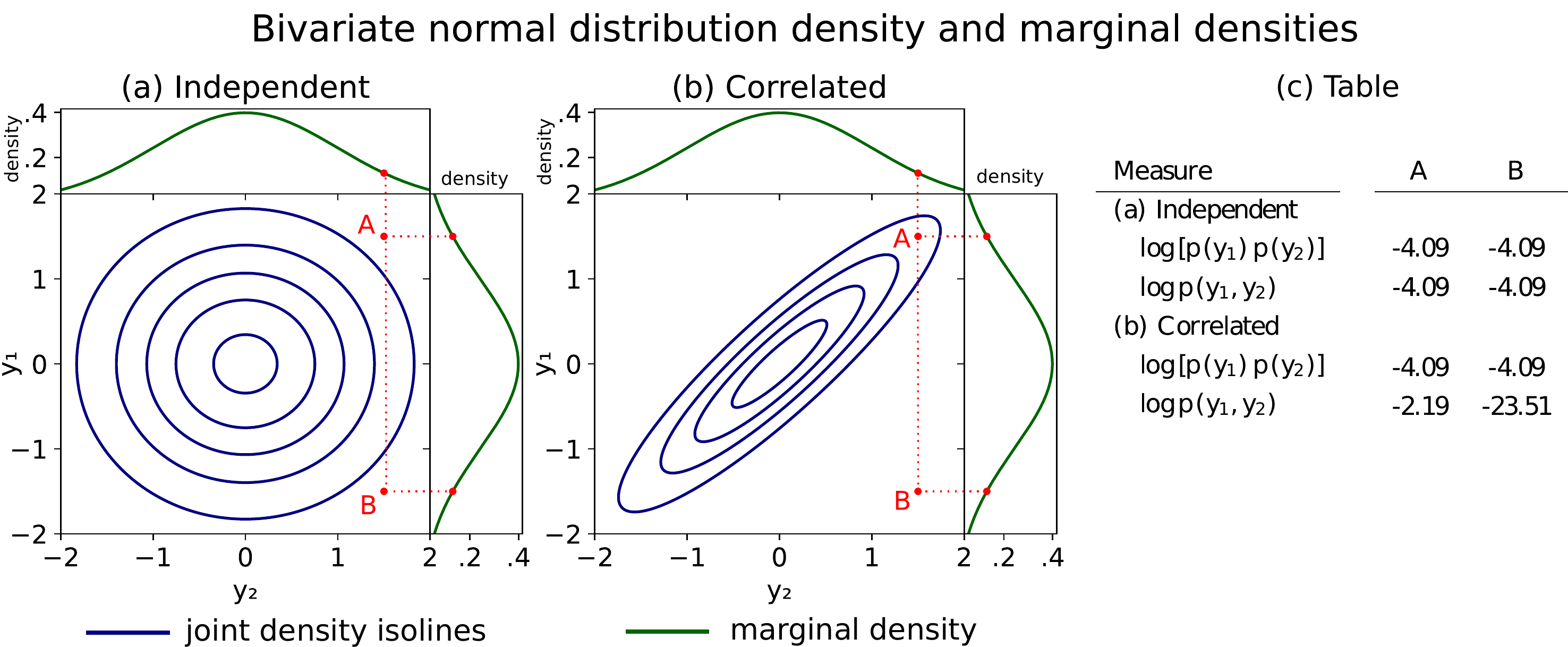}\\
\end{centering}
\caption{Illustration of the distinction between joint and pointwise log density 
measures in correlated models. Both plots show bivariate normal densities centered at the origin with unit marginal variance and correlation coefficients of 0 (Panel (a)) and 0.9 (Panel (b)). 
Panel (c) tabulates joint and pointwise log densities evaluated at the marked points A and B, indicated in red on the plots. In the correlated case, only the joint log density, $\log p\left(y_1, y_2\right)$,
identifies point B as having low log density. In contrast, the pointwise density $\log p\left(y_1, y_2\right)$ is the same for both.}
\label{fig:bivariate-normal}
\end{figure}

We address two important aspects of scoring rule design for models of correlated data. First, whether the scoring rule used for model assessment will be univariate or multivariate. Second, for multivariate scoring rules, whether it will be evaluated jointly (as a multivariate predictive density) or pointwise (as univariate marginal densities). We begin with a demonstration of the importance of the latter, showing improved statistical power of model selection with a jointly-evaluated scoring rule. We continue with a detailed case study of model selection under several popular CV schemes. This comparison includes several specific univariate (pointwise) methods,
and several joint methods. Throughout,
we find that joint methods achieve greater (statistical) efficiency,
measured as lower adverse selection rates,
and the associated CV estimators tend to have have lower variability.

Figure~\ref{fig:bivariate-normal} illustrates the importance of joint multivariate assessments
when data are correlated. The figure shows two bivariate normal distributions. In one
the variates are mutually independent, and in the other they are strongly correlated. 
The table shows that points A and B have identical pointwise log densities under both
distributions, even though point B lies in a region of very low (joint) density in the
correlated case. We conclude that in contrast to the joint approach, when correlation is 
strong the pointwise density fails to detect that point B is in a region of low probability for
this model.

\begin{figure}
\begin{centering}
\includegraphics[width=1\textwidth]{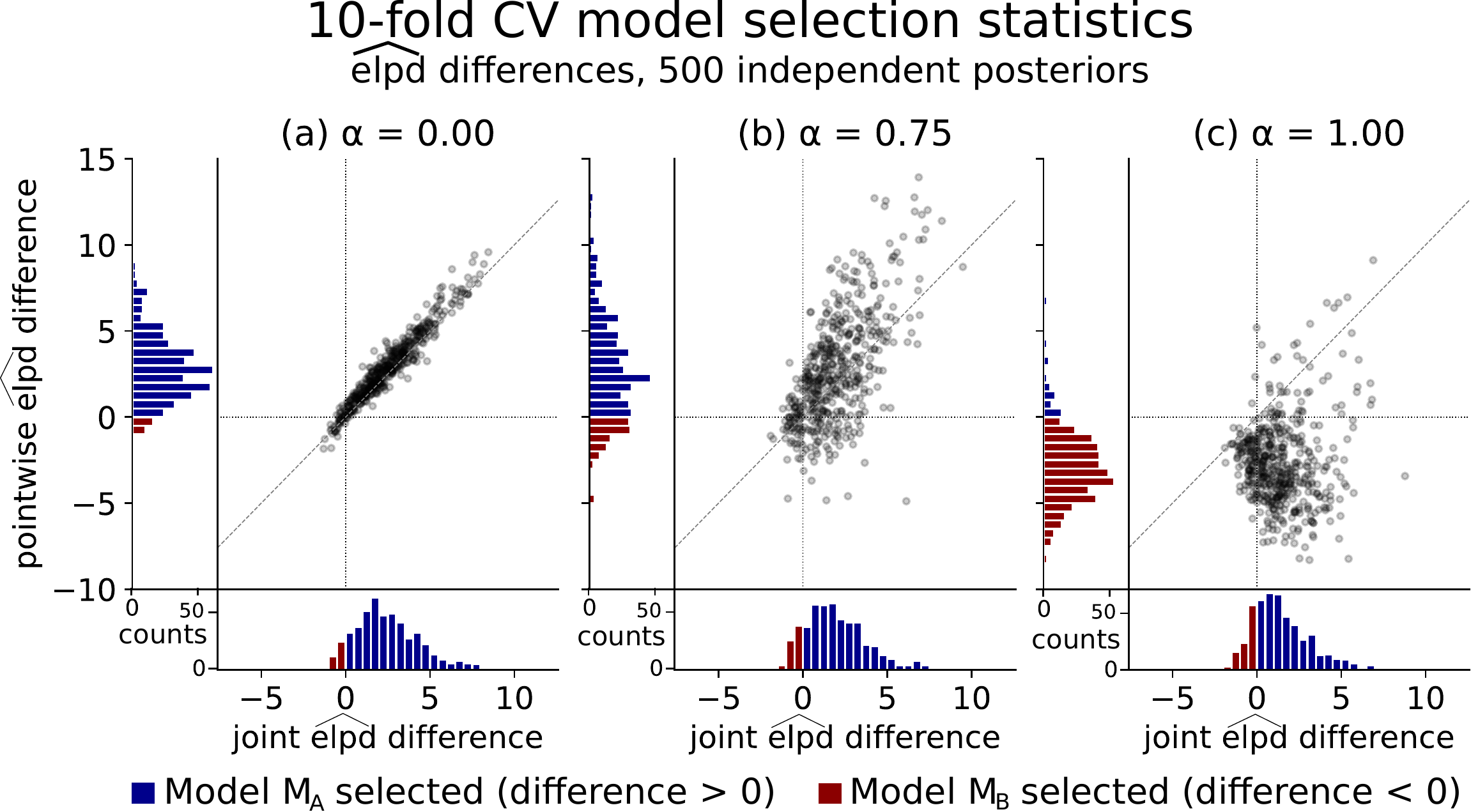}
\par\end{centering}
\caption{Joint log-score differences versus pointwise log score differences, 
computed using 10-fold-CV in a model selection statistics for 500 independent posteriors
of the `hard' case (see Section~\ref{sec:experiment}). The DGP is a stationary $\ARX\left(2,3\right)$
and candidate models are $M_A:\,\ARX\left(1,2\right)$
and $M_B:\,\ARX\left(1,1\right)$. Model selection statistics are expected log pointwise predictive density ($\elpd$)
differences. The DGP has autoregressive parameter $\phi_* = \alpha\left(0.75, 0.2\right)$ so that $\alpha$ selects increasing serial dependence from left to right. Adverse selection increases for the pointwise method (vertical axis) as dependence increases. See Section~\ref{sec:experiment} for
a full description.}
\label{fig:cv-differences}
\end{figure}

To further motivate our approach, Figure~\ref{fig:cv-differences} previews the 
results of a model selection experiment we 
will describe in more detail in Section~\ref{sec:experiment}. The figure compares
the distribution of CV model selection statistics for selecting between two candidate $\ARX$ models. The panels
show increasing degrees of dependence from left to right.
As dependence increases, the pointwise model selection statistic shows an increasing rate of adverse model selection (red bars in the vertical margin). In contrast, there is little change for the jointly-evaluated case (red bars in the horizontal margin). For a full description of this experiment, please refer to Section~\ref{sec:experiment}.

Several CV strategies are available for models of serially dependent data, and there seems to
be little agreement
in the literature about which one practitioners should adopt, especially in the Bayesian
modeling literature. Furthermore, much of the existing literature addresses different blocking
strategies, but does not make a distinction between joint and pointwise evaluation of the scoring rule.
Moreover, we speculate that different 
CV schemes will be useful when assessing different aspects of such a model. Even
when the analytical focus is not the autoregressive component,
joint predictive measures appear to be useful for identifying the regression
components.

In contrast to much of the existing literature on CV for autoregressive
models, including many large-sample consistency results \citep[e.g.,][]{Bergmeir2018,Racine_2000},
our emphasis is
on the frequency properties of the CV estimator in finite samples.
Further to the three problematic scenarios identified by \citet{Sivula2020a},
our results suggest that strong serial dependence and a cross-validatory
objective function that does not capture model dynamics (e.g. pointwise objectives) 
can pose difficulties
for CV-based model selection, even when the goal is not limited to identification of the
autoregressive component of the model. Our results stand as a counterpoint
to large-sample consistency results and suggest that mere consistency of the
estimator is not enough. That is, under certain choices of the CV
scheme the variance of the model selection statistic can be very high,
leading to elevated rates of adverse selection.

\subsection{Contributions}

We present novel results for procedures that use CV methods under the logarithmic scoring function when serial dependence is present. Working with the logarithmic scoring function and focusing on identifying the regression component of the model, we demonstrate that:
\begin{itemize}
\item Under serial dependence, CV schemes should be designed to account for the presumed dependence structure of the data in order to achieve good model selection performance;
\item When serial dependence is strong, performance measures evaluated jointly are much more (statistically) efficient than pointwise counterparts;
\item When the sample size is finite, there is a U-shaped relationship between certain CV scheme hyperparameters and the adverse selection rate;
\item We present novel results on the variability of Bayesian CV procedures for $\ARX(p,q)$ models.  To our knowledge, these are among the first results describing the finite-sample uncertainty of CV methods under serial dependence, particularly in a Bayesian setting.
\end{itemize}
We offer the following advice for practitioners working with models of serially-dependent data. Broadly speaking, since CV methods based on joint scoring rules are usually more 
complex to implement, their improved efficiency should be traded off against 
implementation burden.
Following model criticism of each candidate model ahead of model selection, we recommend:
\begin{itemize}
\item Where measured serial dependence in the data is not very strong, simpler pointwise CV methods (like LOO-CV and LFO-CV) can be used as a first-pass, and relied upon where the results are clear \citep[see][for criteria]{Sivula2020a};
\item Otherwise, if serial dependence is strong or results are unclear then joint CV model selection methods should be implemented instead.
\item Even when the actual predictive task requires a univariate prediction (like a one-step-ahead prediction), for model selection it may be better to use a CV scheme that leaves out multiple observations, combined with a multivariate scoring rule.
\end{itemize}
The remainder of the paper proceeds as follows. In Section~\ref{sec:bg}
we describe the model class and summarize CV-based model selection
and some relevant literature, highlighting some key challenges associated
with CV for dependent data. Section~\ref{sec:experiment} presents a short simulation experiment, and Section~\ref{sec:sel-dep} presents a
detailed case study of CV model selection in a simplified form of
$\ARX$ model, focusing on the properties of CV under dependence
and demonstrating where challenges can arise. Finally, Section~\ref{sec:disc-conc}
discusses the results and concludes. See \url{https://github.com/kuperov/arx}
for code and experiments.
 
\section{Background\label{sec:bg}}

In this section, we briefly review CV model selection and review some
relevant literature. We will suppose we have observed a data vector
$y=\left(y_{1},\dots,y_{T}\right)$, presumed to be drawn from
a joint distribution $p_{\mathrm{true}}$, the (typically unknown) data-generating process
(DGP). Our goal is to construct predictions by first selecting the
best available model $M^{*}$ from some set $\mathcal{M}$ of candidates
(or candidate model families identified up to a parameter). This selection
is made according to candidate models' ability to predict as-yet
unseen realizations of the process, that is, by their out-of-sample predictive
power. 

To simplify our analysis, we will consider only pairwise comparisons
(i.e. $\left|\mathcal{M}\right|=2$), but we do explicitly allow that
$M_{\mathrm{true}}\not\in\mathcal{M}$, i.e. the model associated with $p_{\mathrm{true}}$ is not in $\mathcal{M}$.

\subsection{Autoregressions with exogenous regressors\label{subsec:ARX}}

We will write $\ARX\left(p,q\right)$ for an autoregression with $p$
lags of the dependent variable and $q$ exogenous regressors. Throughout
we assume $p\ll T$ and $q\ll T$, and that $T$ is `small', corresponding to common applied settings where data are limited.

The $\ARX\left(p,q\right)$ class is a key building block for time
series models in a wide range of scientific, policy, and business
applications. For instance, autoregressions underpin the popular vector
autoregression (VAR) models used by macroeconomic policymakers \citep[e.g.,][]{Sims1980}
and spatial epidemiological studies \citep{Lee2011}.

The $\ARX\left(p,q\right)$ model is conditionally normal,
\begin{align}
p\left(y_{t}|\phi,\beta,y_{t-1},\dots,y_{t-p}\right) & =\mathcal{N}\left(y_{t}|\phi_{1}y_{t-1}+\cdots+\phi_{p}y_{t-p}+z_{t}^{\top}\beta,\sigma^{2}\right),
\label{eq:ar:def:scalar}
\end{align}
for $t=1,\dots,T$, where the first element of the $q\times 1$ vector of exogenous variables $z_{t}$ is $1$. For simplicity,
we will initialize the sequence from zero, so that $y_{1-p}=\cdots=y_{0}=0$.

In comparison to the linear regression (LR) model studied by \citet{Sivula2020a},
the dependence structure of the $\ARX$ class substantially complicates the analysis.
Naturally, one could view the lags of $y_{t}$ as explanatory variables which would mean the $\ARX\left(p,q\right)$ model is identical to that of the Gaussian
linear regression (LR) analyzed by \citet{Sivula2020a}. However,
to restrict our attention to models in the stationary regime, we must either
impose informative priors on the autoregressive parameters (as we do in Section~\ref{sec:experiment})
or fix them (as we do in Section~\ref{sec:sel-dep}).
In both Sections\ \ref{sec:experiment} and \ref{sec:sel-dep} we 
we have allowed analytical convenience to guide the choice of prior.

It is worth emphasizing that all the results we present in this paper
depend on $Z$, the $T\times q$ matrix of exogenous covariates. Our
results do not need to make any assumptions about the distribution
of $Z$, since they are assumed known and fixed. In our experiments,
we construct $Z$ by drawing a matrix of independent standard normal
variates, a matrix we keep fixed across all replicates of each experiment.

\subsection{Predictive model selection\label{subsec:Model-selection}}

When the goal of a modeling exercise is prediction, it is natural
to use predictive performance as a measure of model goodness or
`utility'. Predictive
performance can be assessed using a scoring rule, a function that produces
a numerical assessment of a probabilistic prediction against actual
observations \citep{Gneiting2007}. Since the choice of scoring rule
governs the selection of $M^{*}$, an ideal choice for a scoring
rule would be tailored to the modeling task at hand. However, in the absence of a specific
application, general-purpose scoring rules are available.

We focus on the popular logarithmic scoring rule, which enjoys the
mathematical properties of being local and strictly proper and is
closely related to the KL divergence \citep{Gneiting2007,Vehtari2012a,Dawid1984}.
Under the logarithmic score, we call the expected score for some model
$M_{\ell}$ the \emph{expected log joint predictive density }(eljpd),
\begin{align}
\eljpd\left(M_{\ell}|y\right) & =\mathbb{E}_{\tilde{y}\sim p_{\mathrm{true}}}\left[\log\int p\left(\tilde{y}|\theta,M_{\ell}\right)p\left(\theta\,|\,y,M_{\ell}\right)\,\mathrm{d}\theta\right]\label{eq:elpjd-expanded}\\
 & =\mathbb{E}_{\tilde{y}\sim p_{\mathrm{true}}}\left[\log p\left(\tilde{y}|y,M_{\ell}\right)\right]\label{eq:elpjd}
\end{align}
where `joint' refers to the fact that the multivariate predictive
$p\left(\tilde{y}|y,M_{\ell}\right)$ is a joint density. Here, the $T\times 1$ random
variable $\tilde y$ is independent of the data $y$.

If $p_{\mathrm{true}}$ were known or unlimited independent replicates $\tilde{y}\sim p_{\mathrm{true}}$ were available so that \eqref{eq:elpjd-expanded} could be evaluated, the utility-maximizing model $M^{*}$ could
be selected by `external validation' \citep{Gelman2014}
of the model $M_{\ell}$ joint predictive $p\left(\cdot|y,M_{\ell}\right)$
with respect to $p_{\mathrm{true}}$,
\begin{equation}
M^{*}:=\arg\max_{M_{\ell}\in\mathcal{M}}\eljpd\left(M_{\ell}|y\right)=\arg\max_{M_{\ell}\in\mathcal{M}}\mathbb{E}_{\tilde{y}\sim p_{\mathrm{true}}}\left[\log p\left(\tilde{y}|y,M_{\ell}\right)\right].\label{eq:util-max}
\end{equation}
In many cases it is computationally convenient to compute (\ref{eq:elpjd})
in a pointwise fashion, which yields the expected log \emph{pointwise}
predictive density (elppd),
\begin{align}
\elppd\left(M_{\ell}|y\right) & =\mathbb{E}_{\tilde{y}\sim p_{\mathrm{true}}}\left[\log\prod_{t=1}^{T}\int p\left(\tilde{y}_{t}|\theta\right)p\left(\theta\,|\,y,M_{\ell}\right)\,\mathrm{d}\theta\right]\\
 & =\mathbb{E}_{\tilde{y}\sim p_{\mathrm{true}}}\left[\sum_{t=1}^{T}\log p\left(\tilde{y}_{t}|y,M_{\ell}\right)\right],\label{eq:elppd}
\end{align}
The pointwise predictive $p\left(\tilde{y}_t\,|\,\theta,M_\ell \right)$ that
appears in \eqref{eq:elppd} is simply the multivariate predictive with all
but one $\tilde{y}$ element marginalized out,
\begin{equation}
p\left(\tilde{y}_{t}|y,M_{\ell}\right)=\int\!\cdots\!\int p\left(\tilde{y}|y,M_{\ell}\right)\,\prod_{s\neq t}\mathrm{d}\tilde{y}_{s}.
\end{equation}

The resulting utility measure for model $M_{\ell}\in\mathcal{M}$ given
observed data $y$ can be computed using the model joint predictive
density. We will often want to discuss both classes of expected predictive
densities in a generic sense, in which case we will use the umbrella
term \emph{expected log predictive density }(elpd).

We adopt (\ref{eq:util-max}) as our benchmark for the preferred model.
From this perspective, the pointwise density (elppd) is useful
to the extent that it is a computationally convenient approximation
of the joint density (eljpd). It is important to note that while 
elppd and eljpd are both useful for making comparisons against similarly-constructed
measures, they are fundamentally different quantities. See, for instance, \citet{Madiman2010} for inequalities between joint and pointwise densities.

When observations are conditionally independent
given global model parameters, it is often the case that the $\elppd$
and $\eljpd$ are close or even identical. However, under serial
and other forms of dependence, this is rarely the case because the
eljpd captures additional information about serial dependence of the
observations not reflected by the pointwise measure.

Unfortunately, the expected utility maximization framework described above suffers
a crucial drawback: $p_{\mathrm{true}}$ is rarely ever known
in practice, and
thus the elpd must be estimated purely from observed data. While one
might be tempted to simply substitute $p\left(y\,|\,y,M_{\ell}\right)$
into (\ref{eq:elpjd}) or (\ref{eq:elppd}), this will lead to a positively
biased (over-optimistic) estimate due to model overfit \citep{Vehtari2012a,Gelman2014}.
Instead, we need a method for estimating elppd and eljpd using only the available data.

\subsection{Cross-validation\label{subsec:cv}}

CV is a method for estimating the $\textnormal{elpd}$ purely from
observed data by data splitting and repeated re-fits of the model.
Suppose for a moment that independent replicates of the data
$\tilde{y}^{(s)}\sim p_{\mathrm{true}}$, $s=1,\dots,S$,
were available, and the predictive were able to be evaluated pointwise. Then
the utility (\ref{eq:elppd}) under model $M_{\ell}$ could be targeted
by the following Monte Carlo estimator,
\begin{equation}\widehat{\elppd}\left(M_{\ell}|y\right)
=\frac{1}{S}\sum_{s=1}^{S}\sum_{t=1}^{T}\left[\log p\left(\tilde{y}_{t}^{(s)}|y,M_{\ell}\right)\right].
\end{equation}

In applications where such replicates are unavailable, CV estimators
exploit the fact that the data $y$ are distributed according to 
$p_{\textnormal{true}},$ even if $p_{\textnormal{true}}$
is itself unknown. CV proceeds by repeatedly splitting the data into
disjoint testing and training data subsets, estimating the model on
the training set, then constructing an estimator using
pointwise predictions for the testing set. The CV estimator for $\elppd\left(M_{\ell}|y\right)$,
which divides $y$ into $K$ test sets, can be defined as
\begin{align}
\widehat{\elppd}_{CV}\left(M_{\ell}\,|\,y\right) & =\frac{T}{K}\sum_{k=1}^{K}\frac{1}{\left|\mathsf{test}_{k}\right|}\sum_{t\in\mathsf{test}_{k}}\log p\left(y_{t}\,|\,y_{\mathsf{train}_{k}},M_{\ell}\right),\label{eq:def:cv}
\end{align}
where $\mathsf{test}_{k}$ denotes the subset of $y$ to be evaluated
under the predictive, $\mathsf{train}_{k}$ denotes the subset of
$y$ to be used to train the data. The scaling factors normalize the measure to `sum scale'.
The corresponding joint measure is given by
\begin{align}
\widehat{\eljpd}_{CV}\left(M_{\ell}\,|\,y\right) & =\frac{T}{K}\sum_{k=1}^{K}\frac{1}{\left|\mathsf{test}_{k}\right|}\log p\left(y_{\mathsf{test}_{k}}\,|\,y_{\mathsf{train}_{k}},M_{\ell}\right).\label{eq:def:jcv}
\end{align}

We stress that CV schemes with multivariate test sets, like
$hv$-block and $K$-block, can be evaluated in either a joint
or pointwise fashion. In comparison, univariate schemes like
LOO can only be evaluated pointwise.

The CV scheme blocking design is fully described by the triple $\left(K,\left\{ \mathsf{test}_{k}\right\} _{k=1}^{K},\right.$ 
$\left.\left\{ \mathsf{train}_{k}\right\} _{k=1}^{K}\right)$.
Classic LOO, for instance, has $K=T$,
$\mathsf{test}_{k}=\left\{ k\right\} $ and $\mathsf{train}_{k}$
includes all but the $k$th element.

Model selection using cross-validation selects the model with the
greatest estimated utility---or at least the simplest model similar
to the best model. For a pairwise comparison between $\mathcal{M}=\left\{ M_{A},M_{B}\right\} $,
the CV estimate of the utility-maximizing objective is the sign of
the difference
\begin{align}
\widehat{\elppd}_{CV}\left(M_{A},M_{B}\,|\,y\right) & =\widehat{\elppd}_{CV}\left(M_{A}\,|\,y\right)-\widehat{\elppd}_{CV}\left(M_{B}\,|\,y\right).\label{eq:def:emelpd}
\end{align}
We have omitted from this formulation of the model selection objective
the bias correction term that is sometimes included to account for
the fact that there are fewer elements in the training set for each
CV fold than in the full-data posterior. Typically a first-order correction
is used, and it is usually very small \citep[see][]{Gelman2014}.

Under correct model specification, the summands in the CV estimator
(\ref{eq:def:cv}) will usually be weakly correlated. Under relatively
mild regularity conditions $\widehat{\elppd}_{CV}$ should converge
to the expected utility $\elppd$ as $T$ grows large \citep[see][for an analysis of LOO-CV, for instance]{Bergmeir2018}.

\subsection{CV schemes for serial dependence\label{subsec:blocking}}

In models of cross-sectional data where all observations can be assumed
conditionally independent, the data structure imposes relatively few
constraints on the sequence of training and test sets used for CV.

\begin{figure}
\begin{centering}
\includegraphics[width=0.8\textwidth]{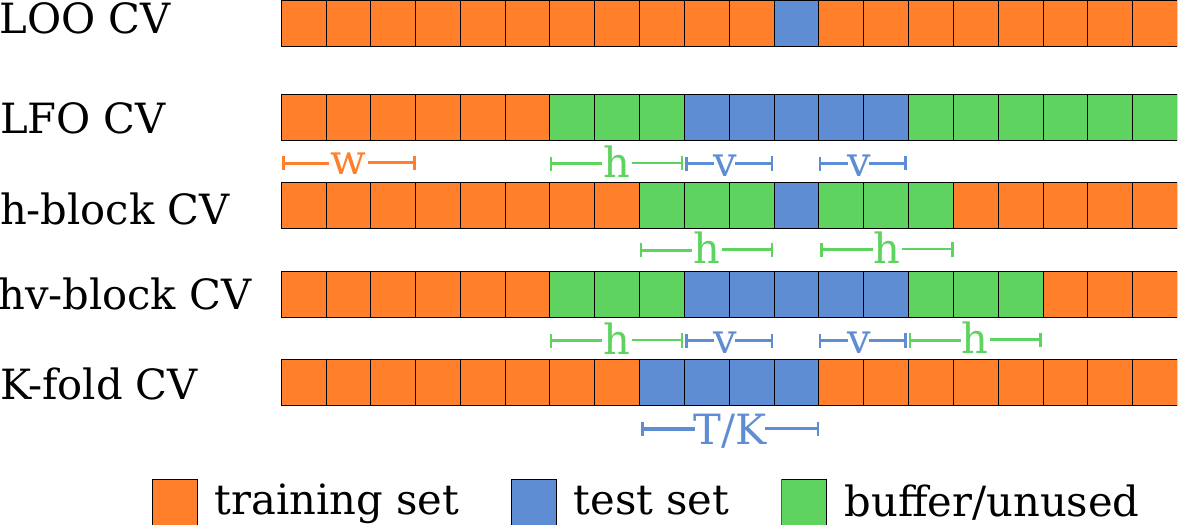}
\par\end{centering}
\caption{Cross-validation blocking schemes described in Section~(\ref{subsec:blocking})
for a sequence of length $T=20$. The various schemes have hyperparameters
$h=3$ (h- and hv-block CV, LFO), $v=2$ (hv-block CV and LFO), $K=5$ (K-fold
CV), and $w=3$ (LFO).}
\label{fig:blocking}
\end{figure}

Under serial dependence, care is needed to ensure the contributions
to (\ref{eq:elppd}) are mutually independent, or at least as independent as we can make them. To this
end, a number of CV schemes have been developed specifically for models of serially dependent data.

A key consideration when selecting a CV scheme is the nature of the
intended prediction task, for instance, whether the model will be used
for one-step-ahead or $M$-step-ahead predictions.

Existing analyses of these schemes
refer to specific contexts that do not necessarily align
with our Bayesian framework. Most use different scoring functions
and all but a few are analyzed with reference to classical models
that yield point predictions. For our purpose, what we take from this
earlier work is the design of the blocking scheme, i.e. the choices
of $\left(K,\left\{ \mathsf{test}_{k}\right\} _{k=1}^{K},\left\{ \mathsf{train}_{k}\right\} _{k=1}^{K}\right)$.
These are summarized below and illustrated in Figure~\ref{fig:blocking}.

\citet{Burman1994} present h-block CV, an adaptation of CV for stationary
dependent sequences. In order to nearly eliminate the bias arising
from dependence between train and test sets in LOO-CV, their procedure
deletes a buffer of size $h$ around the training set, while retaining
just a single test observation (see Figure~\ref{fig:blocking}). This reduces the size of the training
set by $2h$ observations, but still leaves a total of $n$ test sets.
They propose $h$ be a fixed proportion of the data length. Although
this is `conservative' (the authors refer to \citet{Gyorfi1989},
whose results allow consistency if $h/T\longrightarrow0$ so long
as certain conditions on the underlying dependence structure are met),
they argue this is appropriate because in practice the dependence
structure is typically unknown. Since each $h$-block fold has a single
test element, it is by definition a pointwise CV framework.

\citet{Racine_2000} proposes $hv$-block CV as an extension of $h$-block
CV, which increases the test set dimension from $1$ to $2v+1$. The
author claims this provides selection consistency in a wider range
of circumstances, including nested models, which may be of interest in
the case where model identification is the goal. Since $hv$-block
has a multivariate test set, it can be evaluated jointly.

Another blocking scheme specific to serially dependent data is leave-future-out
\citep[LFO-CV;][]{Burkner2020}. LFO-CV trains the model only on past
observations, starting with a warmup period $0<w\ll T$, and leaves
future observations unused. To be comparable to the other methods,
our implementation of LFO mirrors the structure of $hv$-block CV (Figure~\ref{fig:blocking}). That is, we write $\mathrm{LFO}\left(h,v,w\right)$
for a LFO scheme that includes a halo $h\geq0$, size parameter $v$, and
initial buffer $w$. This generalized form of LFO contains
the usual formulation, which is $\mathrm{LFO}(0, 0, w)$.

We have also included $K$-fold CV. This method is not specific to
serial dependence, although it is commonly applied under serial dependence
in the literature \citep{Cerqueira2020,Bergmeir2012,Bergmeir2018}.
We use a variant of $K$-fold CV that partitions the sample into $K$ contiguous sub-blocks each roughly
of size $T/K$. Typical values for $K$ are 5 or 10. The predictive
may be evaluated jointly or pointwise, depending on the context.

\section{A model selection experiment\label{sec:experiment}}

In this section, we illustrate the behavior of CV model selection under
serial dependence by repeatedly performing a model selection experiment
on simulated data. We have chosen this experiment because we believe it is illustrative
of general behaviors of CV for autoregressive models. We use a sequence of experiments, 
where we control the degree of serial dependence.

Consider the following model selection problem.
Let the vector $y=y_{1},\dots,y_{T}$ be distributed according to
an $\ARX\left(2,3\right)$ of the following form,
\begin{equation}
\textnormal{DGP}:\ y_{t}=\alpha\begin{pmatrix}0.75\\
0.2
\end{pmatrix}^{\top}\begin{pmatrix}y_{t-1}\\
y_{t-2}
\end{pmatrix}+\beta_{*1}+\beta_{*2}z_{2t}+\beta_{*3}z_{3t}+\sigma_{*}\varepsilon_{t},\label{eq:ex:dgp-1}
\end{equation}
where $\varepsilon_{t}\overset{\mathrm{iid}}{\sim}\mathcal{N}\left(0,1\right)$.
The fixed parameter $\alpha\in\left[0,1\right]$ allows us to select
the degree of serial dependence. When $\alpha=0$ the observations
are mutually independent, and $\alpha=1$ generates a highly persistent series for the model in (\ref{eq:ex:dgp-1}).
All $\alpha\in[0,1]$ generate stationary series. While the upper bound $\alpha=1$ is arbitrary, corresponding to $\phi_*=(0.75,0.2)$, it is a useful upper limit for our experiments that generates a persistent but nonetheless stationary series. We observe similar results for other choices of the upper bound and ratio between elements of $\phi$. For simplicity, we fix initial conditions $y_{t}=0$ for $t\leq0$.

The experiment selects between two candidate models:
\begin{align}
M_{A}:\ y_{t} & =\phi_{1}y_{t-1}+\beta_{1}+\beta_{2}z_{2t}+\varepsilon_{t} & \ARX\left(1,2\right)\label{eq:ex:mA-1}\\
M_{B}:\ y_{t} & =\phi_{1}y_{t-1}+\beta_{1}+\varepsilon_{t} & \ARX\left(1,1\right).
\end{align}

We choose the analytically-convenient (although non-conjugate) prior,
\begin{equation}
\beta|\sigma^{2}\sim\mathcal{N}\left(\beta_*,\sigma^{2}I_p\right),\quad\sigma^{2}\sim\mathcal{IG}\left(a_{0},b_{0}\right),\quad\phi\sim\mathcal{BE}_{(-1,1)}\left(c_{0},d_{0}\right),\label{eq:full-prior}
\end{equation}
where $a_0=b_0=c_0=d_0=1$. $\mathcal{IG}$ denotes the inverse-gamma density and $\mathcal{BE}_{(-1,1)}$ the beta distribution scaled to have support on $(-1,1)$.

This model is `fully Bayesian' is the sense that we regard all
three parameters ($\beta$, $\sigma^{2}$, and $\phi$) as unknown,
and we allow them all to be estimated. For computational tractability,
we have chosen conjugate priors for $\beta$ and $\sigma^{2}$, and
we will conduct inference only on stationary $\ARX\left(1,q\right)$
models. In this special case, $\phi$ is univariate with support on
the interval $\left(-1,1\right)$. We center the $\beta$ prior on the 
truth $\beta_*$ to avoid distortions as the effective sample
size changes with $\alpha$.

Both candidate models are `misspecified' in the sense that neither
has the same functional form as (\ref{eq:ex:dgp-1}). However, while
both models omit the second $y_{t}$ lag and effect $\beta_{3}$,
the candidate $M_{A}$ is nonetheless the better model in the sense that it is closer in KL divergence to the DGP. This is because it
includes $\beta_{2}$, which is also omitted by $M_{B}$.

We will work with two vectors of true DGP parameters $\beta_{*}$,
distinguished by the relative ease with which CV is able to select
the better model in our experiments:
\[
\textnormal{`easy' case: }\beta_{*}^{\textnormal{easy}}=\left(1,2,1\right),\qquad\textnormal{`hard' case: }\beta_{*}^{\textnormal{hard}}=\left(1,\frac{1}{2},1\right).
\]

These arbitrary parameter values were chosen for convenience in the
context of our experiments and simulated covariates. $\beta_{*}^{\textnormal{easy}}$
is an example of the case where CV has little difficulty separating
$M_{A}$ and $M_{B}$ under logarithmic loss, and under $\beta_{*}^{\textnormal{hard}}$
model identification by CV is much more challenging. Naturally, the
relative difficulty of $\beta_{*}^{\textnormal{easy}}$ and $\beta_{*}^{\textnormal{hard}}$
depends on a range of factors, including the covariates $z_{it}$,
noise variance, and data length. This setup does not make any assumptions
about the distribution of the matrix $Z$ with elements $z_{it}$, other than that it is known. In
our simulations, we have drawn $z_{it}\overset{\mathrm{iid}}{\sim}\mathcal{N}\left(0,1\right)$,
which remain fixed throughout.

Although the posteriors associated with the above candidate models
have no closed form, we are
able to estimate them relatively computationally cheaply using one-dimensional
quadrature or MCMC; see Appendix~\ref{sec:complete} for the complete
computational details. The availability of efficient estimation procedures
is important for our experiments because we perform CV by brute
force for each simulation draw. That is, we avoid 
computational shortcuts like importance sampling to ensure our
results are not being driven by approximation error.

Figure~\ref{fig:cv-differences} summarizes the results of this model
selection experiment. It plots model selection statistic for 500 
independent simulated data sets for
the `hard' model variant, comparing model selection statistics
from two variants of 10-fold CV. The vertical axis shows the CV objective
evaluated pointwise, that is $\widehat{\elppd}_{\mathrm{CV}}\left(M_A,M_B | y\right)$,
and the horizontal axis shows the objective evaluated jointly, that is
$\widehat{\eljpd}_{\mathrm{CV}}\left(M_A,M_B | y\right)$. Adverse selection for
both measures is indicated by negative selection statistics, that is
those that would selection $M_B$. Points lying in the first quadrant,
for instance, represent correct selection by both joint and pointwise
methods.

Some stylized facts are evident in the Figure~\ref{fig:cv-differences}.
First, when data are mutually independent (the dependence parameter $\alpha=0$, left panel) the
marginal distribution and relative performance of both methods is
approximately equal. However, as $\alpha\to 1$ we see that the variance of the
pointwise estimates grows sharply and the location of the distribution shifts
in a negative direction, indicating a sharp increase in the adverse selection rate.
In contrast, the joint estimates are little changed as $\alpha$ varies.
Further note that many LOO estimates fall in the fourth quadrant,
indicating that the wrong model would be selected in those cases.

An equivalent experiment with the `easy' parameter variant (not
shown) displays a similar increase in the variability of the pointwise model
selection statistic, but because the entire distribution lies far
enough from the x-axis there is no appreciable increase in the adverse
selection rate.

\begin{figure}
\begin{centering}
\includegraphics[width=1\textwidth]{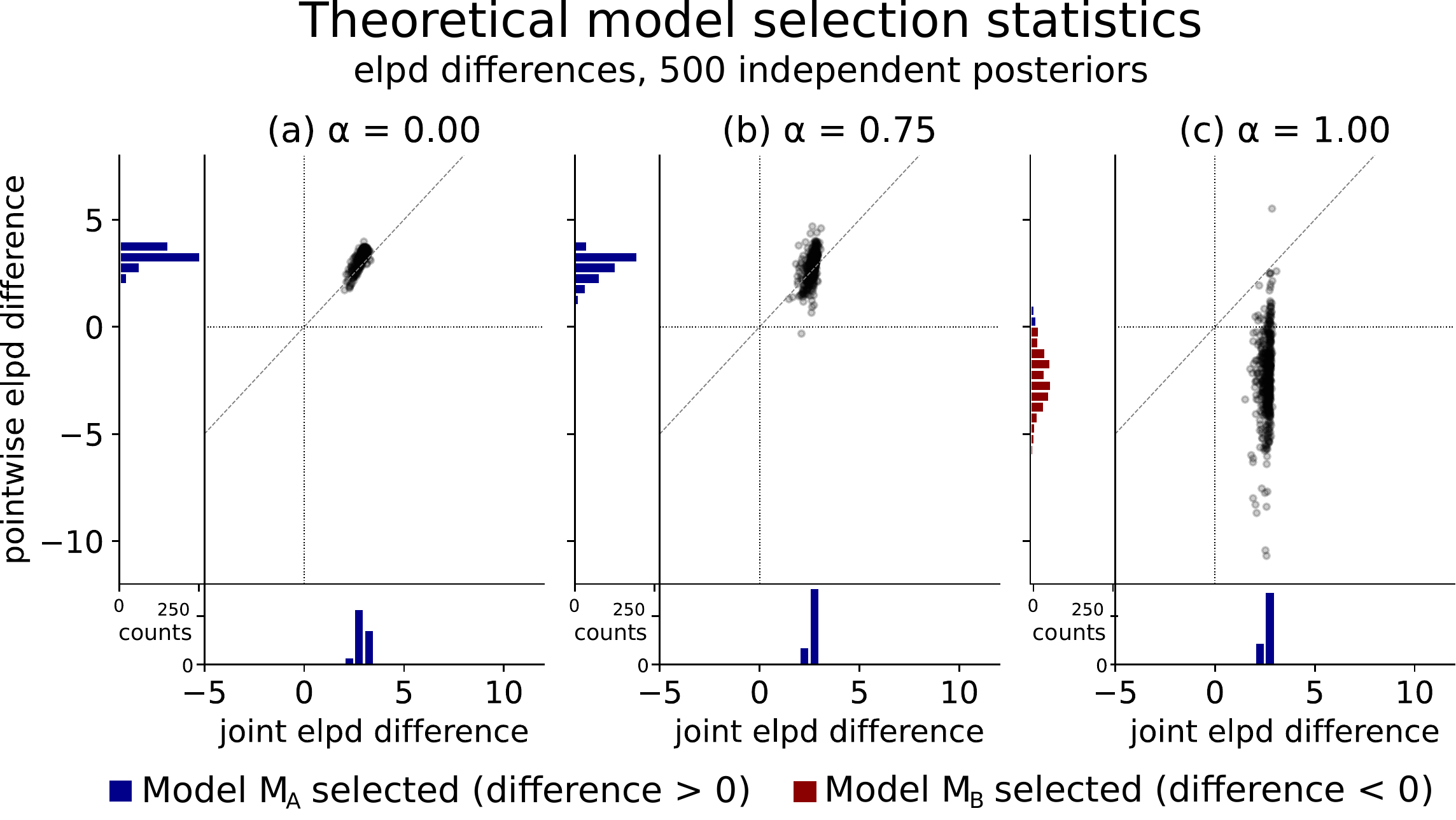}
\par\end{centering}
\caption{Joint and pointwise theoretical model selection statistics under log
score for 500 independent posteriors, as dependence increases from
left to right. This is the `hard'
example described in Section~\ref{sec:sel-dep}: the DGP is an $\ARX\left(2,3\right)$
and the candidates are $M_A:\,\ARX\left(1,2\right)$
and $M_B:\,\ARX\left(1,1\right)$. Positive values select the $M_A$,
the better model. The gray line is the 45 degree
line where both estimators are equal. Marginal histograms are also shown, with red bars indicating adverse selection.}
\label{fig:theoretical-differences}
\end{figure}

Figure~\ref{fig:theoretical-differences} offers one explanation
for the rising variance of the pointwise CV statistic. It plots
the true elppd and eljpd for the 500 posteriors in Figure~\ref{fig:cv-differences}, i.e. 
the underlying quantities that the joint and pointwise CV estimators
are targeting. The change in distributions evident in both figures
is quite similar, suggesting
that it is a difference between the underlying quantities, rather than
some problem with the CV estimators under serial dependence, that
is driving the differences between joint and pointwise 10-fold CV.

\section{Detailed case study\label{sec:sel-dep}}

In this section, we focus on the problem of identifying
the regression component within the $\ARX$ class. We will work with
a simplified version of the $\ARX\left(p,q\right)$, where we regard
only the regression parameter $\beta$ as random, with a Gaussian
prior centered on the truth, $\beta\sim\mathcal{N}\left(\beta_*,\Sigma_{0}\right)$.
This simplification allows us to focus our attention narrowly on the
task of identifying $\beta$. It also makes available analytical expressions
for the posterior distribution, as well as the distribution of the
theoretical $\elppd$ and $\eljpd$, associated CV estimators, and
model selection statistics.

This approach circumvents a key challenge associated with analyzing
the variability of CV procedures: in general there are no closed-form
expressions for the variance of elpd measures \citep{bengio2004}.
However, when $\phi$ and $\sigma^{2}$ are fixed, a closed form does
exist. The existence of closed form expressions allows us to derive
the finite-sample distribution of the elpd for all of the CV procedures
we consider.

\subsection{Utility measures and model selection statistic}

In this simple case, we can obtain exact distributions for the joint
$\eljpd$ and pointwise $\elppd$ for the $\ARX$ class with fixed
$\phi_{*}$ and $\sigma_{*}^{2}$, as well as their associated CV
estimators. The results in this section extend earlier results for the $\elppd$ and
LOO-CV derived by \citet{Sivula2020a}, in the case of i.i.d. Gaussian
linear regressions with fixed variance.

For arbitrary $\ARX$ models $M_A$ and $M_B$, we show that the stochastic
variables $\eljpd\left(M_{A}\,|\,y\right)$ and
$\widehat{\eljpd}_{\text{CV}}\left(M_{A}\,|\,y\right)$ (for all of
the CV schemes listed in Subsection~\ref{subsec:blocking}), as well
as the model selection criteria $\eljpd\left(M_{A},M_{B}\,|\,y\right)$ and  $\widehat{\eljpd}_{\text{CV}}\left(M_{A},M_{B}\,|\,y\right)$,
all have generalized $\chi^{2}$ distributions with parameters that
depend on parameters of the DGP, the posited model, and the exogenous
covariates.

Following the setup in Section~\ref{sec:experiment}, suppose that $y$
is distributed as $\ARX\left(p_{*},q_{*}\right)$ as described above, and suppose
an experimenter posits a candidate $\ARX\left(p_{\ell},q_{\ell}\right)$
model $M_{\ell}$ for $y$. This and other results are proven in Appendix~\ref{app:Proofs}.

\begin{prop}[Quadratic polynomial form of utility measures]
\label{prop:sarx:elpd:gchisq}Let $y$ be distributed according to
an $\ARX\left(p_{*},q_{*}\right)$ process, and let $M_{\ell}$ be the
simplified $\ARX\left(p_{\ell},q_{\ell}\right)$ model described in Section~\ref{subsec:ARX}.
Then the theoretical pointwise and joint measures $\eljpd\left(M_{\ell}|y\right)$
and $\elppd\left(M_{\ell}|y\right)$ respectively defined in \eqref{eq:elpjd}
and \eqref{eq:elppd}, as well as the corresponding CV estimates $\widehat{\eljpd}_{\mathrm{CV}}\left(M_{\ell}|y\right)$
and $\widehat{\elppd}_{\mathrm{CV}}\left(M_{\ell}|y\right)$, can be
expressed as second-degree vector polynomials in $y$,
\begin{equation}
\omega_{\ell}\left(y\right)=y^{\top}A_{\ell}y+y^{\top}b_{\ell}+c_{\ell},\label{eq:poly}
\end{equation}
for nonrandom coefficients $A_{\ell}$ (a $T\times T$ matrix), $b_{\ell}$
(a $T$-vector), and scalar $c_{\ell}$. The coefficients are functions
of $\phi^{\left(\ell\right)},\sigma_{\ell}^{2},Z_{\ell}$, and the CV blocking scheme parameters. All
are defined in Appendix~\ref{sec:sarx}.
\end{prop}

A reviewer pointed out the similarity between the quadratic polynomial form of \eqref{eq:poly} and that of the likelihood ratio test for non-nested models by \citet{Vuong1989}. Although the latter's results
are frequentist and asymptotic in character, the similarity is
nonetheless interesting considering that the intent of these test statistics
are so similar.

\subsection{\textquoteleft Oracle\textquoteright{} plug-in values for fixed parameters\label{subsec:oracle}}

Our candidate models require appropriate choices for the noise variance
parameter $\sigma^{2}$ and autoregressive parameter $\phi$. Especially
when the model is misspecified, it would not necessarily be optimal to
use the true DGP value $\sigma_{\ell}^{2}=\sigma_{*}^{2}$, in the sense
that this choice would not produce
the best possible predictions with respect to log score for the chosen
model class.

Inference requires suitable choices for $\sigma_{\ell}^{2}$ and $\phi^{(\ell)}$ that would correspond as closely as possible
to the behavior of an inference procedure where $\phi$ and $\sigma^2$ are unknown.

Suppose some hypothetical Oracle happens to know the true DGP, and offers to select
the best-performing autoregressive and variance parameters $\phi_{\ell}$
and $\sigma_{\ell}^{2}$ for our particular model and covariates. Naturally,
this choice will be independent of any specific realization of $y$
since we are interested in utility distributions across all potential
values of $y$. Consider two approaches our Oracle
might use for selecting these parameters. She might choose to minimize
the distance (in KL divergence) between the DGP and the model K predictive.
Alternatively, she could directly target the objective function by maximizing the achievable $\mathbb{E}\left[\mathrm{elpd}\left(M_{\ell}\,|\, y\right)\right]$.

Under the logarithmic loss function, it follows from \citet{Dawid1984} that these
two options represent equivalent calculations.
That is, maximizing the loss function,
\begin{equation}
\left(\widehat{\phi_{\ell}},\widehat{\sigma_{\ell}^{2}}\right) :=\arg\max_{\begin{array}{c}
\sigma_{\ell}^{2}\in\mathbb{R}_{+}\\
\phi\in\Phi_{\ell}
\end{array}}\mathbb{E}\left[\mathrm{elpd}\left(M_{\ell}\,|\,y\right)\right].\label{eq:sarx:maxelpd}
\end{equation}
and minimizing the expected KL divergence between the DGP and model predictive,
\begin{equation}\left(\widehat{\phi_{\ell}},\widehat{\sigma_{\ell}^{2}}\right) :=\arg\min_{\begin{array}{c}
\sigma_{\ell}^{2}\in\mathbb{R}_{+}\\
\phi_{\ell}\in\Phi_{\ell}
\end{array}}\mathbb{E}\left[\mathbb{D}\left(p_{\mathrm{true}}\left(\tilde{y}\right)\,\|\,p\left(\tilde{y}\,|\,y,\sigma_{\ell}^{2},\phi^{(\ell)},M_{\ell}\right)\right)\right],\label{eq:sarx:minkld}
\end{equation}
yield the same answer. In the above $\Phi_\ell\subset\mathbb{R}^{p_\ell}$ denotes the  
parameter space for $\phi$ associated with stationary $\ARX\left(p_\ell,\cdot\right)$ models.
In our experiments, we solve the optimization (\ref{eq:sarx:minkld})
using the Nelder-Mead algorithm.

\subsection{Distribution of the model selection statistic}

The purpose of conducting CV on the candidate models is to determine which
candidate has the greatest predictive performance, in our case under log score.
The CV model selection statistic (\ref{eq:def:emelpd}) is the difference
between the estimated scores of the two models.

The form of the model selection statistic used in this experiment is characterized by 
the following corollary, implied by Proposition~\ref{prop:sarx:elpd:gchisq}.

\begin{cor}[Form of model selection objectives and CV estimates]
\label{cor:sarx:elpd:elpd:quadratic}Let $y$ be distributed according
to an $\ARX\left(p_{*},q_{*}\right)$ process, and let both $M_{A}$
and $M_{B}$ be simplified $\ARX\left(p_{A},q_{A}\right)$ and
simplified $\ARX\left(p_{B},q_{B}\right)$, respectively. Then
the theoretical model selection statistics $\eljpd\left(M_{A},M_{B}\,|\,y\right)$
and $\elppd\left(M_{A},M_{B}\,|\,y\right)$, and their corresponding
CV estimators $\widehat{\eljpd}_{\mathrm{CV}}\left(M_{A},M_{B}\,|\,y\right)$
and $\widehat{\elppd}_{\mathrm{CV}}\left(M_{A},M_{B}\,|\,y\right)$,
can be expressed as second-degree polynomials in $y$.
\end{cor}

Proposition~\ref{prop:sarx:elpd:gchisq}
and Corollary~\ref{cor:sarx:elpd:elpd:quadratic} imply that all of the
quantities of interest follow generalized $\chi^{2}$ distributions
(see Definition~\ref{def:gchisq} in Appendix~\ref{subsec:dist-fun}).
Proposition~\ref{prop:dist-omega} describes the mean and variance, and
further states the parameters of this distribution. The associated distribution
function $F_{\omega\left(y\right)}\left(w\right)=\Pr\left(\omega\left(y\right)<w\right)$ 
must be approximated numerically. This can be done by simulation or 
the method of \citet{Davies1973}.

\begin{prop}[Distribution of $\omega\left(y\right)$]
\label{prop:dist-omega}Let $\omega\left(y\right)$ be a quadratic
polynomial in $y$ with coefficients $A$, $b$, and $c$ as described
in Proposition~\ref{prop:sarx:elpd:gchisq} and Corollary~\ref{cor:sarx:elpd:elpd:quadratic},
where $A=A^{\top}$. Then $\omega\left(y\right)$ has mean and covariance:
\begin{align}
\mathbb{E}\left[\omega\left(y\right)\right] & =\sigma_{*}^{2}\mathrm{tr}\!\left(AV_{*}\right)+m_{*}^{\top}Am_{*}+b^{\top}m_{*}+c,\\
\mathrm{var}\!\left(\omega\left(y\right)\right) & =2\sigma_{*}^{4}\mathrm{tr}\!\left(A^{2}V_{*}^{2}\right)+\sigma_{*}^{2}b^{\top}V_{*}b+\sigma_{*}^{2}4b^{\top}V_{*}Am_{*}+4\sigma_{*}^{2}m_{*}^{\top}AV_{*}Am_{*},
\end{align}
for a fixed $T$-vector $m_*$ and fixed $T\times T$ matrix $V_*$. 
Moreover, it has a generalized $\chi^{2}$ distribution,
\[
\omega\left(y\right)\sim\tilde{\chi}^{2}\!\left(\boldsymbol{\lambda},\boldsymbol{1},\boldsymbol{\delta},\mu,\sigma\right),
\]
where \textbf{$\boldsymbol{\lambda}$ }is the vector of $k\leq T$
nonzero eigenvalues in the eigendecomposition $Q\Lambda Q^{-1}$ of
$\sigma_{*}^{2}L_{\phi_{*}}^{-1}AL_{\phi_{*}}^{-\top}$,
$\boldsymbol{1}$ is a $k$-vector of ones, $\boldsymbol{\delta}$
is the $k$-vector with elements $\delta_{j}=\tilde{b}_{j}/\left(2\lambda_{j}\right)$,
where $\tilde{b}=QL_{\phi_{*}}^{-\top}\left(2\sigma_{*}^{2}Am_{*}+\sigma_{*}b\right)$,
$\mu=m_{*}^{\top}Am_{*}+\sigma_{*}m_{*}^{\top}b+c-\frac{1}{4}\sum_{j=1}^{k}\tilde{b}_{j}^{2}\lambda_{j}^{-2}$,
and $\sigma^{2}=\sum_{j=k+1}^{T}\tilde{b}_{j}^{2}$ if $T>k$, or 0 otherwise.
\end{prop}

\begin{figure}
\begin{centering}
\includegraphics[width=1\textwidth]{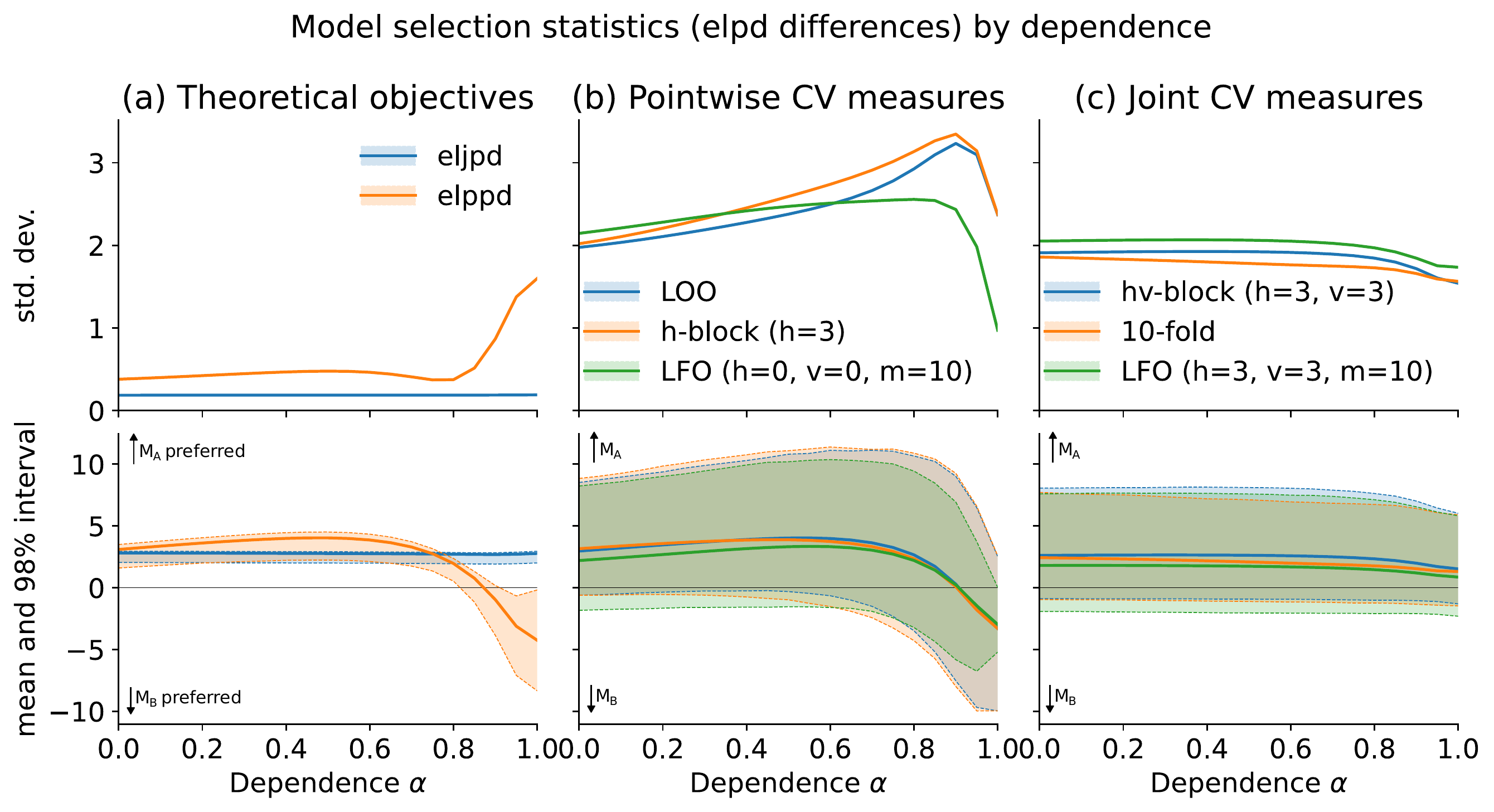}
\par\end{centering}
\caption{Model selection objectives for the `hard' case ($\beta_*=\beta^{\textnormal{hard}}$). Column (a) shows the theoretical model selection objectives,
and columns (b) and (c) the associated CV estimates. The top row plots the standard deviation of the relevant $\omega(y)$ for the corresponding column. The bottom row shows its mean and 98\% interval. The model parameter $\alpha$
governs the degree of serial dependence. Notice that the adverse selection rate for
both joint and pointwise methods is close to zero for all but the strongest dependence.
This model selection experiment compares $M_{A}:\ARX\left(1,2\right)$
vs $M_{B}:\ARX\left(1,1\right)$ under an $\ARX\left(2,3\right)$ DGP,
as described in Section~\ref{sec:sel-dep}. The autoregressive parameter is $\phi_{*}=\alpha\left(0.75,0.2\right)$,
and data length $T$=100. See also Figure~\ref{fig:by-alpha-easy} for the `easy' case.}
\label{fig:by-alpha}
\end{figure}

Several interesting features of the distribution of $\omega(y)$ are evident
in Figure~\ref{fig:by-alpha}, which summarizes the results for the `hard'
($\beta_{*}=\beta_{*}^{\mathrm{hard}}$) case under increasing dependence $\alpha$.
These features are consistent with the findings of Section~\ref{sec:experiment}.
First, notice the striking
difference between the behavior of the pointwise and joint theoretical elpds as $\alpha$ increases (column (a)).
When data are mutually independent ($\alpha=0$), both distributions are basically equal.
For the joint measure, variability and location are little changed as dependence increases.
In contrast, the pointwise
objective exhibits sharply rising variability and shifts in a negative direction as dependence
gains strength. Most of the distribution changes signs entirely to favor the simpler, worse
candidate $M_B$ as $\alpha$ approaches 1, indicating an adverse selection rate near 100\%.

Second, the different profiles exhibited by pointwise and joint CV methods (columns (b) and
(c), respectively). While there are clearly differences between the specific joint and pointwise methods,
whether the method is computed jointly or pointwise clearly has the greatest bearing on its behavior as dependence
increases. The pointwise methods in column (b) exhibit a similar increase in variability and negative
shift as the pointwise objective, while the joint methods are little changed as $\alpha\to 1$.
In addition, the pointwise CV methods (joint methods too, to a lesser extent) display an interesting decrease in
variability as $\alpha$ approaches 1.
A possible explanation for this drop-off is the decrease in effective sample size \citep{Berger2014}
for autoregressive models as dependence grows, all else being equal.

Under stronger dependence, increased variability and downward movement together reduce model
selection power, consistent with the known bias of CV procedures
toward simpler models under small sample sizes. See, for instance
\citet{Burman1989} for an analysis of CV bias and sample size.
In contrast, for the `easy' experiment variant ($\beta_{*}=\beta_{*}^{\mathrm{easy}}$) where the
results are much clearer,
cross-validation generates correct model selections for all but the most strongly dependent data
(see Figure~\ref{fig:by-alpha-easy} in Appendix~\ref{sec:supp-plots}).

\subsection{The cost of an inefficient CV scheme}

The distribution of the model selection statistic in 
Proposition~\ref{prop:dist-omega} provides a direct method for computing the
probability of adverse selection. Noting that a positive selection
statistic indicates the correct model choice (corresponding to $M_{A}$),
it follows that
\begin{equation}
\Pr\left(\textnormal{adverse selection}\right)=\Pr\left(\omega\left(y\right)<0\right)=F_{\omega\left(y\right)}\!\left(0\right).\label{eq:pr-adv-sel}
\end{equation}
The quantity $F_{\omega(y)}\!\left(0\right)$ can be interpreted visually in Figure~\ref{fig:by-alpha}
as the share of the $\omega\left(y\right)$ distribution that falls
below the $x$-axis. Where the probability of adverse selection is
very small we say that the models are \emph{well separated} under
that particular CV scheme. While the infinite support of $\omega\left(y\right)$
means that (\ref{eq:pr-adv-sel}) can never be zero, for
practical purposes we define a small threshold $\gamma$ as the cutoff
for well-separatedness. For the remainder of this paper, we will use
$\gamma=0.01$.
\begin{defn}[Well-separated]
We say the CV model selection procedure defined above is well-separated at level
$\gamma\in\left(0,1\right)$ when $F_{\omega\left(y\right)}\!\left(0\right)<\gamma$.
\end{defn}

A particular CV scheme may be well-separated in one situation and poorly
separated in another. Whether a model selection procedure is well-separated
is determined by all aspects of that procedure, including the details
of the data generating process, candidate models, any hyperparameters
for the procedure, and the values of the exogenous covariates.

We have seen that an inappropriate choice of CV scheme can result
in an elevated probability of adverse selection. In this section, we
attempt to quantify this cost.

Perhaps the simplest way to measure the cost of any model selection
procedure is the rate of adverse model selection, also known as the
`0-1 loss' because it scores all errors equally. The adverse selection
rate is the probability (with respect to repeated samples) that the
selection procedure will select the wrong model.

\begin{figure}
\begin{centering}
\includegraphics[width=1.0\textwidth]{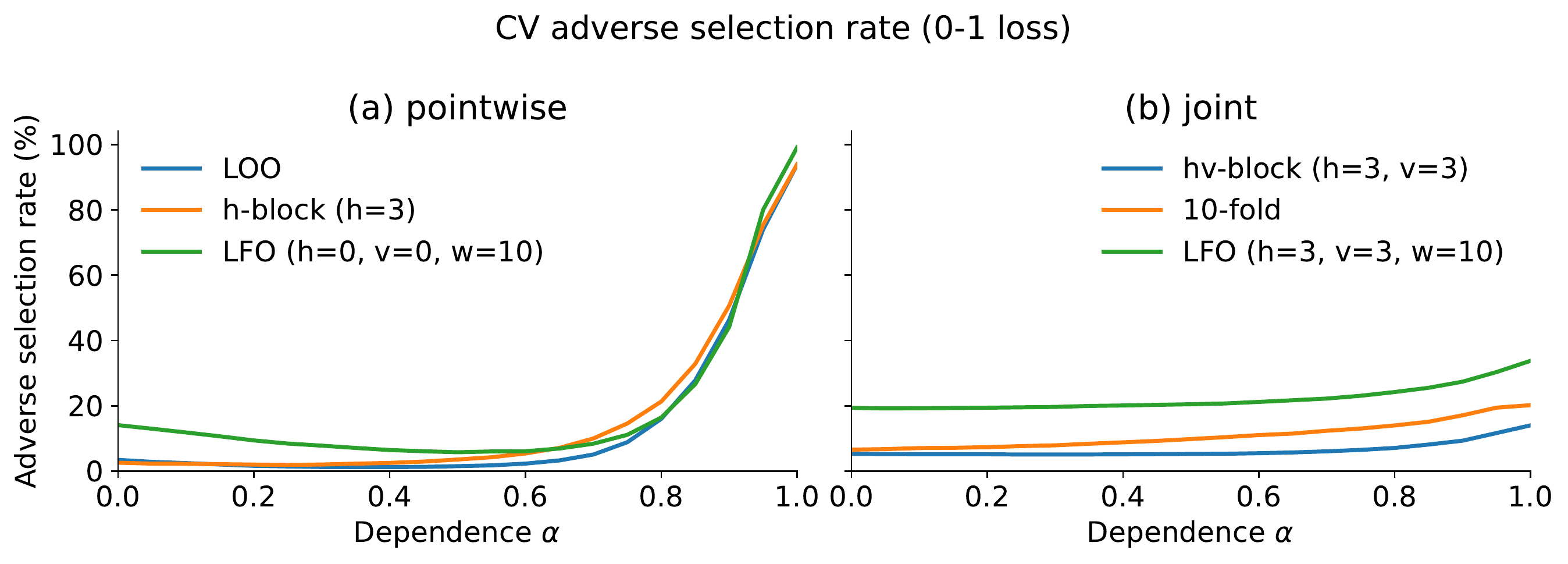}
\par\end{centering}
\caption{Adverse selection rate for pointwise and joint CV methods as data dependence
increases. This experiment compares $M_{A}:\,\ARX\left(1,2\right)$
vs $M_{B}:\,\ARX\left(1,1\right)$, under an $\ARX\left(2,3\right)$
DGP with $\phi_{*}=\alpha\left(0.75,0.2\right)$, for $\alpha\in\left[0,1\right]$
as described in Section~\ref{sec:sel-dep}. Greater values of $\alpha$
denote stronger serial dependence. The \textquoteleft hard\textquoteright{}
case ($\beta^{\textnormal{hard}}$) with $T=100$ is shown.}
\label{fig:loss}
\end{figure}

Framing the loss as a probability over all realizations of $y$ makes
good sense for this paper, since our focus is on the properties of
CV methods for $\ARX$ models \emph{in general}, without reference
to a particular data realization $y$. Panels (a) and (b) of Figure~\ref{fig:loss}
compare the adverse selection rate for joint and pointwise CV methods
for the `hard' variant of our model selection experiment (losses
for the `easy' variant are negligible, and are not shown). The adverse
selection rate picks up as $\alpha\to1$ and for pointwise
procedures reaches almost 100 per cent, indicating that CV incorrectly
prefers the simpler model when dependence is very strong.

The adverse selection rate overlooks a key fact, however: it does
not account for the severity of the prediction error, scoring all
incorrect selections equally. When there is very little difference
between candidates' model predictions, it matters little which model
is chosen. On the other hand, when predictions differ significantly
this should be reflected in the cost of the error.

An alternative measure of the cost of adverse selection is the reduction
in log utility that results from choosing the incorrect model for
a given \textbf{$y\sim p_{\mathrm{true}}$}. Under this formulation,
we use the underlying finite-sample theoretical elpd \emph{for each
$y$} that CV is trying to estimate as the benchmark. This is by its
nature a finite-sample loss function.

That is, in the case where CV does not select the elpd-maximizing
model we can regard the CV utility cost relative to the counterfactual
where the correct model was chosen as the difference between the chosen
and maximal elpd:
\begin{equation}
\mathrm{cost}_{CV}\left(y\right)=\max_{M\in\mathcal{M}}\mathrm{elpd}\left(M|y\right)-\mathrm{elpd}\left(M_{CV}^{*}|y\right),
\end{equation}
where $M_{CV}^{*}=\arg\max_{M\in\mathcal{M}}\widehat{\mathrm{elpd}}_{CV}\left(M|y\right).$

Pointwise elppd and joint eljpd are of course not directly comparable.
To put joint and pointwise CV measures on an equal footing, we specify
the cost measure measured in terms of joint utility for both pointwise
and joint CV procedures,
\begin{equation}
\widetilde{\mathrm{cost}}_{CV}\left(y\right)
=\max_{M\in\mathcal{M}}\eljpd\left(M|y\right)-\eljpd\left(M_{CV}^{*}|y\right),
\end{equation}
where $M_{CV}^{*}=\arg\max_{M\in\mathcal{M}}\widehat{\mathrm{elpd}}_{CV}\left(M|y\right)$
and $\widehat{\mathrm{elpd}}_{CV}\left(\cdot\right)$ can represent
either the joint estimate $\widehat{\eljpd}_{CV}\left(\cdot\right)$
or the pointwise estimate $\widehat{\elppd}_{CV}\left(\cdot\right)$.

Figure~\ref{fig:log-loss} in Appendix~\ref{sec:supp-plots} plots the log loss defined
in the previous display. While excess log loss
is an attractive concept, the resulting relative
measures are practically indistinguishable from the adverse selection
rate. For the remainder of this paper, we will use the adverse selection rate.

\subsection{Joint and pointwise objectives}

Our results suggest that under dependence, joint estimators
usually have lower variability and consequently a lower rate of adverse
selection than their pointwise counterparts. In our experiments, these
differences are typically negligible for independent observations
($\alpha=0$) and are most pronounced as $\alpha$ approaches 1 and
the underlying data become highly persistent.
This comparison is quite evident in Figure~\ref{fig:loss}: when
$\alpha=0$, joint and pointwise estimators perform roughly equally.
As $\alpha\to1$, there is little change in performance for joint
estimators, compared with a dramatic increase in the error rate for the
pointwise estimators.

To construct an apples-to-apples comparison between joint and pointwise
CV methods, abstracting from other details of the CV scheme design,
we construct pointwise analogs to joint designs. For the present experiment
this amounts to diagonalizing the covariance matrix of
the Gaussian predictive distribution, setting $\sigma_{\ell}^{2}V_{\ell}^{pw}=\sigma_{\ell}^{2}\left(I_{T}\odot V_{\ell}\right)$,
for $\odot$ the elementwise product operator. The results confirm
a sharp increase in the error rate for pointwise CV methods under
strong dependence ($\alpha=1$), compared with almost no difference
in the independent case ($\alpha=0$). (See Figure~\ref{fig:joint-pointwise}
in Appendix~\ref{sec:supp-plots}.)

\subsection{Specific CV scheme design considerations\label{subsec:hyperparam}}

We have seen that CV design parameters can have a significant bearing
on the overall efficiency, and therefore performance, of CV model
selection. In this section we look more closely at hyperparameter
choices for specific CV schemes for time-series data.

Note that $hv$-block CV schemes require the choice of a validation block size
$v$ (the total validation set dimension is $2v+1$) and halo size
$h$. Consistent with earlier results, for dependent data we find
that by far the most important choice is to ensure that $v$ is large
enough to capture the dynamic behavior of the model. 
Either parameter can harm CV performance if it is too large,
since both reduce training set size, which imposes
a cost on statistical efficiency, resulting in a U-shaped
relationship between the adverse selection rate and both $h$ and $v$
when dependence is present. (Figure~\ref{fig:halo-dimension-variance} in
Appendix~\ref{sec:supp-plots} compares error rates with various choices
of these parameters.)

The importance of preserving the size of the training set is evident
in the underperformance of LFO, which represents an extreme case for
dealing with contamination by discarding the entire future sample.
Several authors have pointed out that this is unnecessarily conservative
\citep{Bergmeir2018}. Moreover, the analyst should carefully weigh the tradeoff between the bias resulting from the use of future data and the benefit of increasing the training set size. See Figure~\ref{fig:hvblock-lfo} in Appendix~\ref{sec:supp-plots}
for a comparison between LFO and methods that use future data.
In each case, the adverse selection rate for LFO is substantially higher,
a consequence of LFO's reduced training set size.

\subsection{Required sample size\label{subsec:samplesize}}

One important practical consequence of an inefficient CV scheme design
is that a larger sample size is needed for the models under consideration
to be well-separated. This either requires the experimenter to collect
more data than is necessary or for the adverse selection rate to be
higher than it otherwise would be. In this section we demonstrate
that the required sample size increases with dependence, all else
being equal. When the underlying process is highly persistent, the
required sample size can be significantly larger than for the independent
case.

The required sample size for a model to be well-separated goes beyond
the well-known principle of the effective sample size (ESS) for a time
series model \citep[see e.g.][]{Berger2014}. In this section we
demonstrate that properties of the CV procedure---especially whether the
scoring rule is evaluated joint or pointwise---strongly determines the
data length required.

\begin{figure}
\begin{centering}
\includegraphics[width=1.0\textwidth]{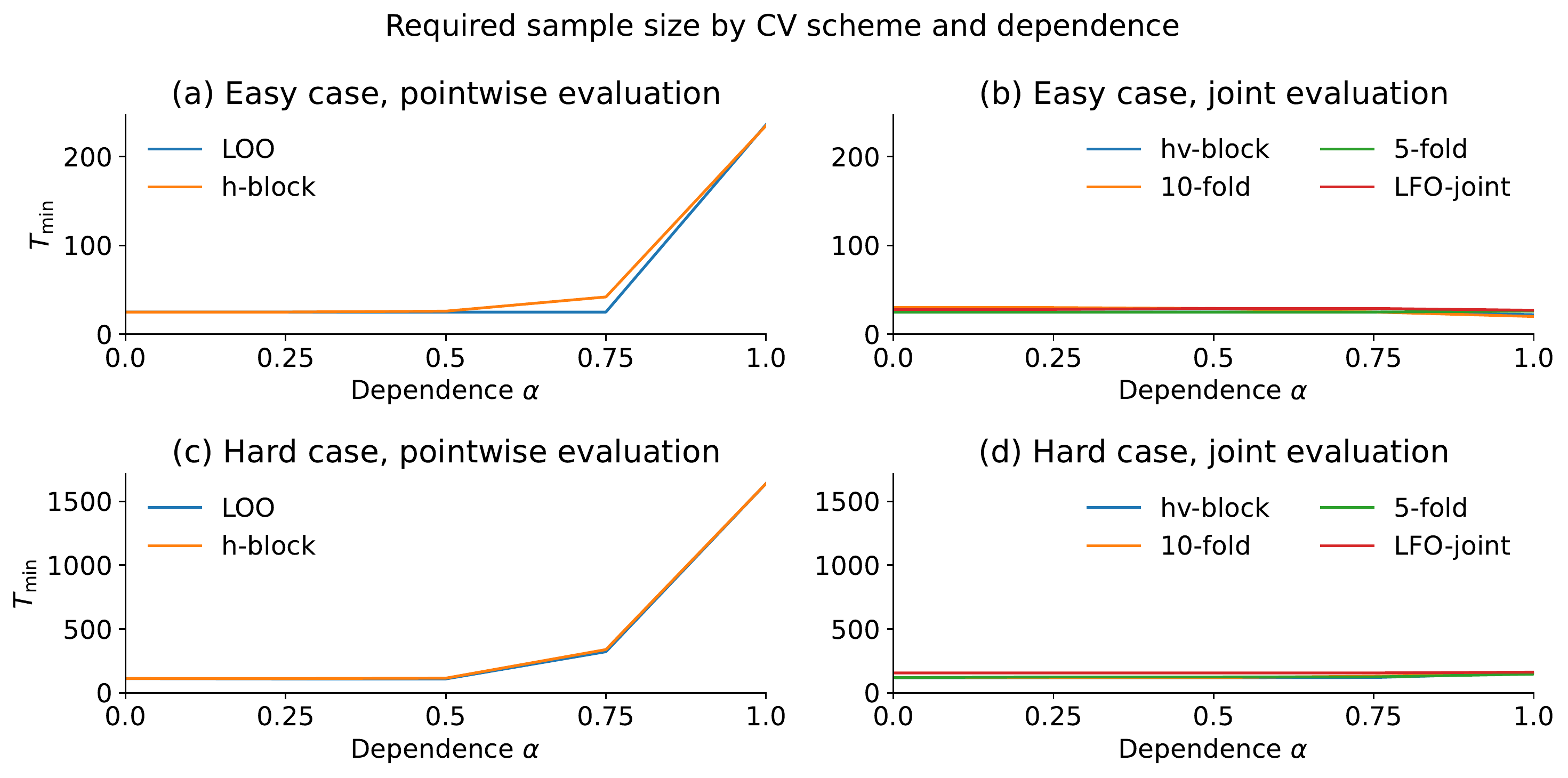}
\par\end{centering}
\caption{Minimum data length $T$ to be well-separated at the 1\% level, found
by binary search over the range 10-2500. This model selection experiment
compares $M_{A}:\ARX\left(1,2\right)$ vs $M_{B}:\ARX\left(1,1\right)$
under an $\ARX\left(2,3\right)$ DGP, as described in Section~\ref{sec:sel-dep}.
The DGP autoregressive parameter is $\phi_{*}=\alpha\left(0.75,0.2\right)$.
See also Figure~\ref{fig:data-length} in Appendix~\ref{sec:supp-plots}.}
\label{fig:sample-size}
\end{figure}

Figure~\ref{fig:sample-size} compares the minimum sample size required
for several more joint and pointwise CV methods. Consistent with earlier 
results, there is little difference required sample size between pointwise
and joint methods in the independent case ($\alpha=0$).
Under stronger dependence, however, the greater variability of pointwise
methods leads to a requirement for larger sample sizes for the two candidate
models to be well-separated, all else being equal. These results underscore
the importance of using efficient CV designs---especially the use of joint
scoring rules---when strong dependence is present. See also
Figure~\ref{fig:data-length} in Appendix~\ref{sec:supp-plots}.

As we might expect, the benefit of using more efficient CV methods
is greatest when candidate models are more challenging to separate.
In Figure~\ref{fig:sample-size}, required sample size is greatest
for the `hard' variant, especially under a high degree of dependence.
In comparison, under the `easy' variant the differences between
joint and pointwise methods are smaller, although there is nonetheless
a pickup in the sample size required for pointwise CV methods.

Figure~\ref{fig:sample-size} also underscores the benefit of using
as much of the available sample as possible, rather than discarding
future observations as in LFO methods. As dependence increases, the
additional statistical efficiency associated with allowing the model
to learn from future data results in a well-separated model with shorter
overall data lengths.

\section{Discussion and conclusion\label{sec:disc-conc}}

We have demonstrated that in settings where serial dependence is present,
appropriate CV procedure design can dramatically improve model
selection performance. Working with the logarithmic scoring rule and the
$\ARX(p,q)$ class of autoregressive models, we have
shown that evaluating the score pointwise can yield highly inefficient
CV estimators that perform poorly when compared with procedures that
target joint densities. Our experiments show that pointwise CV
estimators exhibit greater variability and require larger sample sizes
than joint designs.

We are not the first to compare the performance between joint and
pointwise densities in predictive model assessment. \citet{Osband2021},
for instance, apply model assessment with joint densities in the context
of neural networks.

Our results show that the consequences of using an inefficient CV
procedure can be particularly pernicious under strong serial dependence.
One consequence is that CV procedures become biased toward overly
simple models. In extreme cases where serial dependence is greatest,
this can result in different CV procedures assigning completely different
orderings to candidate models. Viewed another way, the consequence
of the use of inappropriate CV procedures is the need for larger sample
sizes to achieve good separation between models.

Relatively conservative methods like LFO can be a good choice in applied
contexts, especially when CV is able to clearly separate models even without
the use of the full sample for reducing estimator variability. Furthermore,
some authors have advocated for the use of LFO \citep[e.g.,][]{Burkner2020}
when the dependence structure is unknown. While LFO certainly eliminates
the possibility that contamination will bias results, the results
of Section~\ref{subsec:hyperparam} suggest that contamination
arising from the use of future data and training set size does need
to be carefully traded off against the benefit of retaining a
larger training set. In general, it seems unlikely that
the optimal CV design would be the corner solution that excludes all
future observations. Risks associated with
contamination can also be avoided by the use of specification tests on candidate
models before conducting model comparison, including standard Bayesian
model criticism procedures and testing for autocorrelation in the residuals
\citep{Bergmeir2018}. Model assessment is especially
important when the underlying process is highly persistent. Not
only is the need for larger effective sample sizes greatest under
persistence processes, but the adverse impact of contamination is
most pernicious.

Our goal throughout this paper has been
model selection. We are not claiming that exploiting future observations
in CV schemes yields nearly unbiased estimators of the elpd. Instead, here we are
targeting relative measures that are efficient model selection objectives.

This analysis has focused on serial dependence, which most often
appears in time series models. However, we expect that similar
results would apply for other forms of dependence such as spatial
and spatio-temporal data, especially where the autoregressive signal
is relatively strong compared with the global conditional mean.
Naturally, dependence structures in more than one dimension presents
additional analytical challenges, so further research is warranted
for these and other dependence structures. Furthermore, many of our conclusions
are not specific to the logarithmic score and would also apply under other scoring
rules \citep[]{Gneiting2007}.

From a practical standpoint, it should be noted that implementing
joint CV methods tends to be computationally costly when compared
with pointwise procedures like LOO. In situations where the difference
between two models is absolutely clear, as for the `easy' variant
of our examples under weak serial dependence, pointwise CV estimators may be
adequate for performing model selection and are far more convenient to construct.
This is particularly relevant considering the availability
of efficient computational shortcuts for computing pointwise CV procedures,
such as PSIS-LOO \citep{Vehtari_2016,Burkner2020}, and a lack of
similar shortcuts for joint procedures. Although in principle PSIS-LOO can also approximate joint CV procedures, in practice the thick tails of the weight distributions tend to cause importance sampling to fail.

With the complexity of implementing joint procedures in mind, we recommend
the following workflow for model selection under serial dependence.
Begin with a thorough model criticism of each of the candidate models, and iterate
model specification until the candidate models are well-specified.
Where inference results show that serial dependence is relatively
weak, pointwise CV methods such as PSIS-LOO can be used as a first
pass for model selection. If the pointwise CV results show a clear
preference for one candidate, then that candidate can be selected.
Otherwise, a joint CV procedure should be implemented and relied
upon instead.

The present paper represents a first look at the uncertainty of
CV-based model selection under serial dependence, but there is
considerable work remaining. We have focused on identification
of the regression parameter, leaving to one side the tasks of
identifying the autoregressive component, theoretical analysis of these results, choosing suitable
priors for model identification procedures, and constructing
efficient computational methods for CV under serial dependence.
We leave these aspects for future work.

\begin{supplement}
\stitle{Appendixes}
\sdescription{
    This supplement contains additional plots and tables referenced in the main text.
    The supplement also contains derivations and proofs to support the experiments
    in the paper.
}
\end{supplement}

\bibliographystyle{ba}
\bibliography{depcv}

\begin{thebibliography}{30}
\newcommand{\enquote}[1]{``#1''}
\expandafter\ifx\csname natexlab\endcsname\relax\def\natexlab#1{#1}\fi
\expandafter\ifx\csname url\endcsname\relax
  \def\url#1{{\tt #1}}\fi
\expandafter\ifx\csname urlprefix\endcsname\relax\def\urlprefix{URL }\fi
\ifx\endbibitem\undefined \let\endbibitem\relax\fi

\bibitem[{Ahmed and Gokhale(1989)}]{Ahmed1989}
Ahmed, N.~A. and Gokhale, D.~V. (1989).
\newblock \enquote{{Entropy expressions and their estimators for multivariate
  distributions}.}
\newblock {\em IEEE Transactions on Information Theory\/}, 35(3): 688--692.
\endbibitem

\bibitem[{Arlot and Celisse(2010)}]{Arlot2010}
Arlot, S. and Celisse, A. (2010).
\newblock \enquote{{A survey of cross-validation procedures for model
  selection}.}
\newblock {\em Statistics Surveys\/}, 4: 40--79.
\endbibitem

\bibitem[{Bengio and Grandvalet(2004)}]{bengio2004}
Bengio, Y. and Grandvalet, Y. (2004).
\newblock \enquote{{No unbiased estimator of the variance of K-fold
  cross-validation}.}
\newblock {\em Journal of Machine Learning Research\/}, 5: 1089--1105.
\newline\urlprefix\url{https://www.semanticscholar.org/paper/17f4a82822309d4ba0e9b2840afc5dfaa499be97}
\endbibitem

\bibitem[{Berger et~al.(2014)Berger, Bayarri, and Pericchi}]{Berger2014}
Berger, J., Bayarri, M.~J., and Pericchi, L.~R. (2014).
\newblock \enquote{{The Effective Sample Size}.}
\newblock {\em Econometric Reviews\/}, 33(1-4): 197--217.
\endbibitem

\bibitem[{Bergmeir and Ben{\'{i}}tez(2012)}]{Bergmeir2012}
Bergmeir, C. and Ben{\'{i}}tez, J.~M. (2012).
\newblock \enquote{{On the use of cross-validation for time series predictor
  evaluation}.}
\newblock {\em Information Sciences\/}, 191: 192--213.
\newline\urlprefix\url{https://doi.org/10.1016{\%}2Fj.ins.2011.12.028}
\endbibitem

\bibitem[{Bergmeir et~al.(2018)Bergmeir, Hyndman, and Koo}]{Bergmeir2018}
Bergmeir, C., Hyndman, R.~J., and Koo, B. (2018).
\newblock \enquote{{A note on the validity of cross-validation for evaluating
  autoregressive time series prediction}.}
\newblock {\em Computational Statistics and Data Analysis\/}, 120: 70--83.
\newline\urlprefix\url{https://doi.org/10.1016/j.csda.2017.11.003}
\endbibitem

\bibitem[{Billingsley(2008)}]{Billingsley2008}
Billingsley, P. (2008).
\newblock {\em {Probability and measure}\/}.
\newblock John Wiley {\&} Sons.
\endbibitem

\bibitem[{B{\"{u}}rkner et~al.(2020)B{\"{u}}rkner, Gabry, and
  Vehtari}]{Burkner2020}
B{\"{u}}rkner, P.~C., Gabry, J., and Vehtari, A. (2020).
\newblock \enquote{{Approximate leave-future-out cross-validation for Bayesian
  time series models}.}
\newblock {\em Journal of Statistical Computation and Simulation\/}, 1--25.
\endbibitem

\bibitem[{Burman(1989)}]{Burman1989}
Burman, P. (1989).
\newblock \enquote{{A comparative study of ordinary cross-validation, v-fold
  cross-validation and the repeated learning-testing methods}.}
\newblock {\em Biometrika\/}, 76(3): 503--514.
\newline\urlprefix\url{https://doi.org/10.1093/BIOMET/76.3.503}
\endbibitem

\bibitem[{Burman et~al.(1994)Burman, Chow, and Nolan}]{Burman1994}
Burman, P., Chow, E., and Nolan, D. (1994).
\newblock \enquote{{A cross-validatory method for dependent data}.}
\newblock {\em Biometrika\/}, 81(2): 351--358.
\newline\urlprefix\url{https://doi.org/10.1093/BIOMET/81.2.351}
\endbibitem

\bibitem[{Cerqueira et~al.(2020)Cerqueira, Torgo, and Mozetic}]{Cerqueira2020}
Cerqueira, V., Torgo, L., and Mozetic, I. (2020).
\newblock {\em {Evaluating time series forecasting models: an empirical study
  on performance estimation methods}\/}, volume 109.
\newblock Springer US.
\newline\urlprefix\url{https://doi.org/10.1007/s10994-020-05910-7}
\endbibitem

\bibitem[{Davies(1973)}]{Davies1973}
Davies, R.~B. (1973).
\newblock \enquote{{Numerical inversion of a characteristic function}.}
\newblock {\em Biometrika\/}, 60(2): 415--417.
\endbibitem

\bibitem[{Dawid(1984)}]{Dawid1984}
Dawid, A.~P. (1984).
\newblock \enquote{{Statistical Theory: The Prequential Approach}.}
\newblock {\em Journal of the Royal Statistical Society. Series A\/}, 147(2):
  278--292.
\endbibitem

\bibitem[{Duchi(2007)}]{Duchi2007}
Duchi, J. (2007).
\newblock \enquote{{Derivations for Linear Algebra and Optimization}.}
\newblock {\em Berkeley, California\/}, 1--13.
\endbibitem

\bibitem[{Geisser(1975)}]{Geisser1975}
Geisser, S. (1975).
\newblock \enquote{{The predictive sample reuse method with applications}.}
\newblock {\em Journal of the American Statistical Association\/}, 70(350):
  320--328.
\newline\urlprefix\url{https://doi.org/10.1080/01621459.1975.10479865
  https://doi.org/10.1080{\%}2F01621459.1975.10479865}
\endbibitem

\bibitem[{Gelman et~al.(2014)Gelman, Carlin, Stern, Dunson, Vehtari, and
  Rubin}]{Gelman2014}
Gelman, A., Carlin, J.~B., Stern, H.~S., Dunson, D.~B., Vehtari, A., and Rubin,
  D.~B. (2014).
\newblock {\em {Bayesian data analysis}\/}.
\newblock Boca Raton, FL, USA: Chapman {\&} Hall/CRC, 3 edition.
\endbibitem

\bibitem[{Gneiting and Raftery(2007)}]{Gneiting2007}
Gneiting, T. and Raftery, A.~E. (2007).
\newblock \enquote{{Strictly Proper Scoring Rules, Prediction, and
  Estimation}.}
\newblock {\em Journal of the American Statistical Association\/}, 102(477):
  359--378.
\newline\urlprefix\url{http://www.tandfonline.com/doi/abs/10.1198/016214506000001437}
\endbibitem

\bibitem[{Gy{\"{o}}rfi et~al.(1989)Gy{\"{o}}rfi, H{\"{a}}rdle, Sarda, and
  Vieu}]{Gyorfi1989}
Gy{\"{o}}rfi, L., H{\"{a}}rdle, W., Sarda, P., and Vieu, P. (1989).
\newblock {\em {Nonparametric Curve Estimation from Time Series}\/}.
\newblock New York, NY: Springer.
\endbibitem

\bibitem[{Imhof(1961)}]{P.J.1961}
Imhof, P. (1961).
\newblock \enquote{{Computing the distribution of quadratic forms in normal
  variables}.}
\newblock {\em Biometrika\/}, 48(3): 419.
\endbibitem

\bibitem[{Lee(2011)}]{Lee2011}
Lee, D. (2011).
\newblock \enquote{{A comparison of conditional autoregressive models used in
  Bayesian disease mapping}.}
\newblock {\em Spatial and Spatio-temporal Epidemiology\/}, 2(2): 79--89.
\newline\urlprefix\url{http://dx.doi.org/10.1016/j.sste.2011.03.001}
\endbibitem

\bibitem[{Madiman and Tetali(2010)}]{Madiman2010}
Madiman, M. and Tetali, P. (2010).
\newblock \enquote{{Information inequalities for joint distributions, with
  interpretations and applications}.}
\newblock {\em IEEE Transactions on Information Theory\/}, 56(6): 2699--2713.
\endbibitem

\bibitem[{Mathai and Provost(1992)}]{Mathai1992}
Mathai, A. and Provost, S.~B. (1992).
\newblock \enquote{{Quadratic forms}.}
\endbibitem

\bibitem[{Osband et~al.(2022)Osband, Wen, Asghari, Dwaracherla, Lu, Ibrahimi,
  Lawson, Hao, O'Donoghue, and Roy}]{Osband2021}
Osband, I., Wen, Z., Asghari, S.~M., Dwaracherla, V., Lu, X., Ibrahimi, M.,
  Lawson, D., Hao, B., O'Donoghue, B., and Roy, B.~V. (2022).
\newblock \enquote{The Neural Testbed: Evaluating Joint Predictions.}
\newblock In Oh, A.~H., Agarwal, A., Belgrave, D., and Cho, K. (eds.), {\em
  Advances in Neural Information Processing Systems\/}.
\newline\urlprefix\url{https://openreview.net/forum?id=JyTT03dqCFD}
\endbibitem

\bibitem[{Racine(2000)}]{Racine_2000}
Racine, J. (2000).
\newblock \enquote{{Consistent cross-validatory model-selection for dependent
  data: hv-block cross-validation}.}
\newblock {\em Journal of Econometrics\/}, 99(1): 39--61.
\newline\urlprefix\url{https://doi.org/10.1016{\%}2Fs0304-4076{\%}2800{\%}2900030-0}
\endbibitem

\bibitem[{Schad et~al.(2022)Schad, Nicenboim, B{\"{u}}rkner, Betancourt, and
  Vasishth}]{Schad2022}
Schad, D.~J., Nicenboim, B., B{\"{u}}rkner, P.~C., Betancourt, M., and
  Vasishth, S. (2022).
\newblock \enquote{{Workflow Techniques for the Robust Use of Bayes Factors}.}
\newblock {\em Psychological Methods\/}.
\endbibitem

\bibitem[{Sims(1980)}]{Sims1980}
Sims, C.~A. (1980).
\newblock \enquote{{Macroeconomics and Reality}.}
\newblock {\em Econometrica\/}, 48(1): 1--48.
\endbibitem

\bibitem[{Sivula et~al.(2020)Sivula, Magnusson, Matamoros, and
  Vehtari}]{Sivula2020a}
Sivula, T., Magnusson, M., Matamoros, A.~A., and Vehtari, A. (2020).
\newblock \enquote{{Uncertainty in Bayesian Leave-One-Out Cross-Validation
  Based Model Comparison}.}
\newblock ArXiv preprint.
\newline\urlprefix\url{http://arxiv.org/abs/2008.10296}
\endbibitem

\bibitem[{Vehtari et~al.(2017)Vehtari, Gelman, and Gabry}]{Vehtari_2016}
Vehtari, A., Gelman, A., and Gabry, J. (2017).
\newblock \enquote{{Practical Bayesian model evaluation using leave-one-out
  cross-validation and WAIC}.}
\newblock {\em Statistics and Computing\/}, 27(5): 1413--1432.
\newline\urlprefix\url{https://doi.org/10.1007{\%}2Fs11222-016-9709-3
  https://doi.org/10.1007{\%}2Fs11222-016-9696-4}
\endbibitem

\bibitem[{Vehtari and Ojanen(2012)}]{Vehtari2012a}
Vehtari, A. and Ojanen, J. (2012).
\newblock \enquote{{A survey of Bayesian predictive methods for model
  assessment, selection and comparison}.}
\newblock {\em Statistics Surveys\/}, 6(1): 142--228.
\endbibitem

\bibitem[{Ward(2008)}]{Ward2008a}
Ward, E.~J. (2008).
\newblock \enquote{{A review and comparison of four commonly used Bayesian and
  maximum likelihood model selection tools}.}
\newblock {\em Ecological Modelling\/}, 211(1-2): 1--10.
\endbibitem

\end{thebibliography}

\begin{acks}[Acknowledgments]
    The authors are grateful for the input of 
    two anonymous referees and an associate editor. Any
    remaining errors are our own.
    AC's work was supported in part by an
    Australian Government Research Training Program Scholarship.
    AV acknowledges the Academy of Finland Flagship program: Finnish Center
    for Artificial Intelligence, and Academy of Finland project (340721).
    CF acknowledges financial support under National Science Foundation
    Grant SES-1921523. LK gratefully recognises support from the National
    Institutes of Health (5R01AG067149-02).
\end{acks}

\appendix

\section{Additional plots\label{sec:supp-plots}}
This section contains supplementary plots referenced in the main text.

\begin{figure}[H]
\begin{centering}
\includegraphics[width=1\textwidth]{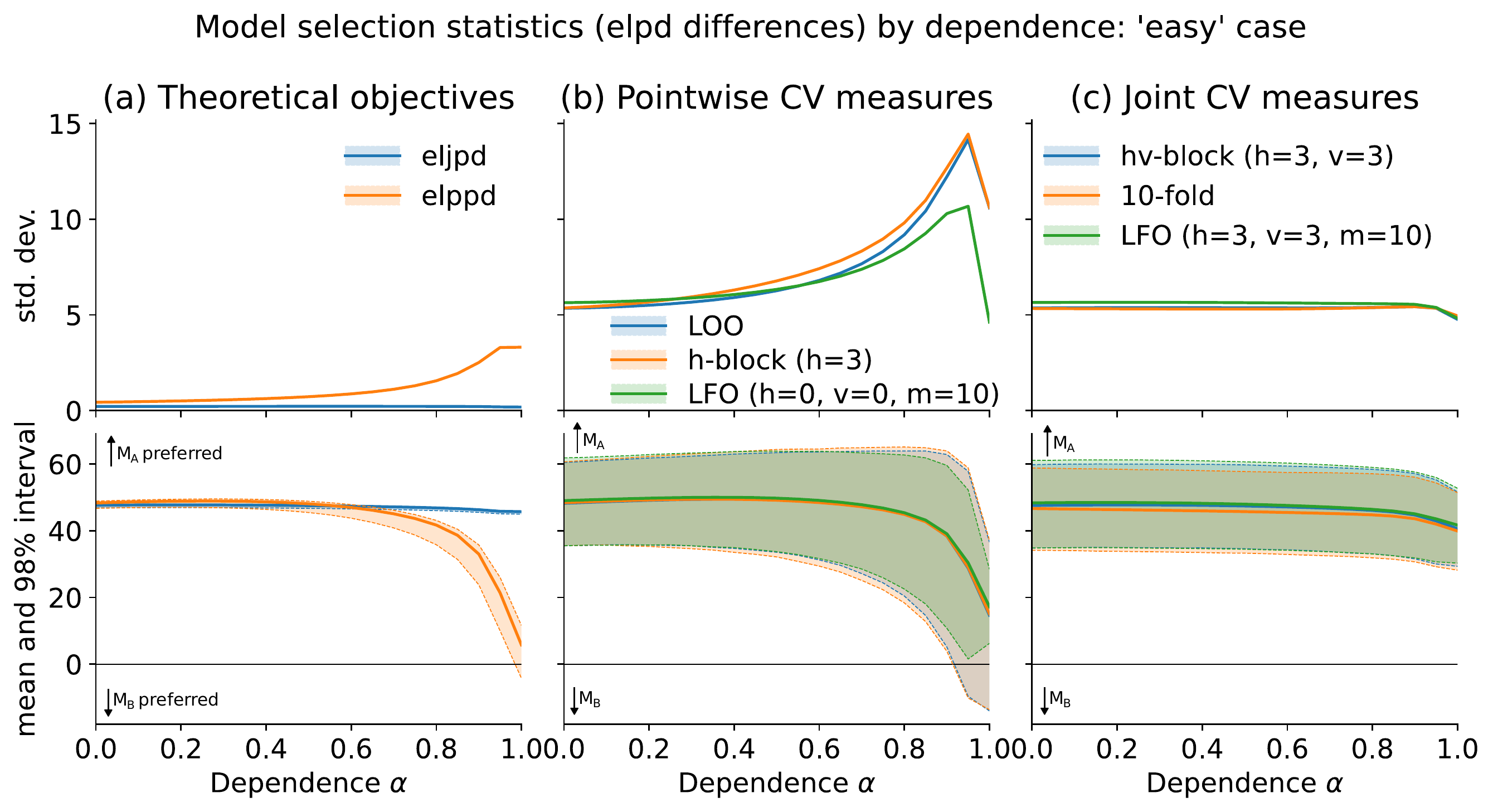}
\par\end{centering}
\caption{Behavior of model selection objectives for the `easy' case, a counterpoint
to Figure~\ref{fig:by-alpha}. Column (a) shows the theoretical model selection objectives,
and columns (b) and (c) the associated CV estimates. The top row plots the standard deviation of the relevant $\omega(y)$ for the corresponding column. The bottom row shows its mean and 98\% interval. The model parameter $\alpha$
governs the degree of serial dependence. Notice that the adverse selection rate for
both joint and pointwise methods is close to zero for all but the strongest dependence.
This model selection experiment compares $M_{A}:\ARX\left(1,2\right)$
vs $M_{B}:\ARX\left(1,1\right)$ under an $\ARX\left(2,3\right)$ DGP,
as described in Section~\ref{sec:sel-dep}. We use the $\beta^{\textnormal{easy}}$
parameter setting, autoregressive parameter $\phi_{*}=\alpha\left(0.75,0.2\right)$,
and data length $T$=100.}

\label{fig:by-alpha-easy}
\end{figure}

\begin{figure}[H]
\begin{centering}
\includegraphics[width=1.0\textwidth]{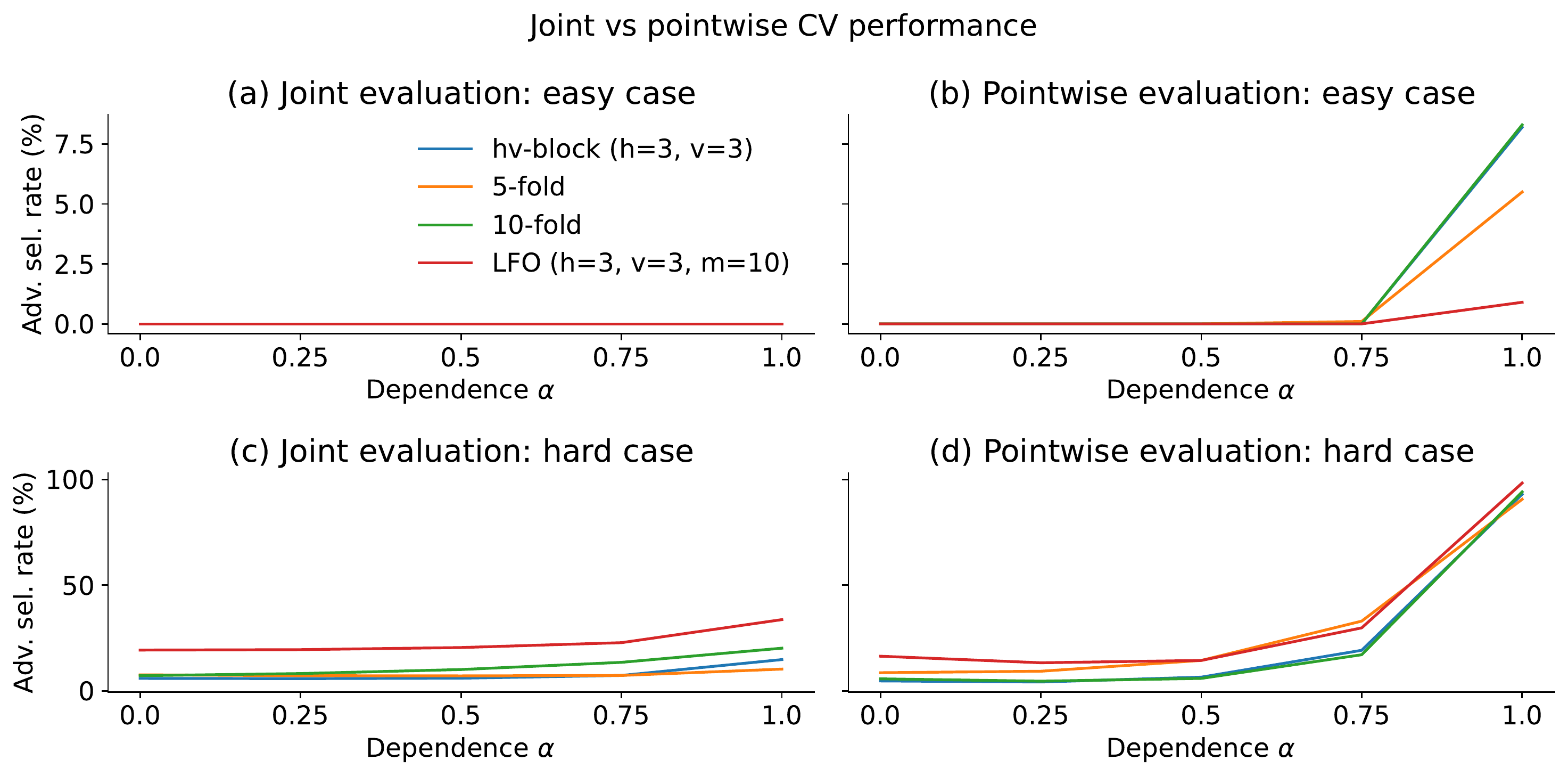}
\par\end{centering}
\caption{Adverse model selection rates for joint CV methods and structurally
identical methods evaluated pointwise. This model selection experiment
compares $M_{A}:\ARX\left(1,2\right)$ vs $M_{B}:\ARX\left(1,1\right)$
under an $\ARX\left(2,3\right)$ DGP, as described in Section~\ref{sec:sel-dep}.
The \textquoteleft easy\textquoteright{} and \textquoteleft hard\textquoteright{}
results respectively use the $\beta^{\textnormal{easy}}$ and $\beta^{\textnormal{hard}}$
parameter settings. The autoregressive parameter is $\phi_{*}=\alpha\left(0.75,0.2\right)$
and data length $T$=100.}
\label{fig:joint-pointwise}
\end{figure}

\begin{figure}[H]
\begin{centering}
\includegraphics[width=1\textwidth]{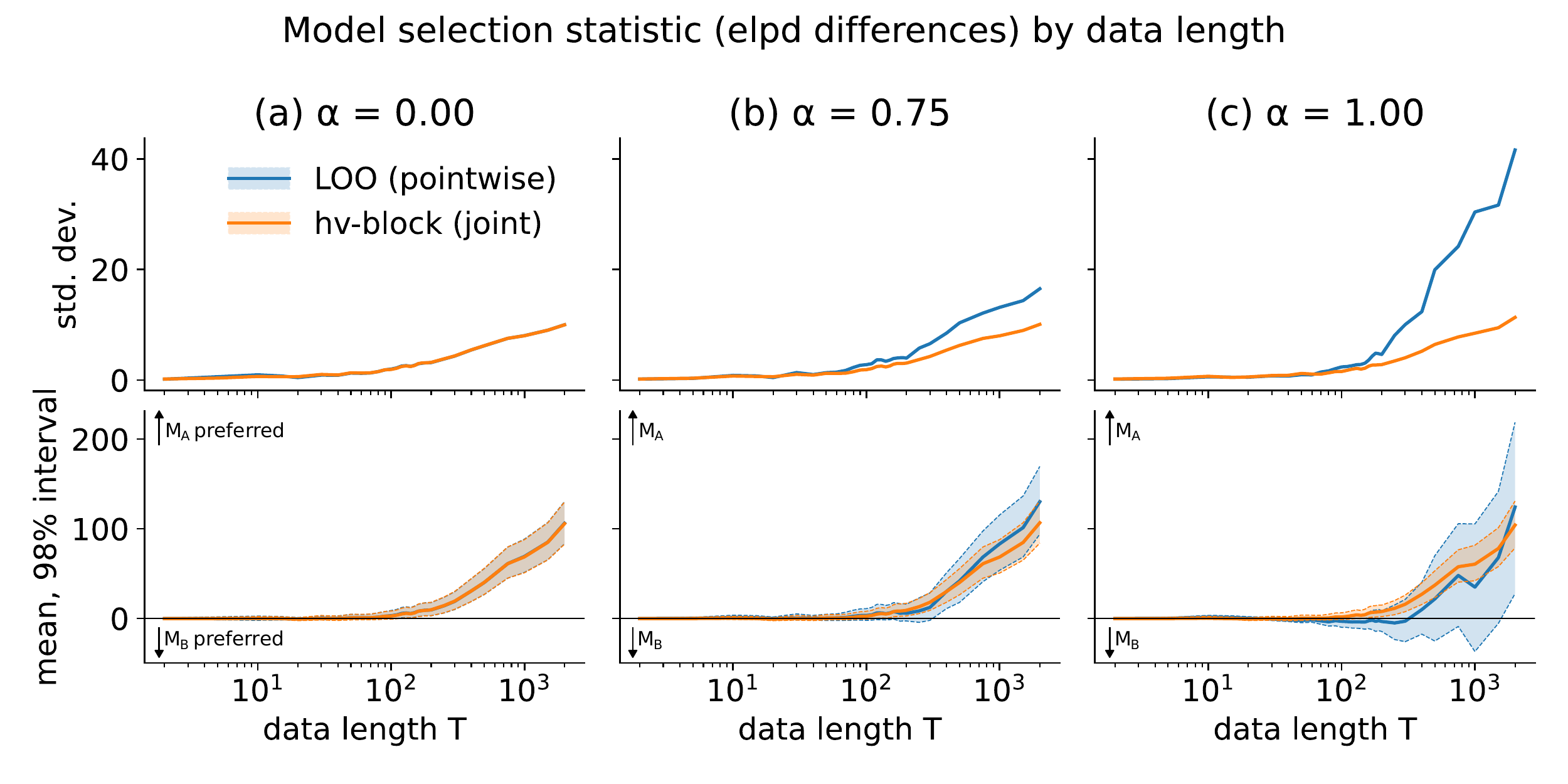}
\par\end{centering}
\caption{CV model selection statistic variability by data length $T$ for LOO-CV
and $hv$-block CV. The degree of data dependence increases across
columns from left to right. The top row shows the model selection
statistic standard deviation and the bottom row its 98\% probability
interval. This model selection experiment compares $M_{A}:\ARX\left(1,2\right)$
vs $M_{B}:\ARX\left(1,1\right)$ under a $\ARX\left(2,3\right)$ dgp,
as described in Section~\ref{sec:sel-dep}. We use the $\beta^{\textnormal{hard}}$
parameter setting, autoregressive parameter $\phi_{*}=\alpha\left(0.75,0.2\right)$,
and data length $T$=100.}
\label{fig:data-length}
\end{figure}

\begin{figure}[H]
\begin{centering}
\includegraphics[width=1.0\textwidth]{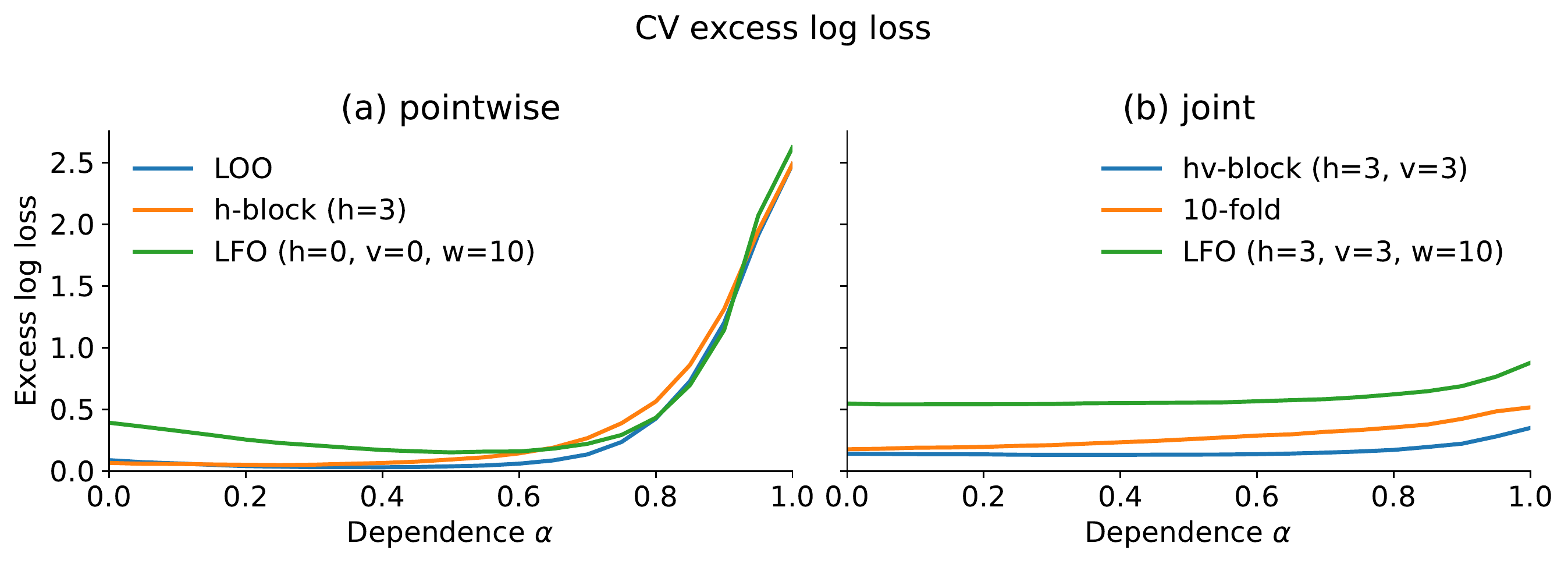}
\par\end{centering}
\caption{Log loss measures for pointwise and joint CV methods as data dependence
increases. This experiment compares $M_{A}:\,\ARX\left(1,2\right)$
vs $M_{B}:\,\ARX\left(1,1\right)$, under an $\ARX\left(2,3\right)$
DGP with $\phi_{*}=\alpha\left(0.75,0.2\right)$, for $\alpha\in\left[0,1\right]$
as described in Section~\ref{sec:sel-dep}. Greater values of $\alpha$
denote stronger serial dependence. T=100. The \textquoteleft hard\textquoteright{}
model variant ($\beta^{\textnormal{hard}}$) is shown.}
\label{fig:log-loss}
\end{figure}

\begin{figure}[H]
\begin{centering}
\includegraphics[width=1.0\textwidth]{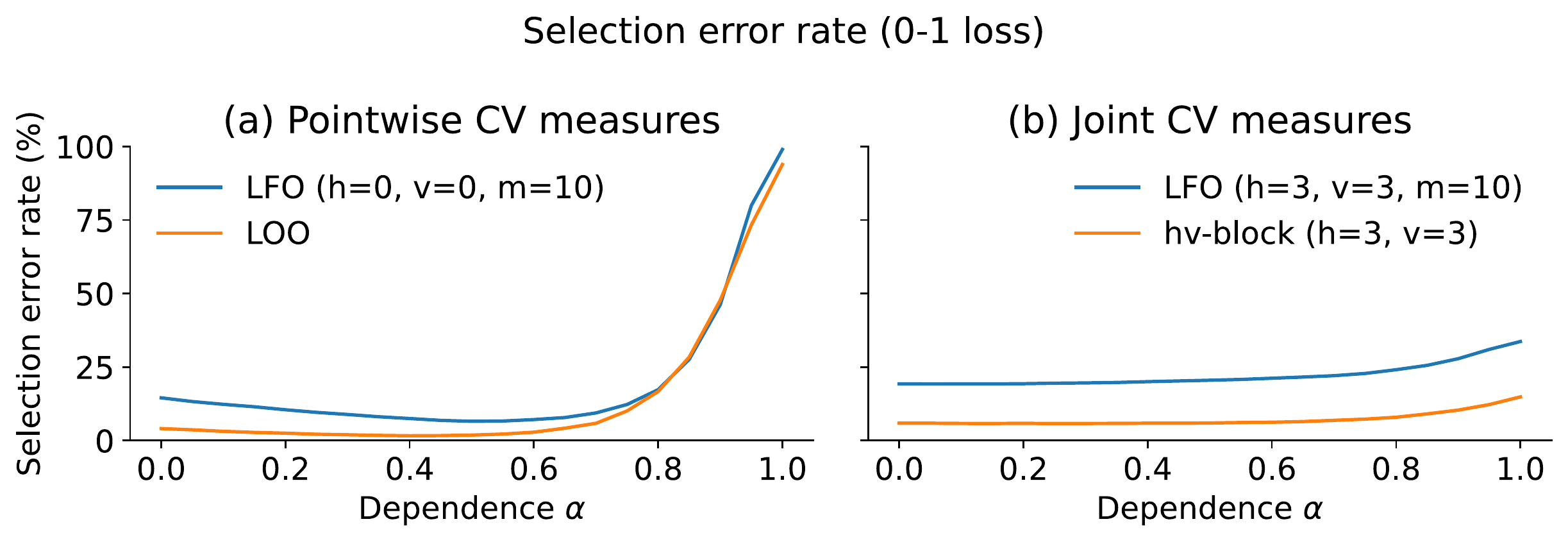}
\par\end{centering}
\caption{Adverse selection rates for LFO and comparable CV schemes that use
future information, by dependence $\alpha$. Notice that in each case,
the LFO error rate (blue) exceeds that of the other CV scheme (orange).
This model selection experiment compares $M_{A}:\ARX\left(1,2\right)$
vs $M_{B}:\ARX\left(1,1\right)$ under an $\ARX\left(2,3\right)$ DGP,
as described in Section~\ref{sec:sel-dep}. We use the $\beta^{\textnormal{hard}}$
parameter setting, autoregressive parameter $\phi_{*}=\alpha\left(0.75,0.2\right)$,
and data length $T$=100.}
\label{fig:hvblock-lfo}
\end{figure}

\begin{figure}[H]
\begin{centering}
\includegraphics[width=1\textwidth]{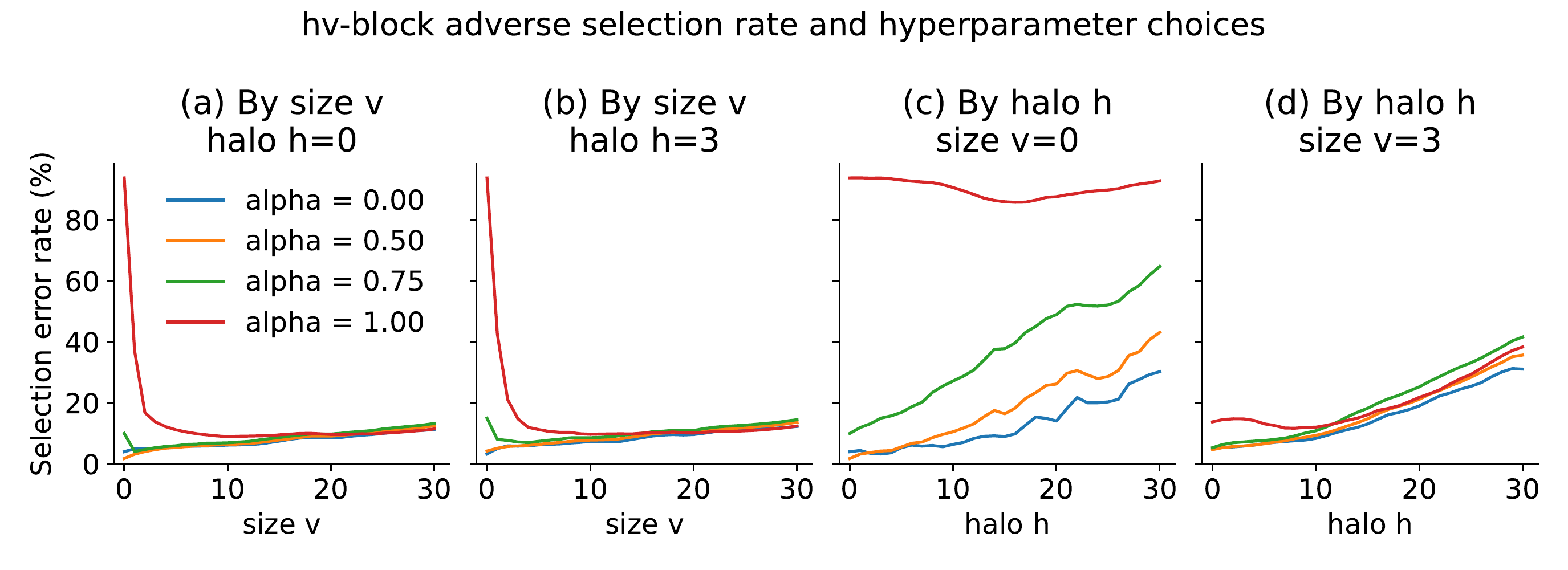}
\par\end{centering}
\caption{Adverse selection rates using $hv$-block CV with different choices
of the halo $h$ and validation set size parameter $v$. Total validation
set dimension is $2v+1$. This model selection experiment compares
$M_{A}:\ARX\left(1,2\right)$ vs $M_{B}:\ARX\left(1,1\right)$ under
a $\ARX\left(2,3\right)$ DGP, as described in Section~\ref{sec:sel-dep}.
We use the $\beta^{\textnormal{hard}}$ parameter setting, autoregressive
parameter $\phi_{*}=\alpha\left(0.75,0.2\right)$, and data length
$T$=100.}
\label{fig:halo-dimension-variance}
\end{figure}

\section{Supplementary experiments\label{sec:experiments}}

The extended example described in Section~\ref{sec:sel-dep} is a
model selection experiment comparing $M_{A}:\,\ARX\left(1,2\right)$
vs $M_{B}:\,\ARX\left(1,1\right)$, where the data are generated by
an $\ARX\left(2,3\right)$ DGP. We chose this experiment and the associated
parameter vectors $\beta_{*}^{\textnormal{easy}}$ and $\beta_{*}^{\textnormal{hard}}$
because we believe they are representative of typical CV problems
faced in experimental settings---that is, neither $M_{A}$ nor $M_{B}$
have the same functional form as the DGP, but one is clearly a better
fit than the other.

This appendix presents the results of additional experiments using
other models from the $\ARX$ class. The goal of these supplementary results
is to demonstrate that the main experiment is representative of typical
cases, and to highlight specific cases where behaviors can differ.
For consistency and simplicity we retain the same values for the parameter
vectors $\beta_{*}^{\textnormal{easy}}$ and $\beta_{*}^{\textnormal{hard}}$.
We choose $LOO$ as an example of a pointwise CV method and $hv-\mathrm{block}(3,3)$
as an example of a joint CV method. In each case the focus is on identifying the
regression parameter, so each candidate model has the same autoregressive
structure.

Let the stationary vector $y=y_{1},\dots,y_{T}$ be distributed according to an $\ARX\left(p_*,q_*\right)$ of the form,
\begin{equation}
\textnormal{DGP}:\ y_{t}=\phi_*^{\top}\begin{pmatrix}y_{t-1}\\ \vdots \\ y_{t-p^*}\end{pmatrix}+\beta_*^\top z_t+\sigma_{*}\varepsilon_{t},\label{eq:ex:dgp-supp}
\end{equation}
where $\varepsilon_{t}\overset{\mathrm{iid}}{\sim}\mathcal{N}\left(0,1\right)$. The experiment selects between two candidate models,
\begin{align}
M_{A}:\ y_{t} & =\widehat{\phi_A} \begin{pmatrix}y_{t-1}\\ \vdots \\y_{t-p_A}\end{pmatrix}+\beta_A^\top z_{t}+\widehat{\sigma_A}\varepsilon_{t} & \ARX\left(p_A,q_A\right)\label{eq:ex:mA-supp}\\
M_{B}:\ y_{t} & = \widehat{\phi_B}\begin{pmatrix}y_{t-1}\\ \vdots \\y_{t-p_B}\end{pmatrix}+\beta_B^\top z_t+\widehat{\sigma_B}\varepsilon_{t} & \ARX\left(p_B,q_B\right).
\end{align}

Table~\ref{tbl:examples} lists the additional experiments. In each
case, the DGP is parameterized by $\alpha\in\left[0,1\right]$, which
controls the degree of dependence. Each experiment is run twice, once
each with $\beta_{*}^{\textnormal{easy}}$ and $\beta_{*}^{\textnormal{hard}}$.
The DGP parameter $\alpha\in\left[0,1\right]$ is varied to select
the degree of serial dependence. All DGPs are stationary. The
random noise component is distributed as $\varepsilon_{t}\overset{iid}{\sim}N\left(0,1\right)$,
the fixed per-period covariates $z_{1t}$ and $z_{2t}$ are drawn
independently from the standard normal distribution, and priors are
$\beta_{i}\sim\mathcal{N}\left(\beta_{*i},1\right)$, for $i=1,2,3$.
Initial values are $y_{t}=0$ for $t\leq0$. For all examples in this
section, the data length is $T=100$.
As in Section~\ref{sec:sel-dep}, we solve \eqref{eq:sarx:minkld} to find optimal noise variance and
autoregressive parameter values.

The results of the additional experiments are broadly similar to Experiment~1:
joint and pointwise methods perform similarly under no dependence, and under 
stronger dependence both the variability and error rates of the pointwise estimators
are higher than that of the joint estimators. Experiment~4, for which
the candidate models both have no autoregressive component at all, is an exception,
and both methods perform terribly under strong dependence. It should perhaps not
be surprising that such badly misspecified models would perform so poorly,
which underscores the need for thorough model criticism for all candidates before
conducting CV.

\begin{table}[H]
{\small{}}%
\begin{tabular}{|>{\raggedright}m{0.04\textwidth}|>{\raggedright}m{0.9\textwidth}|}
\multicolumn{2}{c}{\textbf{Primary experiment (main text)}}\tabularnewline
\hline 
\centering{}\textbf{\small{}\#} & \textbf{\small{}DGP and candidate models}\tabularnewline
\hline 
\noalign{\vskip0.1cm}
\multirow{3}{0.04\textwidth}{\centering{}\textbf{1}\textbf{\small{}.}} & {\small{}$\ARX\left(2,3\right)\,\mathrm{DGP}:y_{t}=\alpha\begin{pmatrix}0.75\\
0.2
\end{pmatrix}^{\top}\begin{pmatrix}y_{t-1}\\
y_{t-2}
\end{pmatrix}+\beta_{*1}+\beta_{*2}z_{2t}+\beta_{*3}z_{3t}+\varepsilon_{t}$}\tabularnewline
 & {\small{}$\ARX\left(1,2\right)\,M_{A}:y_{t}=\phi_{1}y_{t-2}+\beta_{1}+\beta_{2}z_{2t}+\varepsilon_{t}$}\tabularnewline
 & {\small{}$\ARX\left(1,1\right)\,M_{B}:y_{t}=\phi_{1}y_{t-1}+\beta_{1}+\varepsilon_{t}$}\tabularnewline
\hline 
\multicolumn{2}{c}{}\tabularnewline
\multicolumn{2}{c}{\textbf{Supplementary experiments (this appendix)}}\tabularnewline
\hline 
\centering{}\textbf{\small{}\#} & \textbf{\small{}DGP and candidate models}\tabularnewline
\hline 
\multirow{3}{0.04\textwidth}{\centering{}\textbf{\small{}2.}} & {\small{}$\ARX\left(1,3\right)\,\mathrm{DGP}:y_{t}=\alpha\left(0.95\right)y_{t-1}+\beta_{*1}+\beta_{*2}z_{2t}+\beta_{*3}z_{3t}+\varepsilon_{t}$}\tabularnewline
 & {\small{}$\ARX\left(1,3\right)\,M_{A}:y_{t}=\phi_{1}y_{t-1}+\beta_{1}+\beta_{2}z_{2t}+\beta_{3}z_{3t}+\varepsilon_{t}$}\tabularnewline
 & {\small{}$\ARX\left(1,2\right)\,M_{B}:y_{t}=\phi_{1}y_{t-1}+\beta_{1}+\beta_{2}z_{2t}+\varepsilon_{t}$}\tabularnewline
\hline 
\multirow{3}{0.04\textwidth}{\centering{}\textbf{3.}} & {\small{}$\ARX\left(1,3\right)\,\mathrm{DGP}:y_{t}=\alpha\left(0.95\right)y_{t-1}+\beta_{*1}+\beta_{*2}z_{2t}+\beta_{*3}z_{3t}+\varepsilon_{t}$}\tabularnewline
 & {\small{}$\ARX\left(1,2\right)\,M_{A}:y_{t}=\phi_{1}y_{t-1}+\beta_{1}+\beta_{2}z_{2t}+\varepsilon_{t}$}\tabularnewline
 & {\small{}$\ARX\left(1,1\right)\,M_{B}:y_{t}=\phi_{1}y_{t-1}+\beta_{1}+\varepsilon_{t}$}\tabularnewline
\hline 
\multirow{3}{0.04\textwidth}{\centering{}\textbf{\small{}4.}} & {\small{}$\ARX\left(1,3\right)\,\mathrm{DGP}:y_{t}=\alpha\left(0.95\right)y_{t-1}+\beta_{*1}+\beta_{*2}z_{2t}+\beta_{*3}z_{3t}+\varepsilon_{t}$}\tabularnewline
 & {\small{}$\ARX\left(0,2\right)\,M_{A}:y_{t}=\beta_{1}+\beta_{2}z_{2t}+\varepsilon_{t}$}\tabularnewline
 & {\small{}$\ARX\left(0,1\right)\,M_{B}:y_{t}=\beta_{1}+\varepsilon_{t}$}\tabularnewline
\hline 
\noalign{\vskip0.1cm}
\multirow{3}{0.04\textwidth}{\centering{}\textbf{\small{}5.}} & {\small{}$\ARX\left(2,3\right)\,\mathrm{DGP}:y_{t}=\alpha\begin{pmatrix}0.75\\
0.2
\end{pmatrix}^{\top}\begin{pmatrix}y_{t-1}\\
y_{t-2}
\end{pmatrix}+\beta_{*1}+\beta_{*2}z_{2t}+\beta_{*3}z_{3t}+\varepsilon_{t}$}\tabularnewline
 & {\small{}$\ARX\left(1,3\right)\,M_{A}:y_{t}=\phi_{1}y_{t-1}+\beta_{1}+\beta_{2}z_{2t}+\beta_{3}z_{3t}+\varepsilon_{t}$}\tabularnewline
 & {\small{}$\ARX\left(1,1\right)\,M_{B}:y_{t}=\phi_{1}y_{t-2}+\beta_{1}+\varepsilon_{t}$}\tabularnewline
\hline 
\end{tabular}{\small\par}

\caption{Additional model selection experiments. Experiment~1 corresponds to the DGP and candidate models in Section~\ref{sec:sel-dep}. Each experiment in this Appendix compares two commonly-deployed CV methods: LOO (which has a univariate scoring rule, hence evaluated pointwise), and $hv$-block CV with $h=3$ and $v=3$, evaluated jointly.}

\label{tbl:examples}
\end{table}

\begin{table}[H]
\begin{centering}
\begin{tabular}{lcccccccccc}
\multicolumn{11}{c}{\textbf{Supplementary experiments: adverse selection rates}}\tabularnewline
\multicolumn{11}{c}{Probability of adverse selection (\%)}\tabularnewline[0.4cm]
\hline 
\multirow{3}{*}{} &  & \multicolumn{4}{c}{`Easy' variant $\left(\beta^{\textnormal{easy}}\right)$} &  & \multicolumn{4}{c}{`Hard' variant $\left(\beta^{\textnormal{hard}}\right)$}\tabularnewline
\cline{3-6} \cline{4-6} \cline{5-6} \cline{6-6} \cline{8-11} \cline{9-11} \cline{10-11} \cline{11-11} 
 &  & \multicolumn{4}{c}{$\alpha$} &  & \multicolumn{4}{c}{$\alpha$}\tabularnewline
 &  & $0$ & $\frac{1}{2}$ & $\frac{3}{4}$ & $1$ &  & $0$ & $\frac{1}{2}$ & $\frac{3}{4}$ & $1$\tabularnewline
\cline{1-1} \cline{3-6} \cline{4-6} \cline{5-6} \cline{6-6} \cline{8-11} \cline{9-11} \cline{10-11} \cline{11-11} 
\noalign{\vskip0.3cm}
\multicolumn{11}{l}{Experiment 1}\tabularnewline
\quad{}LOO &  & 0.0 & 0.0 & 0.0 & 9.1 &  & 4.1 & 1.8 & 10.1 & 93.9\tabularnewline
\quad{}$hv$-block(3,3) &  & 0.0 & 0.0 & 0.0 & 0.0 &  & 5.9 & 6.0 & 7.3 & 14.8\tabularnewline
\noalign{\vskip0.3cm}
Experiment 2 &  &  &  &  &  &  &  &  &  & \tabularnewline
\quad{}LOO &  & 0.0 & 0.0 & 0.0 & 1.5 &  & 0.0 & 0.0 & 0.0 & 1.0\tabularnewline
\quad{}$hv$-block(3,3) &  & 0.0 & 0.0 & 0.0 & 0.0 &  & 0.0 & 0.0 & 0.0 & 0.0\tabularnewline
\noalign{\vskip0.3cm}
Experiment 3 &  &  &  &  &  &  &  &  &  & \tabularnewline
\quad{}LOO &  & 0.0 & 0.0 & 0.0 & 2.5 &  & 4.1 & 1.5 & 8.7 & 82.9\tabularnewline
\quad{}$hv$-block(3,3) &  & 0.0 & 0.0 & 0.0 & 0.0 &  & 5.9 & 4.8 & 4.6 & 7.8\tabularnewline
\noalign{\vskip0.3cm}
Experiment 4 &  &  &  &  &  &  &  &  &  & \tabularnewline
\quad{}LOO &  & 0.0 & 0.0 & 0.0 & 85.7 &  & 4.5 & 6.3 & 28.8 & 98.5\tabularnewline
\quad{}$hv$-block(3,3) &  & 0.0 & 0.0 & 0.0 & 92.8 &  & 5.0 & 8.6 & 32.9 & 98.7\tabularnewline
\noalign{\vskip0.3cm}
Experiment 5 &  &  &  &  &  &  &  &  &  & \tabularnewline
\quad{}LOO &  & 0.0 & 0.0 & 0.0 & 0.2 &  & 0.0 & 0.0 & 0.0 & 0.1\tabularnewline
\quad{}$hv$-block(3,3) &  & 0.0 & 0.0 & 0.0 & 0.0 &  & 0.0 & 0.0 & 0.0 & 0.0\tabularnewline
\hline 
\end{tabular}
\par\end{centering}
\caption{Adverse selection rates for experiments in Table~\ref{tbl:examples}.}
\label{tbl:examples-adv-sel}
\end{table}

\begin{table}[H]
\begin{centering}
\begin{tabular}{lcccccccccc}
\multicolumn{11}{c}{\textbf{Supplementary experiments: variability}}\tabularnewline
\multicolumn{11}{c}{Model selection statistic standard deviation}\tabularnewline[0.4cm]
\hline 
\multirow{3}{*}{} &  & \multicolumn{4}{c}{`Easy' variant $\left(\beta^{\textnormal{easy}}\right)$} &  & \multicolumn{4}{c}{`Hard' variant $\left(\beta^{\textnormal{hard}}\right)$}\tabularnewline
\cline{3-6} \cline{4-6} \cline{5-6} \cline{6-6} \cline{8-11} \cline{9-11} \cline{10-11} \cline{11-11} 
 &  & \multicolumn{4}{c}{$\alpha$} &  & \multicolumn{4}{c}{$\alpha$}\tabularnewline
 &  & $0$ & $\frac{1}{2}$ & $\frac{3}{4}$ & $1$ &  & $0$ & $\frac{1}{2}$ & $\frac{3}{4}$ & $1$\tabularnewline
\cline{1-1} \cline{3-6} \cline{4-6} \cline{5-6} \cline{6-6} \cline{8-11} \cline{9-11} \cline{10-11} \cline{11-11} 
\noalign{\vskip0.3cm}
\multicolumn{11}{l}{Experiment 1}\tabularnewline
\quad{}LOO &  & 5.35 & 6.26 & 8.32 & 10.59 &  & 1.98 & 2.38 & 2.78 & 2.37\tabularnewline
\quad{}$hv$-block(3,3) &  & 5.36 & 5.37 & 5.38 & 4.77 &  & 1.91 & 1.92 & 1.88 & 1.54\tabularnewline
\noalign{\vskip0.3cm}
Experiment 2 &  &  &  &  &  &  &  &  &  & \tabularnewline
\quad{}LOO &  & 6.13 & 7.68 & 9.99 & 11.76 &  & 6.11 & 7.54 & 9.59 & 10.46\tabularnewline
\quad{}$hv$-block(3,3) &  & 6.07 & 6.23 & 6.33 & 5.89 &  & 6.06 & 6.19 & 6.27 & 5.81\tabularnewline
\noalign{\vskip0.3cm}
Experiment 3 &  &  &  &  &  &  &  &  &  & \tabularnewline
\quad{}LOO &  & 5.35 & 6.47 & 8.74 & 10.90 &  & 1.98 & 2.50 & 3.02 & 2.83\tabularnewline
\quad{}$hv$-block(3,3) &  & 5.36 & 5.38 & 5.42 & 5.02 &  & 1.91 & 1.98 & 2.00 & 1.75\tabularnewline
\noalign{\vskip0.3cm}
Experiment 4 &  &  &  &  &  &  &  &  &  & \tabularnewline
\quad{}LOO &  & 0.0 & 0.0 & 0.0 & 85.7 &  & 4.5 & 6.3 & 28.8 & 98.5\tabularnewline
\quad{}$hv$-block(3,3) &  & 0.0 & 0.0 & 0.0 & 92.8 &  & 5.0 & 8.6 & 32.9 & 98.7\tabularnewline
\noalign{\vskip0.3cm}
Experiment 5 &  &  &  &  &  &  &  &  &  & \tabularnewline
\quad{}LOO &  & 7.04 & 7.86 & 9.60 & 11.23 &  & 6.27 & 7.32 & 8.72 & 6.60\tabularnewline
\quad{}$hv$-block(3,3) &  & 7.04 & 6.97 & 6.87 & 6.12 &  & 6.19 & 6.16 & 6.05 & 5.24\tabularnewline
\hline 
\end{tabular}
\par\end{centering}
\caption{CV model selection statistic standard deviation for experiments in
Table~\ref{tbl:examples}.}
\label{tbl:examples-std}
\end{table}

\section{Autoregressive dependence\label{subsec:Q}}

Start with an $\ARX(p,q)$ with common noise variance $\sigma^{2}>0$,
where
\begin{equation}
y_{t}\,|\,y_{\left(t-p\right):\left(t-1\right)},\phi,\sigma^2\sim \mathcal{N}\left(\sum_{i=1}^{p}\phi_{i}y_{t-i}+z_{t}^{\top}\beta,\sigma^{2}\right),\quad t=1,\dots,T.
\end{equation}
We denote $y_{r:s}=\left(y_r,y_{r+1},\dots,y_{s-1},y_s\right).$ For simplicity, we will fix the $p$ initial values $y_{0}=\dots=y_{1-p}=0$,
\begin{equation}
y_{t}\,|\,y_{\mathrm{max}\left(1,t-p\right):\left(t-1\right)},\phi,\sigma^2\sim \mathcal{N}\left(\sum_{i=1}^{\mathrm{min}\left(p,t-1\right)}\phi_{i}y_{t-i}+z_{t}^{\top}\beta,\sigma^{2}\right),\quad t=1,\dots,T.\label{eq:cond:zero}
\end{equation}
It follows that the joint distribution of the vector $y$ is multivariate
normal, as summarized by Lemma~\ref{lem:Vdef}.
\begin{lem}
\label{lem:Vdef}Let $y=\left(y_1,\dots,y_T\right)$ be distributed according to (\ref{eq:cond:zero})
with $y_{0}=\dots=y_{1-p}=0$. Then its joint density is multivariate
normal,
\begin{equation}
y\ |\ \phi,\beta,\sigma^{2}\sim\mathcal{N}\left(L_{\phi}^{-1}Z\beta,\sigma^{2}W_{\phi}\right),
\end{equation}
where the $T\times T$ covariance matrix $W_{\phi}=\left(L_{\phi}^{\top}L_{\phi}\right)^{-1}$
is positive definite with $\left|W_{\phi}\right|=1$. The $T\times T$
banded lower triangular coefficient matrix $L_{\phi}$ has entries
\begin{equation}
\left(L_{\phi}\right)_{s,t}=\begin{cases}
1 & \textnormal{for }s=t\\
-\phi_{s-t} & \textnormal{for }1\leq s-t\leq p\\
0 & \textnormal{otherwise}.
\end{cases}\label{eq:def:L}
\end{equation}
\end{lem}

\begin{proof}
First note that $L_{\phi}$ is lower-triangular with $1$s on the
diagonal, so $\left|L_{\phi}\right|=1$ and the inverse $L_{\phi}^{-1}$
exists. The $T\times1$ vector $\eta$ with $t$th entry
\begin{equation}
\eta_{t}=y_{t}-\sum_{i=1}^{p}\boldsymbol{1}\!\left\{t-i\geq1\right\}\phi_{i}y_{t-i}-z_{t}^{\top}\beta
\end{equation}
can equivalently be written in vector form as $\eta=L_{\phi}y-Z\beta$, where $Z$
is the matrix with rows $\left(z_{t}^{\top}\right)$ and $\boldsymbol{1}\!\left\{\cdot\right\}$ is
the indicator function. Then, the joint density of $y$ is
\begin{align}
p\left(y\,|\,\phi,\beta,\sigma^2\right) & =p\left(y_{1}\,|\,\phi,\beta,\sigma^2\right)\prod_{t=2}^{T}p\left(y_{t}\,|\,y_{\mathrm{max}\left(1,t-p\right):\left(t-1\right)},\phi,\beta,\sigma^2\right)\\
 & =\left(2\pi\sigma^{2}\right)^{-T/2}\exp\left\{ -\frac{1}{2\sigma^{2}}\eta^{\top}\eta\right\} \\
 & =\left(2\pi\sigma^{2}\right)^{-T/2}\exp\left\{ -\frac{1}{2\sigma^{2}}\left(L_{\phi}y-Z\beta\right)^{\top}\left(L_{\phi}y-Z\beta\right)\right\} \\
 & =\left(2\pi\sigma^{2}\right)^{-T/2}\exp\left\{ -\frac{1}{2\sigma^{2}}\left(y-L_{\phi}^{-1}Z\beta\right)^{\top}L_{\phi}^{\top}L_{\phi}\left(y-L_{\phi}^{-1}Z\beta\right)\right\} \\
 & =\mathcal{N}\left(y\,|\,L_{\phi}^{-1}Z\beta,\sigma^{2}W_{\phi}\right),
\end{align}
as required. To see that $W_{\phi}$ is positive definite, recognize
that $\left|W_{\phi}\right|=\left|L_{\phi}\right|^{-2}=1$, completing
the proof.
\end{proof}
Note that the matrix $L_{\phi}$ is not the Cholesky factor of $V_{\phi}^{-1}$.
Recall that the lower-left Cholesky factor $A$ for $V_{\phi}^{-1}$
is the unique solution to $AA^{\top}=V_{\phi}^{-1}$. In contrast,
we have $L_{\phi}^{\top}L_{\phi}=V_{\phi}^{-1}$.

\section{Simplified model\label{sec:sarx}}

This section derives the simplified $\ARX(p,q)$ model examined in Section~\ref{sec:sel-dep}.

To abstract away the details of particular CV blocking schemes,
in mathematical expressions we will use selection matrixes to construct
subvectors of the training data $y$ and replicate data vector $\tilde{y}$,
as well as parameters of marginal distributions. 
Consider first just a single CV fold. Let the ordered sets
$\mathsf{train}$ and $\mathsf{test}$ be subsets of $\left\{1,2,\cdots,T\right\}$, retaining only the indexes of the
desired test and training subsets. For CV estimators, $\mathsf{train}$ and $\mathsf{test}$ are mutually disjoint. Define the $n\times T$
matrix $\mathring{S}$ and $v\times T$ matrix $\tilde{S}$ such that
$y_{\mathsf{train}}=\mathring{S}y$ and $\tilde{y}_{\mathsf{test}}=\tilde{S}\tilde{y}$.
The elements of $\mathring{S}$ and $\tilde{S}$ are given by
\begin{equation}
\left(\mathring{S}\right)_{j,t}:=\boldsymbol{1}\!\left\{ t=\mathsf{train}{}_{j}\right\} \quad\textnormal{and}\quad\left(\tilde{S}\right)_{k,t}:=\boldsymbol{1}\!\left\{ t=\mathsf{test}{}_{\ell}\right\} ,
\end{equation}
where $\boldsymbol{1}\!\left\{ \cdot\right\} $ denotes the indicator
function, $\mathsf{train}{}_{j}$ and $\mathsf{test}{}_{k}$ respectively
denote the $j$th and $k$th elements of $\mathsf{train}$ and $\mathsf{test}$,
$t=1,\cdots,T$, $j=1,\cdots,n$, and $k=1,\cdots,v$.

\subsection{Posterior and predictive distributions}

Lemma~\ref{lem:Vdef} and (\ref{eq:ar:def:scalar}) imply that the model 
for $y$ given $\phi$, $\beta$, and $\sigma^{2}$ can be written
as
\begin{equation}
p\left(y_{\mathsf{train}}\,|\,\phi,\beta,\sigma^{2}\right)=\mathcal{N}\left(\mathring{S}y\,|\,\mathring{S}L_{\phi}^{-1}Z\beta,\sigma^{2}\mathring{S}W_{\phi}\mathring{S}^{\top}\right)\label{eq:arx-lhood}
\end{equation}
where $y=\left(y_{1},y_{2},\dots,y_{T}\right)$ and $y_{\mathsf{train}}=\mathring{S}y$
by definition. The $T\times q$ matrix of exogenous regressors $Z$
includes a column of ones, and the $T\times T$ lower-triangular coefficient
matrix $L_{\phi}$ is as defined in (\ref{eq:def:L}).

We require the posterior density conditional on fixed values for $\phi$
and $\sigma^{2}$, which we define in Section~\ref{subsec:oracle},
below. The posterior density can be derived analytically in the usual
way for models with conjugate priors, as proportional to the joint
density,
\begin{align}
p\left(\beta\,|\,y_{\mathsf{train}},\sigma^{2},\phi\right) & \propto p\left(y_{\mathsf{train}}\,|\,\beta,\sigma^{2},\phi\right)p\left(\beta\right)\\
 & =\mathcal{N}\left(\mathring{S}y\,|\,\mathring{S}L_{\phi}^{-1}Z\beta,\sigma^{2}\mathring{S}W_{\phi}\mathring{S}^{\top}\right)\mathcal{N}\left(\beta\,|\,\mu_{\beta},\sigma^{2}\Sigma_{\beta}\right)\\
 & \propto\exp\left\{ -\frac{1}{2\sigma^{2}}\left(\beta-\Sigma_{\beta}\mu_{\beta}\right)^{\top}\Sigma_{\beta}^{-1}\left(\beta-\Sigma_{\beta}\mu_{\beta}\right)\right\}.
\intertext{
In the above, we have completed the square and defined
}
\Sigma_{\beta} & =\left[Z^{\top}L_{\phi}^{-\top}\mathring{S}^{\top}\left(\mathring{S}W_{\phi}\mathring{S}^{\top}\right)^{-1}\mathring{S}L_{\phi}^{-1}Z+\Sigma_{0}^{-1}\right]^{-1}\label{eq:sarx:post:cov}\\
\mu_{\beta} & =\Sigma_{\beta}\left[Z^{\top}L_{\phi}^{-\top}\mathring{S}^{\top}\left(\mathring{S}W_{\phi}\mathring{S}^{\top}\right)^{-1}\mathring{S}y+\Sigma_{0}^{-1}\mu_{0}\right].\label{eq:sarx:post:mean}
\end{align}
In the case where the full training set is used so that $\mathring{S}=I_{T}$, we can exploit the fact that $W_\phi^{-1}=L_\phi^\top L_\phi$ to simplify \eqref{eq:sarx:post:cov} as $\Sigma_{\beta}=\left(Z^{\top}Z+\Sigma_{0}^{-1}\right)^{-1}$.

The resulting model $M_{\ell}$ posterior density for $\beta$ (conditional
on $\sigma^{2}$ and $\phi$) is multivariate normal,
\begin{align}
p\left(\beta\,|\,y_{\mathsf{train}},\sigma^{2},\phi\right) & =\mathcal{N}\left(\beta\,|\,\mu_{\beta},\sigma^{2}\Sigma_{\beta}\right).\label{eq:sarx:post}
\end{align}
The posterior predictive density for new data $\tilde{y}_{test}=\tilde{S}\tilde{y}$
can be derived by integrating out the parameter $\beta$ from the
joint distribution of new data and $\beta$ given $y_{\mathsf{train}}$,
\begin{align}
 & p\left(\tilde{y}_{\mathsf{test}}|y_{\mathsf{train}},\sigma^{2},\phi\right)\nonumber \\
 & =\int p\left(\tilde{y}_{\mathsf{test}}\,|\,\beta,\sigma^{2},\phi\right)p\left(\beta\,|\,y_{\mathsf{train}},\sigma^{2},\phi\right)\mathrm{d}\beta\\
 & =\int\mathcal{N}\left(\tilde{S}y\,|\,\tilde{S}L_{\phi}^{-1}Z_{\ell}\beta,\sigma^{2}\tilde{S}W_{\phi}\tilde{S}^{\top}\right)\mathcal{N}\left(\beta\,|\,\mu_{\beta},\sigma^{2}\Sigma_{\beta}\right)\mathrm{d}\beta \\
 & =\mathcal{N}\Bigl(\tilde{S}\tilde{y}\,|\,\tilde{S}\underbrace{L_{\phi}^{-1}Z_{\ell}\mu_{\beta}}_{m_{\ell}},\sigma^{2}\tilde{S}\Bigl(\underbrace{W_{\phi}+L_{\phi}^{-1}Z_{\ell}\Sigma_{\beta}Z_{\ell}^{\top}L_{\phi}^{-\top}}_{V_{\ell}}\Bigr)\tilde{S}^{\top}\Bigr).\label{eq:sarx:pred}
\end{align}

It is worth emphasizing that in our framework the DGP (\ref{eq:ex:dgp-1}) and model
$M_{\ell}$ can represent different underlying models. In
other words, we allow $M_{\ell}$ to be misspecified. This means that
the theoretical and cross-validatory utility measures will be a function
of both the DGP parameters (with $*$ subscripts) and the model $M_{\ell}$
parameters (with $\ell$ subscripts).

\subsection{Theoretical expected log predictive density\label{subsec:sarx-eljd}}

The theoretical log utility under external validation \eqref{eq:elpjd} is simply the
expectation of the log predictive density under the true DGP. In the
simplified version of the model, both the DGP and predictives are
multivariate Gaussian and an exact solution is available analytically
as a consequence of Lemma~\ref{lem:mvn:cross-entropy} (Appendix \ref{app:Proofs}). \citet{Sivula2020a} found that the $\elppd$ for linear 
regressions with flat priors can be expressed
as a second-degree polynomial in $y$.  We find that this is
also true for both the $\eljpd$ and $\elppd$ for our simplified
$\ARX$ model with conjugate priors.

\subsubsection{Expected log joint predictive density (eljpd)\label{subsec:eljpd}}

We will derive the $\eljpd$ in a form that includes selection matrixes
for both testing and training data. This form is general enough that it
can be used to analyze full-data $\elpd$s as well as to construct
cross-validation folds with subsets of the
data. Let the selection matrixes $\mathring{S}$ and $\tilde{S}$
respectively define the training and testing subsets of $y$ and $\tilde{y}$.
Then, dropping the conditioning $\sigma^{2}$ and $\phi$ from the
notation for brevity, the joint theoretical utility for model $M_{\ell}$ given $y$
is,
\begin{align}
 & \eljpd\left(M_{\ell}\,|\,y,\tilde{S}\right)\nonumber \\
 & =\mathbb{E}_{\tilde{y}_{\mathsf{test}}\sim p_{\mathrm{true}}}\left[\log p\left(\tilde{y}_{\mathsf{test}}\,|\,y_{\mathsf{train}},M_{\ell}\right)\right] \\
 & =\int\mathcal{N}\left(\tilde{S}\tilde{y}\,|\,\tilde{S}m_{*},\sigma_{*}^{2}\tilde{S}V_{*}\tilde{S}^{\top}\right)\log\mathcal{N}\left(\tilde{S}\tilde{y}\,|\,\tilde{S}m_{\ell},\sigma_{\ell}^{2}\tilde{S}V_{\ell}\tilde{S}^{\top}\right)\,\mathrm{d}\tilde{y} \\
 & =-\frac{1}{2}\Biggl\{\log\left|2\pi\sigma_{\ell}^{2}\tilde{S}V_{\ell}\tilde{S}^{\top}\right|+\frac{\sigma_{*}^{2}}{\sigma_{\ell}^{2}}\mathrm{tr}\left[\left(\tilde{S}V_{\ell}\tilde{S}^{\top}\right)^{-1}\tilde{S}V_{*}\tilde{S}^{\top}\right]\nonumber \\
 & \qquad+\left(m_{\ell}-m_{*}\right)^{\top}\tilde{S}^{\top}\left(\sigma_{\ell}^{2}\tilde{S}V_{\ell}\tilde{S}^{\top}\right)^{-1}\tilde{S}\left(m_{\ell}-m_{*}\right)\Biggr\}\label{eq:eljpd:expanded}
\end{align}
where we have used (\ref{eq:sarx:ptrue}) and (\ref{eq:sarx:pred})
and Lemma~\ref{lem:mvn:cross-entropy}. Now note that $m_{\ell}$ is
linear in $y$,
\begin{align}
m_{\ell} & =L_{\phi}^{-1}Z_{\ell}\mu_{\beta}=D_{\ell}y+e_{\ell}\label{eq:eljpd:simpler-term}\\
\intertext{\textnormal{where we have used \eqref{eq:arx-lhood} to define}}D_{\ell} & =L_{\phi_{\ell}}^{-1}Z_{\ell}\Sigma_{\beta_{\ell}}Z_{\ell}^{\top}L_{\phi_{\ell}}^{-\top}\mathring{S}^{\top}\left(\mathring{S}W_{\phi}\mathring{S}^{\top}\right)^{-1}\mathring{S},\label{eq:sarx:eljpd:D}\\
e_{\ell} & =L_{\phi_{\ell}}^{-1}Z_{\ell}\Sigma_{\beta_{\ell}}\Sigma_{0}^{-1}\mu_{0}.\label{eq:sarx:eljpd:e}
\end{align}
Observe that in the case where the full training set is used ($\mathring{S}=I_{T}$),
we can use the fact that $W_{\phi}^{-1}=L_{\phi}^{\top}L_{\phi}$
to simplify $D_{\ell}=L_{\phi_{\ell}}^{-1}Z_{\ell}\Sigma_{\beta_{\ell}}Z_{\ell}^{\top}L_{\phi_{\ell}}.$

Finally, rearranging (\ref{eq:eljpd:expanded}) and substituting (\ref{eq:eljpd:simpler-term})
yields the second-degree polynomial in $y$,
\begin{align}
\eljpd\left(M_{\ell}\,|\,y,\tilde{S}\right) & =y^{\top}A_{\ell}y+y^{\top}b_{\ell}+c_{\ell}.\label{eq:eljpd-qf}
\end{align}
The $T\times T$ matrix $A_\ell$, $T\times1$ vector $b_\ell$, and scalar
$c_\ell$ are nonlinear functions of $\phi_{*}$ and $\phi_{\ell}$ (respectively
via $V_{*}$ and $V_{\ell}$) but are free of $y$ and hence nonrandom.
Their values are
\begin{align}
A_{\ell} & =-\frac{1}{2\sigma_{\ell}^{2}}D_{\ell}^{\top}\tilde{S}^{\top}\left(\tilde{S}V_{\ell}\tilde{S}^{\top}\right)^{-1}\tilde{S}D_{\ell}\\
b_{\ell} & =-\frac{1}{\sigma_{\ell}^{2}}D_{\ell}^{\top}\tilde{S}^{\top}\left(\tilde{S}V_{\ell}\tilde{S}^{\top}\right)^{-1}\tilde{S}\left(e_{\ell}-m_{*}\right)\\
c_{\ell} & =-\frac{1}{2}\log\left|2\pi\sigma_{\ell}^{2}\tilde{S}V_{\ell}\tilde{S}^{\top}\right|-\frac{\sigma_{*}^{2}}{2\sigma_{\ell}^{2}}\mathrm{tr}\left[\left(\tilde{S}V_{\ell}\tilde{S}^{\top}\right)^{-1}\tilde{S}V_{*}\tilde{S}^{\top}\right]\nonumber\\
 & \qquad-\frac{1}{2\sigma_{\ell}^{2}}\left(e_{\ell}-m_{*}\right)^{\top}\tilde{S}^{\top}\left(\tilde{S}V_{\ell}\tilde{S}^{\top}\right)^{-1}\tilde{S}\left(e_{\ell}-m_{*}\right).
\end{align}
It can be seen from the form of (\ref{eq:eljpd:expanded}) and positive
definiteness of $\tilde{S}^{\top}\left(\sigma_{\ell}^{2}\tilde{S}V_{\ell}\tilde{S}^{\top}\right)^{-1}\tilde{S}$
that for a given draw of $y$, the log utility measure $\eljpd\left(M_{\ell}\,|\,y\right)$
is larger the closer the posterior predictive mean $m_{\ell}$
is to the DGP mean $m_{*}$, thus assigning a greater score to better
predictions.

\subsubsection{Expected log pointwise predictive density (elppd)\label{subsec:elppd}}

The $\elppd$ can be derived from the $\eljpd$. We use the fact that
the marginal distribution of a multivariate normal distribution is
univariate normal, where $m_{\ell,t}$ denotes the $t$th element of
$m_{\ell}$ and $V_{\ell,t,t}$ the $t$th diagonal element of $V_{\ell}.$
Recognizing that 
\begin{equation}
\sum_{t=1}^{T}\log\mathcal{N}\left(\tilde{y}_{t}\,|\,m_{\ell,t},\sigma_{\ell}^{2}V_{\ell,t,t}\right)=\log\mathcal{N}\left(\tilde{y}\,|\,m_{\ell},\sigma_{\ell}^{2}\left(I_{T}\odot V_{\ell}\right)\right),\label{eq:marg-pw-gauss}
\end{equation}
for $\odot$ the elementwise product operator, we can rewrite the
elppd in a form similar to (\ref{eq:eljpd-qf}) as
\begin{align}
\elppd\left(M_{\ell}|y\right) & =-\frac{1}{2}\left[\log\left|2\pi\sigma_{\ell}^{2}I_{T}\odot V_{\ell}\right|+\mathrm{tr}\left(\left(I_{T}\odot V_{\ell}\right)^{-1}V_{*}\right)\right.\nonumber \\
 & \qquad\left.+\left(m_{\ell}-m_{*}\right)^{\top}\left(\sigma_{\ell}^{2}I_{T}\odot V_{\ell}\right)^{-1}\left(m_{\ell}-m_{*}\right)\right]\\
 & =y^{\top}\bar{A}_{\ell}y+y^{\top}\bar{b}_{\ell}+\bar{c}_{\ell},\label{eq:sarx-elpd-quad}
\end{align}
where
\begin{align}
\bar{A}_{\ell} & =-\frac{1}{2\sigma_{\ell}^{2}}D_{\ell}^{\top}\tilde{S}^{\top}\left(\tilde{S}\left(I_{T}\odot V_{\ell}\right)\tilde{S}^{\top}\right)^{-1}\tilde{S}D_{\ell}\\
\bar{b}_{\ell} & =-\frac{1}{\sigma_{\ell}^{2}}D_{\ell}^{\top}\tilde{S}^{\top}\left(\tilde{S}\left(I_{T}\odot V_{\ell}\right)\tilde{S}^{\top}\right)^{-1}\tilde{S}\left(e_{\ell}-m_{*}\right)\\
\bar{c}_{\ell} & =-\frac{1}{2}\log\left|2\pi\sigma_{\ell}^{2}\tilde{S}\left(I_{T}\odot V_{\ell}\right)\tilde{S}^{\top}\right|-\frac{\sigma_{*}^{2}}{2\sigma_{\ell}^{2}}\mathrm{tr}\left[\left(\tilde{S}\left(I_{T}\odot V_{\ell}\right)\tilde{S}^{\top}\right)^{-1}\tilde{S}V_{*}\tilde{S}^{\top}\right]\nonumber\\
 & \qquad-\frac{1}{2\sigma_{\ell}^{2}}\left(e_{\ell}-m_{*}\right)^{\top}\tilde{S}^{\top}\left(\tilde{S}\left(I_{T}\odot V_{\ell}\right)\tilde{S}^{\top}\right)^{-1}\tilde{S}\left(e_{\ell}-m_{*}\right).
\end{align}

\subsection{Cross-validation estimators\label{subsec:CV-eljd}}

This section derives CV estimators for the $\eljpd$ and $\elppd$.
We will keep the notation generic and construct expressions that apply
for all the CV described in Section~\ref{subsec:blocking}.
The results of this section can be adapted for LOO, LFO, K-fold, h-block,
and hv-block simply by substituting method-specific formulations of
$\mathsf{train}_{k}$ and $\mathsf{test}_{k}$ via the selection
matrixes $\mathring{S}_{k}$ and $\tilde{S}_{k}$ for each fold $k=1,\cdots,K$.

Given the CV method parameters $\left(K,\left\{ \mathsf{test}_{k}\right\} _{k=1}^{K},\left\{ \mathsf{train}_{k}\right\} _{k=1}^{K}\right)$,
the joint and pointwise CV estimators for a simplified $\ARX$ model can
be constructed by substituting (\ref{eq:sarx:pred}) into (\ref{eq:def:cv})
and (\ref{eq:def:jcv}), putting $\tilde{y}_{\mathsf{test}}=y_{\mathsf{test}}$:
\begin{align*}
\widehat{\eljpd}_{CV}\left(M_{\ell}\,|\,y\right) & =\frac{T}{K}\sum_{k=1}^{N}\frac{1}{v_{k}}\log p\left(y_{\mathsf{test}}\,|\,y_{\mathsf{train}},\sigma^{2},\phi,M_{\ell}\right),
\end{align*}
where $v_k = \left|\mathsf{test}_k\right|$ is the size of the $k$th test set.

Let $m_{\ell\left[k\right]}$ and $V_{\ell\left[k\right]}$ denote the predictive
parameters (\ref{eq:sarx:pred}) associated with the $k$th fold.
Then
\begin{align}
\widehat{\eljpd}_{CV}\left(M_{\ell}\,|\,y\right) & =\frac{T}{K}\sum_{k=1}^{K}\frac{1}{v_{k}}\log\mathcal{N}\left(\tilde{S}_{k}y\,|\,\tilde{S}_{k}m_{\ell\left[k\right]},\sigma_{\ell}^{2}\tilde{S}_{k}V_{\ell\left[k\right]}\tilde{S}_{k}^{\top}\right)\\
 & =-\frac{T}{2K}\sum_{k=1}^{K}\frac{1}{v_{k}}\Biggl[\log\left|2\pi\sigma_{\ell}^{2}\tilde{S}_{k}V_{\ell\left[k\right]}\tilde{S}_{k}^{\top}\right|+\nonumber \\
 & \quad\left.\left(y-m_{\ell\left[k\right]}\right)^{\top}\tilde{S}_{k}^{\top}\left(\sigma_{\ell}^{2}\tilde{S}_{k}V_{\ell\left[k\right]}\tilde{S}_{k}^{\top}\right)^{-1}\tilde{S}_{k}\left(y-m_{\ell\left[k\right]}\right)\right].\label{eq:eljpdhat:expanded}
\end{align}
We can write the difference that appears in (\ref{eq:eljpdhat:expanded})
as
\begin{align}
y-m_{\ell\left[k\right]} & =y-L_{\phi}^{-1}Z_{\ell}\mu_{\beta}=\left(I_{T}-D_{\ell\left[k\right]}\right)y-e_{\ell\left[k\right]}
\end{align}
where $D_{\ell\left[k\right]}$ and $e_{\ell\left[k\right]}$ denote fold-$k$-specific
values of $D_{\ell}$ (\ref{eq:sarx:eljpd:D}) and $e_{\ell}$ (\ref{eq:sarx:eljpd:e}),
\begin{align}
D_{\ell\left[k\right]} & =L_{\phi_{\ell}}^{-1}Z_{\ell}\Sigma_{\beta_{\ell}\left[k\right]}Z_{\ell}^{\top}L_{\phi_{\ell}}^{-\top}\mathring{S}^{\top}\left(\mathring{S}W_{\phi}\mathring{S}^{\top}\right)^{-1}\mathring{S},\\
e_{\ell\left[k\right]} & =L_{\phi_{\ell}}^{-1}Z_{\ell}\Sigma_{\beta_{\ell}\left[k\right]}\Sigma_{0}^{-1}\mu_{0}.
\end{align}
Rearranging (\ref{eq:eljpdhat:expanded}) yields a second-degree polynomial
\begin{equation}
\widehat{\eljpd}_{CV}\left(M_{\ell}\,|\,y\right)=y^{\top}\widehat{A_{\ell}}y+y^{\top}\widehat{b_{\ell}}+\widehat{c_{\ell}}.\label{eq:eljpdhat:qf}
\end{equation}
with coefficients
\begin{align}
\widehat{A_{\ell}} & =-\frac{T}{2K\sigma_{\ell}^{2}}\sum_{k=1}^{K}\frac{1}{v_{k}}\left(I_{T}-D_{k\left[k\right]}\right)^{\top}\tilde{S}_{k}^{\top}\left(\tilde{S}_{k}V_{\ell\left[k\right]}\tilde{S}_{k}^{\top}\right)^{-1}\tilde{S}_{k}\left(I_{T}-D_{\ell\left[k\right]}\right),\\
\widehat{b_{\ell}} & =\frac{T}{K\sigma_{\ell}^{2}}\sum_{k=1}^{K}\frac{1}{v_{i}}\left(I_{T}-D_{\ell\left[k\right]}\right)^{\top}\tilde{S}_{k}^{\top}\left(\tilde{S}_{k}V_{\ell\left[k\right]}\tilde{S}_{k}^{\top}\right)^{-1}\tilde{S}_{k}e_{\ell\left[k\right]},\\
\widehat{c_{\ell}} & =-\frac{T}{2}\log\left(2\pi\sigma_{\ell}^{2}\right)-\frac{T}{2K}\sum_{k=1}^{K}\frac{1}{v_{k}}\log\left|\tilde{S}_{k}V_{\ell\left[k\right]}\tilde{S}_{k}^{\top}\right|\nonumber\\
 & \qquad-\frac{T}{2N\sigma_{\ell}^{2}}\sum_{k=1}^{K}\frac{1}{v_{k}}e_{\ell\left[k\right]}^{\top}\tilde{S}_{k}^{\top}\left(\tilde{S}_{k}V_{\ell\left[k\right]}\tilde{S}_{k}^{\top}\right)^{-1}\tilde{S}_{k}e_{\ell\left[k\right]}.
\end{align}
Recall that in the context of these expressions, the various CV blocking
patterns---LOO, LFO, $h$-block, $hv$-block, and $K$-fold---differ
only by the definitions of $\mathring{S}_{k}$ and $\tilde{S}_{k}$.

\subsubsection{Pointwise evaluation of CV estimators\label{subsec:elppd-cv}}

Pointwise CV estimators are special cases of joint estimators. They
can be constructed in a similar manner to the pointwise $\elppd$
as
\begin{equation}
\widehat{\elppd}_{CV}\left(M_{\ell}\,|\,y\right)=\frac{T}{K}\sum_{k=1}^{K}\frac{1}{v_{k}}\sum_{t\in\mathsf{test}_k}\log p\left(y_{t}|y_{\mathsf{train}},\sigma^{2},\phi,M_{\ell}\right).
\end{equation}
For this specific model, we can again exploit multivariate normality
of the fold predictives. The pointwise predictive $p\left(y_{t}|y_{\mathsf{train}},\sigma^{2},\phi,M_{\ell}\right)$
that appears in $\widehat{\elppd}_{CV}\left(M_{\ell}\,|\,y\right)$,
is constructed by marginalizing out unneeded variables from \eqref{eq:sarx:pred}.
The identity \eqref{eq:marg-pw-gauss} shows that this is the same as
diagonalizing the predictive covariance matrix. It follows that,
\begin{equation}
\widehat{\elppd}_{CV}\left(M_{\ell}\,|\,y\right)=y^{\top}\widetilde{A_{\ell}}y+y^{\top}\widetilde{b_{\ell}}+\widetilde{c_{\ell}}.\label{eq:elppdhat:qf}
\end{equation}
with coefficients
\begin{align}
\widetilde{A_{\ell}} & =-\frac{T}{2K\sigma_{\ell}^{2}}\sum_{k=1}^{K}\frac{1}{v_{k}}\left(I_{T}-D_{\ell\left[k\right]}\right)^{\top}\tilde{S}_{k}^{\top}\left(\tilde{S}_{k}\left(V_{\ell\left[k\right]}\odot I_T\right)\tilde{S}_{k}^{\top}\right)^{-1}\tilde{S}_{k}\left(I_{T}-D_{\ell\left[k\right]}\right),\\
\widetilde{b_{\ell}} & =\frac{T}{K\sigma_{\ell}^{2}}\sum_{k=1}^{K}\frac{1}{v_{k}}\left(I_{T}-D_{\ell\left[k\right]}\right)^{\top}\tilde{S}_{k}^{\top}\left(\tilde{S}_{k}\left(V_{\ell\left[k\right]}\odot I_T\right)\tilde{S}_{k}^{\top}\right)^{-1}\tilde{S}_{k}e_{\ell\left[k\right]},\\
\widetilde{c_{\ell}} & =-\frac{T}{2}\log\left(2\pi\sigma_{\ell}^{2}\right)-\frac{T}{2K}\sum_{k=1}^{K}\frac{1}{v_{k}}\log\left|\tilde{S}_{k}\left(V_{\ell\left[k\right]}\odot I_T\right)\tilde{S}_{k}^{\top}\right|\nonumber\\
 & \qquad-\frac{T}{2K\sigma_{\ell}^{2}}\sum_{k=1}^{K}\frac{1}{v_{k}}e_{\ell\left[k\right]}^{\top}\tilde{S}_{k}^{\top}\left(\tilde{S}_{k}\left(V_{\ell\left[k\right]}\odot I_T\right)\tilde{S}_{k}^{\top}\right)^{-1}\tilde{S}_{k}e_{\ell\left[k\right]}.
\end{align}

\subsection{Distribution function for quadratic polynomials\label{subsec:dist-fun}}

Proposition~\ref{prop:dist-omega} shows that the quadratic polynomials
$\omega\left(y\right)$ have a generalized $\chi^{2}$ distribution.
This section provides a formal definition for this distribution class.
\begin{defn}[Generalized $\chi^{2}$ distribution]
\label{def:gchisq}If the law of a variable $\omega$ can be expressed
as 
\begin{equation}
\omega\sim X_{0} + \sum_{j=1}^{k}\lambda_{j}X_{j}\label{eq:poly-sum-indep}
\end{equation}
 where the variables $X_{0},X_{1},\dots,X_{k}$ are mutually independent,
$X_{0}\sim \mathcal{N}\!\left(\mu,\sigma^{2}\right)$, and the remaining are noncentral
$\chi^{2}$ variables, with $r_j$ degrees of freedom and noncentrality $\delta^2_j$,
$X_{j}\sim\chi^{2}\left(r_{j},\delta^2_{j}\right)$, for $j=1,\dots,k$.
Then, we say the distribution of $\omega$ is generalized
$\chi^{2}$, and write $\omega\sim\tilde{\chi}^{2}\left(\boldsymbol{\lambda},\boldsymbol{r},\boldsymbol{\delta},\mu,\sigma\right)$.
\end{defn}

Quantiles of the generalized $\chi^{2}$ distribution can be approximated by simulation or
from its CDF (Lemma~\ref{lem:genchisq}) using the method of \citet{Davies1973}.
\begin{lem}
\label{lem:genchisq}Let $\omega$ follow a generalized $\chi^{2}$
distribution, 
\begin{equation}
\omega\sim\tilde{\chi}^{2}\left(\boldsymbol{\lambda},\boldsymbol{r},\boldsymbol{\delta},\mu,\sigma\right).
\end{equation}
Then the cumulative distribution function $F_{\omega}\left(w\right)=\mathrm{Pr}\left(w<\omega\right)$ is
\begin{equation}
F_{\omega}\left(w\right)=\frac{1}{2}+\int_{\mathbb{R}}\frac{\exp\left\{ i\mu t-iwt-\sigma^{2}t^{2}/2+it\sum_{j=1}^{k}\frac{\lambda_j\delta_{j}^{2}}{\left(1-2i\lambda_{j}t\right)}\right\} }{\pi it\prod_{j=1}^{k}\left(1-2i\lambda_{j}t\right)^{r_j/2}}\,\mathrm{d}t,
\end{equation}
where $i=\sqrt{-1}$ is the imaginary unit vector.
\end{lem}

\section{Complete model\label{sec:complete}}

This section derives the `full Bayes' $\ARX\left(1,q\right)$ model
used in the experiment of Section~\ref{sec:experiment}.
The quantity of interest is the elpd for a given data vector $y$, 
\begin{align*}
\mathrm{elpd}\left(M_\ell | y\right) & =\int p\left(\tilde{y}_{\mathsf{test}}\,|\,\phi_{*},\beta_{*},\sigma_{*}^{2}\right)\log p\left(\tilde{y}_{\mathsf{test}}\,|\,y_{\mathsf{train}}, M_\ell\right)\,\mathrm{d}\tilde{y}_{\mathsf{test}},
\end{align*}
where we have suppressed the conditioning exogenous covariates $Z_{\ell}$
from the notation.

To simplify calculations, we impose priors $\beta\sim\mathcal{N}\left(\mu_{0},\sigma^{2}\Sigma_{0}\right)$
and $\sigma^{2}\sim\mathcal{IG}\left(a_{0},b_{0}\right)$. To ensure
covariance stationarity the prior on the univariate variable $\phi$
has support only on the one-dimensional open unit disc and is beta
distributed, $\phi\sim\mathcal{BE}_{(-1,1)}\left(c_{0},d_{0}\right)$,
where $\mathcal{BE}_{(l,u)}\left(\phi\,|\,c,d\right):=\mathcal{BE}\left(\frac{\phi-l}{u-l}\,|\,c,d\right)$
denotes the standard beta distribution scaled to have support $\left(l,u\right)$.

The DGP is multivariate normal with a covariance structure that depends
on $\phi_{*}$ (via $W_{\phi_{*}}$) and $\sigma_{*}^{2}$:
\begin{align}
p_{\mathrm{true}}\left(y\right) & =\mathcal{N}\left(y\,|\,m_{*},\sigma_{*}^{2}V_{*}\right),\quad\textnormal{where}\ m_{*}=L_{*}^{-1}Z_{*}\beta_{*}\ \textnormal{and}\ V_{*}=W_{\phi_{*}}.\label{eq:sarx:ptrue}
\end{align}

Now consider a candidate model $M_{\ell}$, with $T\times p_{\ell}$ matrix
of covariates $Z_{\ell}$ and autoregressive lag structure $\phi_{\ell}$,
where $\phi_{\ell}$ is $p_{\ell}\times1$. 


The joint density for the combined parameter and the training data
is given by
\begin{align}
p\left(\beta,\phi,\sigma^{2},y_{\mathsf{train}}\right) & =p\left(\phi\right)p\left(\beta\,|\,\sigma^{2}\right)p\left(\sigma^{2}\right)p\left(y_{\mathsf{train}}\,|\,\beta,\phi,\sigma^{2}\right) \\
 & =\mathcal{BE}_{(-1,1)}\left(\phi\,|\,c_{0},d_{0}\right)\mathcal{N}\left(\beta\,|\,\mu_{0},\sigma^{2}\Sigma_0\right)\mathcal{IG}\left(\sigma^{2}\,|\,a_{0},b_{0}\right)\times\nonumber \\
 & \qquad\qquad\mathcal{N}\left(\mathring{S}y\,|\,\mathring{S}L_{\phi}^{-1}Z\beta,\sigma^{2}\mathring{S}W_{\phi}\mathring{S}^{\top}\right).\label{eq:arx:joint}
\end{align}
The resulting posterior density is not available in closed form, but
can be estimated with MCMC, one-dimensional numerical integration, or
Laplace approximation. To make computations tractable, we estimate
cross-validatory estimates $\widehat{elpd}_{\mathrm{CV}}$ by quadrature
for each fold, and we estimate the (theoretical) elpd using MCMC.

The posterior density follows from an application of Bayes' rule,
\begin{equation}
p\left(\beta,\phi,\sigma^{2}\,|\,y_{\mathsf{train}}\right)=\frac{p\left(\beta,\phi,\sigma^{2},y_{\mathsf{train}}\right)}{p\left(y_{\mathsf{train}}\right)}=\frac{p\left(\beta,\phi,\sigma^{2},y_{\mathsf{train}}\right)}{\int\negthinspace\int\negthinspace\int p\left(y_{\mathsf{train}},\beta,\phi,\sigma^{2}\right)\,\mathrm{d}\beta\,\mathrm{d}\sigma^{2}\,\mathrm{d}\phi}.\label{eq:arx:post:bayes}
\end{equation}
Evaluating \eqref{eq:arx:post:bayes} therefore requires the marginal likelihood
$p\left(y_{\mathsf{train}}\right)$.

\subsection{Marginal likelihood\label{subsec:full:post-pred}}

The marginal likelihood can be obtained by integration,
\begin{align}
p\left(y_{\mathsf{train}}\right)
& =\int\negthinspace\int\negthinspace\int p\left(y_{\mathsf{train}},\beta,\phi,\sigma^{2}\right)\,\mathrm{d}\beta\,\mathrm{d}\sigma^{2}\,\mathrm{d}\phi \\
& =\int\negthinspace\int\negthinspace\int 
p\left(y_{\mathsf{train}},\beta\,|\,\sigma^{2},\phi\right)
p\left(\sigma^2 | \phi\right)
p\left(\phi\right)
\,\mathrm{d}\beta\,\mathrm{d}\sigma^{2}\,\mathrm{d}\phi\\
& =\int\negthinspace\int\negthinspace\int 
p\left(y_{\mathsf{train}},\beta\,|\,\sigma^{2},\phi\right)\,\mathrm{d}\beta\,
p\left(\sigma^2\right)\,\mathrm{d}\sigma^{2}\,
p\left(\phi\right)\,\mathrm{d}\phi.
\end{align}

We first compute $p\left(y_{\mathsf{train}}\,|\,\sigma^{2},\phi\right)$
by marginalizing out $\beta$ from the joint density of training
data, testing data, and $\beta$ conditional on $\phi$ and $\sigma^2$,
which is jointly multivariate normal. We have 
\begin{equation}
\begin{pmatrix}y_{\mathsf{train}}\\
\tilde{y}_{\mathsf{test}}\\
\beta
\end{pmatrix}\equiv\begin{pmatrix}\mathring{S}y\\
\tilde{S}\tilde{y}\\
\beta
\end{pmatrix} | \phi, \sigma^2 \sim\mathcal{N}\left(\begin{pmatrix}\mu_{y}\\
\mu_{\tilde{y}}\\
\mu_{\beta}
\end{pmatrix},\sigma^{2}\begin{pmatrix}\Sigma_{yy} & \Sigma_{y\tilde{y}} & \Sigma_{y\beta}\\
\Sigma_{\tilde{y}y} & \Sigma_{\tilde{y}\tilde{y}} & \Sigma_{\tilde{y}\beta}\\
\Sigma_{\beta y} & \Sigma_{\beta\tilde{y}} & \Sigma_{\beta\beta}
\end{pmatrix}\right)\label{eq:joint:train-test-param}
\end{equation}
where the mean is
\begin{equation}
\begin{pmatrix}\mu_{y}\\
\mu_{\tilde{y}}\\
\mu_{\beta}
\end{pmatrix}=\begin{pmatrix}\mathring{S}L_{\phi}^{-1}Z\mu_{0}\\
\tilde{S}L_{\phi}^{-1}Z\mu_{0}\\
\mu_{0}
\end{pmatrix}
\end{equation}
and the $3\times3$ block-partitioned covariance matrix in \eqref{eq:joint:train-test-param}
is
\[
\begin{pmatrix}\mathring{S}L_{\phi}^{-1}\left(I_{T}+Z\Sigma_0Z^{\top}\right)L_{\phi}^{-\top}\mathring{S}^{\top} & \mathring{S}L_{\phi}^{-1}Z\Sigma_0Z^{\top}L_{\phi}^{-\top}\tilde{S}^{\top} & \mathring{S}L_{\phi}^{-1}Z\Sigma_0\\
\tilde{S}L_{\phi}^{-1}Z\Sigma_0Z^{\top}L_{\phi}^{-\top}\mathring{S}^{\top} & \tilde{S}L_{\phi}^{-1}\left(I_{T}+Z\Sigma_0Z^{\top}\right)L_{\phi}^{-\top}\tilde{S}^{\top} & \tilde{S}L_{\phi}^{-1}Z\Sigma_0\\
\Sigma_0Z^{\top}L_{\phi}^{-\top}\mathring{S}^{\top} & \Sigma_0Z^{\top}L_{\phi}^{-\top}\tilde{S}^{\top} & \Sigma_0
\end{pmatrix}.
\]
It follows that
\begin{align}
p\left(y_{\mathsf{train}}\,|\,\sigma^{2},\phi\right) & =\int p\left(y_{\mathsf{train}},\beta\,|\,\sigma^{2},\phi\right)\,d\beta\\
 & =\mathcal{N}\Bigl(\mathring{S}y\,|\,\underbrace{\mathring{S}L_{\phi}^{-1}Z\mu_{0}}_{\mu_{y|\phi}},\sigma^{2}\underbrace{\mathring{S}\left(W_{\phi}+L_{\phi}^{-1}Z\Sigma_{0}Z^{\top}L_{\phi}^{-\top}\right)\mathring{S}^{\top}}_{\Sigma_{y|\phi}}\Bigr).
\end{align}
Integrating out $\sigma^{2}$ shows that $p\left(y_{\mathsf{train}}\,|\,\phi\right)$
is a multivariate normal-inverse-gamma density:
\begin{align}
p\left(y_{\mathsf{train}}\,|\,\phi\right) & =\int p\left(y_{\mathsf{train}}\,|\,\sigma^{2},\phi\right)p\left(\sigma^{2}\,|\,\phi\right)\,\mathrm{d}\sigma^{2}\\
 & =\int N\left(\mathring{S}y\,|\,\mu_{y|\phi},\sigma^{2}\Sigma_{y|\phi}\right)\mathcal{IG}\left(\sigma^{2}\,|\,a_{0},b_{0}\right)\,\mathrm{d}\sigma^{2}\\
 & =\mathcal{NIG}\left(\mathring{S}y\,|\,\mu_{y|\phi},\Sigma_{y|\phi},a_{0},b_{0}\right),
\end{align}
where $\mathcal{NIG}\left(x\,|\,\mu,\Sigma,a,b\right)=\frac{\Gamma\left(\frac{p+2a}{2}\right)b^{a}}{\Gamma\left(a\right)\left(2\pi\right)^{\frac{p}{2}}\left|\Sigma\right|^{\frac{1}{2}}}\left[b+\frac{1}{2}(x-\mu)^{\top}\Sigma^{-1}(x-\mu)\right]^{-\frac{p+2a}{2}}.$
Finally, $p\left(y_{\mathsf{train}}\right)$ obtains from univariate
integration over the support of $\phi$,
\begin{align}
p\left(y_{\mathsf{train}}\right) & =\int_{-1}^{1}p\left(y_{\mathsf{train}}\,|\,\phi\right)p\left(\phi\right)\,\mathrm{d}\phi\\
 & =\int_{-1}^{1}\mathcal{NIG}\left(\mathring{S}y\,|\,\mu_{y|\phi},\Sigma_{y|\phi},a_{0},b_{0}\right)\mathcal{BE}_{(-1,1)}\left(\phi\,|\,c_0,d_0\right)\,\mathrm{d}\phi.\label{eq:univ-int}
\end{align}

To evaluate \eqref{eq:univ-int} we adopt two approaches: adaptive quadrature and Laplace approximation.
For quadrature we use adaptive quadrature. Laplace approximation requires the mode of the
joint density,
\begin{equation}
\hat{\phi}=\arg\max_{\phi\in\left(-1,1\right)}p\left(y_{\mathsf{train}}\,|\,\phi\right)p\left(\phi\right),
\end{equation}
which can be found by standard univariate optimization methods e.g.
BFGS. We then estimate
\begin{equation}
p\left(y_{\mathsf{train}}\right)=\sqrt{2\pi H\left(\hat{\phi}\right)}\exp\left(p\left(y_{\mathsf{train}},\hat{\phi}\right)\right),
\end{equation}
where the Hessian $H\left(\phi\right)=\nabla_{\phi}^{2}\log p\left(y_{\mathsf{train}},\phi\right)$.
The Laplace approximation seems very accurate in the particular
context of our experiments.

\subsection{Predictive density}

To evaluate CV estimators, we require the log predictive density 
$\log p\left(\tilde{y}\,|\,y\right)$. This quantity is also only
available numerically. Let $v=\sum_{i,j}\tilde{S}_{ij}$ denote the
number of observations in $\tilde{y}_{\mathsf{test}}$. Then,
\begin{align}
p\left(\tilde{y}_{\mathsf{test}}\,|\,y_{\mathsf{train}}\right) & =\int_{-1}^{1}\!\int_{0}^{\infty}\!\int_{\mathbb{R}^{q}}p\left(\tilde{y}_{\mathsf{test}}\,|\,\beta,\sigma^{2},\phi\right)\frac{p\left(\beta,\sigma^{2},\phi,y_{\mathsf{train}}\right)}{p\left(y_{\mathsf{train}}\right)}\,\mathrm{d}\beta\,\mathrm{d}\sigma^{2}\,\mathrm{d}\phi \\
 & =\frac{1}{p\left(y_{\mathsf{train}}\right)}\int_{-1}^{1}\!\int_{0}^{\infty}\!\int_{\mathbb{R}^{q}}p\left(\tilde{y}_{\mathsf{test}},y_{\mathsf{train}}\,|\,\beta,\sigma^{2},\phi\right)p\left(\beta|\sigma^{2}\right)\,\mathrm{d}\beta
 \nonumber \\
 & \qquad 
p\left(\sigma^{2}\right)\,\mathrm{d}\sigma^{2}\,p\left(\phi\right)\,\mathrm{d}\phi.
\end{align}

As in Section~\ref{subsec:full:post-pred}, we begin with the inner integral over
$\beta$. Marginalizing $\beta$ out of the joint distribution in
(\ref{eq:joint:train-test-param}) yields the joint density of test
and training data, conditional on $\sigma^{2}$ and $\phi$---
\begin{align}
p\left(\tilde{y}_{\mathsf{test}},y_{\mathsf{train}}\,|\,\sigma^{2},\phi\right) & =\int_{\mathbb{R}^{q}}p\left(\tilde{y}_{\mathsf{test}},y_{\mathsf{train}},\beta\,|\,\sigma^{2},\phi\right)\,\mathrm{d}\beta\\
 & =\mathcal{N}\left(\begin{pmatrix}\mathring{S}y\\
\tilde{S}\tilde{y}
\end{pmatrix}\,|\,\mu_{y\tilde{y}|\phi},\sigma^{2}\Sigma_{y\tilde{y}|\phi}\right).
\end{align}
The parameters are taken from the first two blocks of \eqref{eq:joint:train-test-param},
\begin{align}
\mu_{y\tilde{y}|\phi} & =\begin{pmatrix}\mathring{S}\\
\tilde{S}
\end{pmatrix}L_{\phi}^{-1}Z\mu_{0},\\
\Sigma_{y\tilde{y}|\phi} & =\begin{pmatrix}
\mathring{S}L_{\phi}^{-1}\left(I_T+Z\Sigma_0Z^{\top}\right)L_{\phi}^{-\top}\mathring{S}^{\top} & \mathring{S}L_{\phi}^{-1}Z\Sigma_0Z^{\top}L_{\phi}^{-\top}\tilde{S}^{\top}\\
\tilde{S}L_{\phi}^{-1}Z\Sigma_0Z^{\top}L_{\phi}^{-\top}\mathring{S}^{\top} & \tilde{S}L_{\phi}^{-1}\left(I_T+Z\Sigma_0Z^{\top}\right)L_{\phi}^{-\top}\tilde{S}^{\top}
\end{pmatrix}.
\end{align}
Computing the integral over $\sigma^{2}$ yields a normal-inverse-gamma
density,
\begin{align}
p\left(\tilde{y}_{\mathsf{test}},y_{\mathsf{train}}\,|\,\phi\right) & =\int_{0}^{\infty}p\left(\tilde{y}_{\mathsf{test}},y_{\mathsf{train}}\,|\,\sigma^{2},\phi\right)p\left(\sigma^{2}\right)\,\mathrm{d}\sigma^{2}\\
 & =\int_{0}^{\infty}\mathcal{N}\left(\begin{pmatrix}\mathring{S}y\\
\tilde{S}\tilde{y}
\end{pmatrix}\,|\,\mu_{y\tilde{y}|\phi},\sigma^{2}\Sigma_{y\tilde{y}|\phi}\right)\mathcal{IG}\left(\sigma^{2}\,|\,a_{0},b_{0}\right)\,\mathrm{d}\sigma^{2}\\
 & =\mathcal{NIG}\left(\begin{pmatrix}\mathring{S}y\\
\tilde{S}\tilde{y}
\end{pmatrix}\,|\,\mu_{y\tilde{y}|\phi},\Sigma_{y\tilde{y}|\phi},a_{0},b_{0}\right).
\end{align}
Finally, the predictive density can be evaluated as a one-dimensional
integral,
\begin{align}
p\left(\tilde{y}_{\mathsf{test}}\,|\,y_{\mathsf{train}}\right) & =\frac{1}{p\left(y_{\mathsf{train}}\right)}\int_{-1}^{1}p\left(\tilde{y}_{\mathsf{test}},y_{\mathsf{train}}\,|\,\phi\right)p\left(\phi\right)\,\mathrm{d}\phi\\
 & =\frac{1}{p\left(y_{\mathsf{train}}\right)}\int_{-1}^{1}\mathcal{NIG}\left(\begin{pmatrix}\mathring{S}y\\
\tilde{S}\tilde{y}
\end{pmatrix}\,|\,\mu_{y\tilde{y}|\phi},\Sigma_{y\tilde{y}|\phi},a_{0},b_{0}\right)\nonumber\\
& \qquad \mathcal{BE}_{(-1,1)}\left(\phi\,|\,c_{0},d_{0}\right)\,\mathrm{d}\phi.
\end{align}
Integration can be performed either by numerical quadrature or Laplace
approximation. In our experiments, we numerically stabilize quadrature by
first scaling the integrand by the reciprocal of the Laplace approximation,
and then dividing the resulting integral by this scaling factor. This ensures
the integrand to be evaluated is close in value to 1, which limits round-off error.

The Laplace approximation for $\log p\left(\tilde{y}_{\mathsf{test}}\,|\,y_{\mathsf{train}}\right)$
is given by
\begin{equation}
\log p\left(\tilde{y}_{\mathsf{test}}\,|\,y_{\mathsf{train}}\right)=\frac{1}{2}\log\left(2\pi\right)-H\left(\hat{\phi}\right)-\log p\left(\tilde{y}_{\mathsf{test}},\hat{\phi}\,|\,y_{\mathsf{train}}\right),
\end{equation}
where the mode $\hat{\phi}=\arg\max_{\phi\in\left(-1,1\right)}\log p\left(\tilde{y}_{\mathsf{test}},\phi\,|\,y_{\mathsf{train}}\right)$
and the Hessian $H\left(\phi\right)=\nabla_{\phi}^{2}\log p\left(\tilde{y}_{\mathsf{test}},\phi\,|\,y_{\mathsf{train}}\right)$.

\subsection{elpd estimates by MCMC}

Finally, in simulations when the DGP $p_{\mathrm{true}}$ is known, the $\mathrm{elpd}$
for a given model can be estimated by Monte Carlo integration using
independent draws $\left(\tilde{y}^{s}\right)_{s=1}^S$ from the DGP
and draws $\left(\theta^s\right)_{s=1}^S$ from the posterior.
\begin{align}
\mathrm{elpd\left(M_{\ell}\,|\,y\right)} & =\int p_{\mathrm{true}}\left(\tilde{y}\right)\log p\left(\tilde{y}\,|\,y,M_{\ell}\right)\,\mathrm{d}\tilde{y}\\
 & = \frac{1}{S}\sum_{s=1}^{S} \log p\left(\tilde{y}^{s}\,|\,\theta^s\right)+O_{p}\left(\sqrt{1/S}\right).
\end{align}
For the theoretical full-data elpd, this approach requires only one
MCMC run per observed data vector $y$, and is therefore fast
to compute.

\section{Additional results and proofs\label{app:Proofs}}

The proof of Proposition~\ref{prop:sarx:elpd:gchisq} simply summarizes
the derivations in Section~\ref{sec:sarx}.
\begin{proof}[Proof of Proposition~\ref{prop:sarx:elpd:gchisq}]
We have shown that the following utility measures have quadratic
polynomial forms:
\begin{itemize}
\item $\eljpd\left(M_{\ell}\,|\,y,\tilde{S}\right)=y^{\top}A_{\ell}y+y^{\top}b_{\ell}+c_{\ell}$,
where $A_{\ell}$, $b_{\ell}$, and $c_{\ell}$ are defined in Section~\ref{subsec:eljpd}.
Since $\tilde{S}$ is arbitrary, this form holds when $\tilde{S}=I_{T}$.
\item $\elppd\left(M_{\ell}|y\right)=y^{\top}\bar{A}_{\ell}y+y^{\top}\bar{b}_{\ell}+\bar{c}_{\ell}$,
where $\bar{A}_{\ell}$, $\bar{b}_{\ell}$, and $\bar{c}_{\ell}$ are defined
in Section~\ref{subsec:elppd}.
\end{itemize}
And similarly, we have shown that for any CV scheme that can be described
by the triple $\left(K,\left\{ \mathsf{test}_{k}\right\} _{k=1}^{K},\left\{ \mathsf{train}_{k}\right\} _{k=1}^{K}\right)$,
the distribution of the CV estimator also has a quadratic polynomial
form. Specifically,
\begin{itemize}
\item $\widehat{\eljpd}_{CV}\left(M_{\ell}\,|\,y\right)=y^{\top}\widehat{A_{\ell}}y+y^{\top}\widehat{b_{\ell}}+\widehat{c_{\ell}},$
where $\widehat{A_{\ell}}$, $b_{\ell}$, and $\widehat{c_{\ell}}$ are defined
in Section~\ref{subsec:CV-eljd}, and
\item $\widehat{\elppd}_{CV}\left(M_{\ell}\,|\,y\right)=y^{\top}\widetilde{A_{\ell}}y+y^{\top}\widetilde{b_{\ell}}+\widetilde{c_{\ell}},$
for $\widetilde{A_{\ell}}$, $\widetilde{b_{\ell}}$, and $\widetilde{c_{\ell}}$
defined in Section~\ref{subsec:elppd-cv}.
\end{itemize}
From the form of these expressions it can be verified that each are
functions of $\phi^{*},\phi^{\left(\ell\right)},\sigma_{*}^{2},\sigma_{\ell}^{2},Z_{*}$
and $Z_{\ell}$ as well as the CV scheme parameters.
\end{proof}
The proof of Corollary~\ref{cor:sarx:elpd:elpd:quadratic} follows
from the closure of polynomial vector spaces under subtraction.
\begin{proof}[Proof of Corollary~\ref{cor:sarx:elpd:elpd:quadratic}]
Let $\omega_{A-B}\left(y\right)=\omega_{A}\left(y\right)-\omega_{B}\left(y\right)$
denote a model selection objective or CV statistic for comparing model
$M_{A}$ with $M_B$. Since by Proposition~\ref{prop:sarx:elpd:gchisq}
$\omega_A$ and $\omega_B$ are second-degree polynomials,
we have
\[
\omega_{A-B}\left(y\right)=\omega_{A}\left(y\right)-\omega_{B}\left(y\right)=y^{\top}\left(A_{A}-A_{B}\right)y+y^{\top}\left(b_{A}-b_{B}\right)+\left(c_{A}-c_{B}\right).
\]
\end{proof}
\begin{proof}[Proof of Proposition~\ref{prop:dist-omega}]
Since $A=A^{\top}$, we can follow \citet[Appendix D.5]{Sivula2020a}
and directly apply Theorem~3.2b.3 of \citet[3.2b.3]{Mathai1992}.
Substituting $\mu_{*}=m_{*}$, and $\Sigma_{*}=\sigma_{*}^{2}V_{*}$, we have
\begin{align}
\mathbb{E}\left[\omega\left(y\right)\right] & =\mathrm{tr}\!\left(\sigma_{*}V_{*}^{1/2}A\sigma_{*}V_{*}^{1/2}\right)+c+b^{\top}m_{*}+m_{*}^{\top}Am_{*}\\
 & =\sigma_{*}^{2}\mathrm{tr}\!\left(AV_{*}\right)+c+b^{\top}m_{*}+m_{*}^{\top}Am_{*}\\
\mathrm{var}\!\left(\omega\left(y\right)\right) & =2\mathrm{tr}\!\left(\left[\sigma_{*}V_{*}^{1/2}A\sigma_{*}V_{*}^{1/2}\right]^{2}\right)+b^{\top}\left(\sigma^{2}V_{*}\right)b+4b^{\top}\left(\sigma^{2}V_{*}\right)Am_{*}\nonumber\\
 & \qquad\qquad+4m_{*}^{\top}A\left(\sigma_{*}^{2}V_{*}\right)Am_{*}\\
 & =2\sigma_{*}^{4}\mathrm{tr}\!\left(A^{2}V_{*}^{2}\right)+\sigma_{*}^{2}b^{\top}V_{*}b+4\sigma_{*}^{2}b^{\top}V_{*}Am_{*}+4\sigma_{*}^{2}AV_{*}Am_{*}
\end{align}
where we have used that $\mathrm{tr}\!\left(CD\right)=\mathrm{tr}\!\left(DC\right)$.

Now we must show that the quadratic form satisfies Definition~\ref{def:gchisq}.
First, apply the transformation $\varepsilon=\sigma_{*}^{-1}QL_{\phi_{*}}\left(y-m_{*}\right)$,
which is a linear transformation of a jointly Gaussian variable. So
$\varepsilon$ is jointly Gaussian with distribution
\begin{align}
\varepsilon & \sim\sigma_{*}^{-1}QL_{\phi_{*}}\mathcal{N}\left(m_{*},\sigma_{*}^{2}L_{\phi_{*}}^{-1}L_{\phi_{*}}^{-\top}\right)-\sigma_{*}^{-1}QL_{\phi_{*}}m_{*}\\
 & \sim\mathcal{N}\left(\sigma_{*}^{-1}QL_{\phi_{*}}m_{*}-\sigma_{*}^{-1}QL_{\phi_{*}}m_{*},\sigma_{*}^{2}\left(\sigma_{*}^{-1}QL_{\phi_{*}}\right)L_{\phi_{*}}^{-1}L_{\phi_{*}}^{-\top}\left(\sigma_{*}^{-1}QL_{\phi_{*}}\right)^{\top}\right)\\
 & \sim\mathcal{N}\left(0,I_T\right),
\end{align}
that is, a vector of independent standard normal variables. Applying the
inverse transformation by substituting $y=m_{*}+\sigma_{*}L_{\phi_{*}}^{-1}Q^{-1}\varepsilon$,
we have
\begin{align}
\omega\left(y\right) & =\left(m_{*}+\sigma_{*}L_{\phi_{*}}^{-1}Q^{-1}\varepsilon\right)^{\top}A\left(m_{*}+\sigma_{*}L_{\phi_{*}}^{-1}Q^{-1}\varepsilon\right)+\left(m_{*}+\sigma_{*}L_{\phi_{*}}^{-1}Q^{-1}\varepsilon\right)^{\top}b+c\\
 & =\varepsilon^{\top}\sigma_{*}^{2}QL_{\phi_{*}}^{-\top}AL_{\phi_{*}}^{-1}Q^{-1}\varepsilon+\varepsilon^{\top}\underbrace{QL_{\phi_{*}}^{-\top}\left(2\sigma_{*}^{2}Am_{*}+\sigma_{*}b\right)}_{\tilde{b}}\nonumber\\
 & \qquad+\underbrace{m_{*}^{\top}Am_{*}+\sigma_{*}m_{*}^{\top}b+c}_{\tilde{c}}\\
 & =\varepsilon^{\top}\Lambda\varepsilon+\varepsilon^{\top}\tilde{b}+\tilde{c},
\end{align}
where we have used that $\Lambda=Q^{-1}L_{\phi_{*}}^{-1}AL_{\phi_{*}}^{-\top}Q$,
which is diagonal and hence symmetric. We now complete the square
for the $k$ nonzero eigenvalues,
\begin{equation}
\omega\left(y\right)=\sum_{j=1}^{k}\lambda_{j}\underbrace{\left(\varepsilon_{j}+\frac{\tilde{b_j}}{2\lambda_{j}}\right)^{2}}_{\chi^{2}\left(1,\frac{\tilde{b}_j^{2}}{2^2\lambda_{j}^{2}}\right)}+\underbrace{\sum_{j=k+1}^{T}\tilde{b}_{j}\varepsilon_{j}+\tilde{c}-\frac{1}{4}\sum_{j=1}^{k}\frac{\tilde{b}_{j}^{2}}{\lambda_{j}^{2}}}_{\mathcal{N}\left(\mu,\sigma^{2}\right)},
\end{equation}
where we implicitly assumed the diagonal of $\Lambda$ orders the
zero eigenvalues last. Because the $\varepsilon_{j}$ are independent
standard normal variables, the $\left(\varepsilon_{j}+\frac{\tilde{b}_j}{2\lambda_{j}}\right)^{2}$
terms are independent noncentral $\chi^{2}\left(1,\frac{\tilde{b}_j^{2}}{2^2\lambda_{j}^{2}}\right)$
variables. The normally-distributed remainder has mean
$\mu=\tilde{c}-\frac{1}{4}\sum_{j=1}^{k}\tilde{b}_{j}^{2}\lambda_{j}^{-2}$,
and variance $\sigma^{2}=\sum_{j=k+1}^{T}\tilde{b}_{j}^{2}$, as required.
\end{proof}
\begin{proof}[Proof of Lemma~\ref{lem:genchisq}]
To derive the CDF of $\omega\left(y\right)$, we begin with its characteristic
function $\psi\left(t\right)$. Recall that the characteristic function
of a $\chi^{2}\left(r,\delta^{2}\right)$ variable is $\psi_{\chi^{2}\left(r,\delta^{2}\right)}\left(t\right)=\exp\left(\frac{i\delta^{2}t}{1-2it}\right)\left(1-2it\right)^{-r/2}$
and that of a $\mathcal{N}\left(\mu,\sigma^{2}\right)$ variable is
$\psi_{\mathcal{N}\left(\mu,\sigma^{2}\right)}\left(t\right)=\exp\left(i\mu t-\frac{\sigma^{2}t^{2}}{2}\right)$.
By independence of the random variables in (\ref{eq:poly-sum-indep})
and \citet[eq 26.12]{Billingsley2008}, the characteristic function
of $\omega\left(y\right)$ is 
\begin{align}
\psi_{\omega\left(y\right)}\left(t\right) & =\psi_{\mathcal{N}\left(\mu,\sigma^{2}\right)}\left(t\right)\prod_{j=1}^{k}\psi_{\chi^{2}\left(r_j,\delta_j^2\right)}\left(\lambda_{j}t\right)\\
 & =\frac{\exp\left\{ i\mu t-\frac{\sigma^{2}t^{2}}{2}+it\sum_{j=1}^{k}\frac{\lambda_{j}^{2}}{1-2i\lambda_{j}t}\right\} }{\prod_{j=1}^{k}\left(1-2i\lambda_{j}t\right)^{r_j/2}}.
\end{align}
The CDF obtains from the inversion theorem \citep{P.J.1961},
\begin{align}
F_{\omega\left(y\right)}\left(w\right) & =Pr\left(\omega\left(y\right)<w\right)\\
 & =\frac{1}{2}-\int_{\mathbb{R}}\frac{\exp\left(-itw\right)}{-i\pi t}\psi_{\omega\left(y\right)}\left(t\right)\,\mathrm{d}t\\
 & =\frac{1}{2}+\int_{\mathbb{R}}\frac{\exp\left\{ i\mu t-iwt-\sigma^{2}t^{2}/2+it\sum_{j=1}^{k}\frac{\lambda_j\delta_j^2}{1-2i\lambda_{j}t}\right\} }{\pi it\prod_{j=1}^{k}\left(1-2i\lambda_{j}t\right)^{r_j/2}}\,\mathrm{d}t.
\end{align}

~
\end{proof}
The following integral is useful for evaluating the theoretical utility
measures $\mathrm{elppd}$ and $\mathrm{elpjd}$.
\begin{lem}
\label{lem:mvn:cross-entropy}Let $x$ be a vector and define the
densities $p\left(x\right)=\mathcal{N}\left(x\,|\,\mu_{p},\Sigma_{p}\right)$
and $q\left(x\right)=\mathcal{N}\left(x\,|\,\mu_{q},\Sigma_{q}\right)$.
Then
\begin{equation}
\int p(x)\log q\left(x\right)\mathrm{d}x=-\frac{1}{2}\left[\log\left|2\pi\Sigma_{q}\right|+\mathrm{tr}\left(\Sigma_{q}^{-1}\Sigma_{p}\right)+\left(\mu_{q}-\mu_{p}\right)^{\top}\Sigma_{q}^{-1}\left(\mu_{q}-\mu_{p}\right)\right].
\end{equation}
\end{lem}

\begin{proof}
We will used the well-known formulas for the Kullback--Leibler divergence
of two normal densities \citep{Duchi2007}
\begin{align}
\mathbb{D}\left(p\,\|\,q\right) & =\int p\left(x\right)\log\frac{p\left(x\right)}{q\left(x\right)}\,\mathrm{d}x\\
 & =\frac{1}{2}\left\{ \mathrm{tr}\left(\Sigma_{q}^{-1}\Sigma_{p}\right)+\left(\mu_{q}-\mu_{p}\right)^{\top}\Sigma_{q}^{-1}\left(\mu_{q}-\mu_{p}\right)-k+\log\frac{\left|\Sigma_{q}\right|}{\left|\Sigma_{p}\right|}\right\} \label{eq:mvn-kld}
\end{align}
and the entropy of the multivariate normal distribution \citep{Ahmed1989}
\begin{align}
\mathbb{H}\left(p\right) & = -\int p\left(x\right)\log p\left(x\right)\,\mathrm{d}x=\frac{1}{2}\log\left|2\pi e\Sigma_{p}\right|.
\end{align}
Combining the above,

\begin{align}
\int p(x)\log q\left(x\right)\,\mathrm{d}x & =-\mathbb{H}\left(p\right)-\mathbb{D}\left(p\,||\,q\right)\\
 & =-\frac{1}{2}\log\left|2\pi e\Sigma_{p}\right|-\frac{1}{2}\left\{ \mathrm{tr}\left(\Sigma_{q}^{-1}\Sigma_{p}\right)+\log\left|\Sigma_{q}\right|-\log\left|\Sigma_{p}\right|\right.\nonumber \\
 & \qquad\qquad\left.-k+\left(\mu_{q}-\mu_{p}\right)^{\top}\Sigma_{q}^{-1}\left(\mu_{q}-\mu_{p}\right)\right\} .
\end{align}
Rearranging this expression yields the stated result.
\end{proof}
\begin{lem}[Expected KL divergence between predictive and DGP]
\label{lem:sarx:ekld}Let $p_{\mathrm{true}}\left(\tilde{y}\right)=\mathcal{N}\left(\tilde{y}\,|\,m_{*},\sigma_{*}^{2}V_{*}\right)$
be the DGP and let model $M_{\ell}$ be a simplified $\ARX\left(p_{\ell},q_{\ell}\right)$
with fixed $\phi_{\ell}$ and $\sigma_{\ell}^{2}$, and predictive $p\left(\tilde{y}|y,M_{\ell}\right)=\mathcal{N}\left(\tilde{y}|m_{\ell},\sigma_{\ell}^{2}V_{\ell}\right)$.
Then the expected KL divergence between the model $M_{\ell}$ predictive
and the DGP is given by
\begin{align}
\mathbb{E}_{y}\left[\mathbb{D}\left(p_{\mathrm{true}}\,\|\,p\left(\cdot\,|\,y,M_{\ell}\right)\right)\right] & =\frac{1}{2}\log\frac{\left|\sigma_{\ell}^{2}V_{\ell}\right|}{\left|e\sigma_{*}^{2}V_{*}\right|}+\frac{\sigma_{*}^{2}}{2\sigma_{\ell}^{2}}\mathrm{tr}\left(V_{\ell}^{-1}\left(V_{*}+D_{\ell}V_{*}D_{\ell}^{\top}\right)\right)\nonumber\\
 & \quad+\frac{1}{2\sigma_{\ell}^{2}}\left[\left(D_{\ell}-I_{T}\right)m_{*}+e_{\ell}\right]^{\top}V_{\ell}^{-1}\left[\left(D_{\ell}-I_{T}\right)m_{*}+e_{\ell}\right],
\end{align}
where $D_{\ell}$ and $e_{\ell}$ are respectively defined in (\ref{eq:sarx:eljpd:D})
and (\ref{eq:sarx:eljpd:e}).
\end{lem}

\begin{proof}
Substitute the expression for the KL divergence in (\ref{eq:mvn-kld})
and note that all but the $M_{\ell}$ terms in the quadratic form are
free of $y$,

\begin{align}
 & \mathbb{E}_{y}\left[\mathbb{D}\left(p_{\mathrm{true}}\,\|\,p\left(\cdot\,|\,y,M_{\ell}\right)\right)\right]\nonumber \\
 & =\frac{1}{2}\mathbb{E}_{y}\left[\mathrm{tr}\left(\left(\sigma_{\ell}^{2}V_{\ell}\right)^{-1}\sigma_{*}^{2}V_{*}\right)-T+\log\frac{\left|\sigma_{\ell}^{2}V_{\ell}\right|}{\left|\sigma_{*}^{2}V_{*}\right|}+\left(m_{\ell}-m_{*}\right)^{\top}V_{\ell}^{-1}\left(m_{\ell}-m_{*}\right)\right] \\
 & =\frac{\sigma_{*}^{2}}{2\sigma_{\ell}^{2}}\mathrm{tr}\left(V_{\ell}^{-1}V_{*}\right)+\frac{1}{2}\log\frac{\left|\sigma_{\ell}^{2}V_{\ell}\right|}{\left|e\sigma_{*}^{2}V_{*}\right|}+\frac{1}{2\sigma_{\ell}^{2}}\mathbb{E}_{y}\left[\left(m_{\ell}-m_{*}\right)^{\top}V_{\ell}^{-1}\left(m_{\ell}-m_{*}\right)\right].\label{eq:lem:sarx:ekld:proof:quad}
\end{align}
We now just need to evaluate the expectation of the random quadratic
form. Recall from (\ref{eq:eljpd:simpler-term}) that $m_{\ell}=D_{\ell}y+e_{\ell}$,
where $D_{\ell}$ and $e_{\ell}$ are respectively defined in (\ref{eq:sarx:eljpd:D})
and (\ref{eq:sarx:eljpd:e}). Since by (\ref{eq:sarx:ptrue}) $y$
is distributed as $y\sim\mathcal{N}\left(y\,|\,m_{*},\sigma_{*}^{2}V_{*}\right)$,
the difference $\left(m_{\ell}-m_{*}\right)$ follows the law
\begin{align}
\left(M_{\ell}-m_{*}\right) & \sim\mathcal{N}\left\{ \left(D_{\ell}-I_{T}\right)m_{*}+e_{\ell},\sigma_{*}^{2}D_{\ell}V_{*}D_{\ell}^{\top}\right\} .
\end{align}
Applying Proposition~\ref{prop:dist-omega} with $A=\sigma_{*}^{2}D_{\ell}V_{*}D_{\ell}^{\top}$,
$b=0$, and $c=0$, the expectation of the quadratic form in (\ref{eq:lem:sarx:ekld:proof:quad})
evaluates to
\begin{equation}
\mathrm{\sigma_{*}^{2}tr}\left[V_{\ell}^{-1}D_{\ell}V_{*}D_{\ell}^{\top}\right]+\left(\left(D_{\ell}-I_{T}\right)m_{*}+e_{\ell}\right)^{\top}V_{\ell}^{-1}\left(\left(D_{\ell}-I_{T}\right)m_{*}+e_{\ell}\right).
\end{equation}
Substituting this value into (\ref{eq:lem:sarx:ekld:proof:quad})
and rearranging yields the desired expression.
\end{proof}

\end{document}